\documentclass[a4paper,fleqn,usenatbib]{mnras}

\usepackage[T1]{fontenc}
\usepackage{ae,aecompl}

\usepackage{graphicx}	
\usepackage{amsmath}	
\usepackage{amssymb}	
\usepackage{lipsum}
\usepackage{pdflscape}  
\usepackage[dvipsnames]{color}
\usepackage{longtable}

\newcommand{\mrm}[1]{\mathrm{#1}}	
\newcommand{\vect}[1]{\mathit{\mathbf{#1}}}	
\newcommand{\DRSun}[1]{$\mrm{DR_{\odot}}$} 
\newcommand{\ERSun}[1]{$\mrm{ER_{\odot}}$} 
\newcommand{\MFSun}[1]{$\mrm{MF_{\odot}}$} 
\newcommand{\ellsum}[1]{\ell_{\Sigma}} 
\newcommand{\lmod}[1]{$\ell_{\Sigma}$-modes}

\title[Solar cycle identification with ZDI]{Identifying solar-like magnetic cycles with Zeeman-Doppler-Imaging (ZDI)}

\author[L. T. Lehmann et al.]{
L. T. Lehmann,$^{1,2}$\thanks{E-mail: lisa.lehmann@irap.omp.eu}
G. A. J. Hussain,$^{2,3,4}$
A. A. Vidotto,$^{5}$
M. M. Jardine,$^{1}$
D. H. Mackay$^{6}$
\\
$^{1}$SUPA, School of Physics and Astronomy, University of St Andrews, St Andrews KY16 9SS, UK \\
$^{2}$Institut de Recherche en Astrophysique et Plan\'etologie, Universit\'e de Toulouse, UPS-OMP, 31400 Toulouse, France\\
$^{3}$European Southern Observatory, Karl-Schwarzschild-Str. 2, 85748 Garching bei M\"unchen, Germany\\
$^{4}$Science Division, Directorate of Science, European Space Research  and  Technology  Centre  (ESA/ESTEC),  Keplerlaan  1, 2201AZ, \\ Noordwijk, The Netherlands\\
$^{5}$School of Physics, Trinity College Dublin, The University of Dublin, Dublin-2, Ireland\\
$^{6}$School of Mathematics and Statistics, University of St Andrews, St Andrews KY16 9SS, UK\\
}

\date{Accepted XXX. Received YYY; in original form ZZZ}

\pubyear{2018}

\begin{document}
\label{firstpage}
\pagerange{\pageref{firstpage}--\pageref{lastpage}}
\maketitle

\begin{abstract}
We are reaching the point where spectropolarimetric surveys have run for long enough to reveal solar-like magnetic activity cycles. In this paper we investigate what would be the best strategy to identify solar-like magnetic cycles and ask which large-scale magnetic field parameters best follow a solar-type magnetic cycle and are observable with the Zeeman-Doppler-Imaging (ZDI) technique.
We approach these questions using the 3D non-potential flux transport simulations of \cite{Yeates2012} modelling the solar vector magnetic field over 15 years (centred on solar cycle 23). The flux emergence profile was extracted from solar synoptic maps and used as input for a photospheric flux transport model in combination with a non-potential coronal evolution model. We synthesise spectropolarimetric data from the simulated maps and reconstruct them using ZDI. The ZDI observed solar cycle is set into the context of other cool star observations and we present observable trends of the magnetic field topology with time, sunspot number and S-index.
We find that the axisymmetric energy fraction is the best parameter of the ZDI detectable large-scale field to trace solar-like cycles. Neither the surface averaged large-scale field or the total magnetic energy is appropriate. ZDI seems also to be able to recover the increase of the toroidal energy with S-index. We see further that ZDI might unveil hints of the dynamo modes that are operating and of the global properties of the small-scale flux emergence like active latitudes.
\end{abstract}




\section{Introduction}

It is well-known that the solar activity shows cyclic behaviour. At the solar photosphere, the most obvious signature of the activity is the cyclic evolution of the sunspots. Sunspots are foot points of strong magnetic flux tubes breaking through the solar surface, that appear over a range of mid latitude regions. The most prominent cycle at the Sun is the 11-yr Schwabe cycle, \citep{Schwabe1849}. The sunspot number (SSN) increases until the activity maximum is reached, while the latitude of spot emergence decreases. At the activity maximum, the solar dipolar field is near zero and switches polarity and strengthens as the SSN decreases again. During the SSN minimum, the solar dipolar field is strongest. To see the same polarity at the solar poles one needs to wait for two Schwabe activity cycles, so that the solar magnetic cycle is $\approx$22\,years long.

Stars other than the Sun also show activity cycles. The Zeeman-Doppler-Imaging technique (ZDI, \citealt{Semel1989, DonatiBrown1997}) is capable of mapping the large-scale magnetic fields of stars and unveiling global magnetic field reversals that are in phase with the chromospheric activity cycle in a way that is similar to our Sun for three stars: the F7~dwarf $\tau$~Boo \citep{Jeffers2018} (age = $0.9\pm0.5$\,Gyr, \citet{Borsa2015}), the K~dwarf 61~Cyg~A \citep{BoroSaikia2016,BoroSaikia2018} (age = 6\,Gyr, \citet{Kervella2008}) and $\iota$~Hor (Alvarado-Gomez et al., in prep.) (age = 0.88\,Gyr, \citet{Stanke2006}). Their magnetic cycle periods are shorter than the solar 22-yr magnetic cycle with $\approx 240$~days and $\approx 14$~years for $\tau$~Boo and 61~Cyg~A, respectively.
In particular, the slightly older and cooler K dwarf 61~Cyg~A shows a solar-like behaviour in its dipolar field being strongest at activity minimum and weakest at activity maximum \citep{BoroSaikia2018}. \cite{BoroSaikia2016} showed also that the axisymmetry is in anti-phase with the chromospheric activity and the magnetic field topology becomes more complex during the activity maximum.
In general, \cite{See2016} showed that stars having high temporal variations have significant toroidal fields otherwise their large-scale field is predominately poloidal like our Sun. 

We are starting to observe small scale star spot distributions using exoplanet transits as tracers across the stellar surface, e.g. \cite{Morris2017}. \cite{Netto2020} were able to produce a butterfly similar diagram of the spot distribution using this data. \cite{Yu2019}  analysed the low latitude magnetic and brightness spot distribution using Doppler Imaging and ZDI of the T~Tauri star V410 Tau over many years and discovered that the latitude of spot emergence increases over the 8~yr of observations.

This paper aims to examine if a solar-like cycle can be observed with ZDI, given that this technique only detects the large-scale field for very slow rotating and low active stars like our Sun. We aim to determine the best strategy to detect such a cycle, and which parameters are most sensitive to cycle changes. A simple approach is to look for polarity reversals in the reconstructed magnetic field maps. However, a solar-like cycle is not only characterised by the polarity reversals. It has many other characteristic signatures like the cyclic variation of certain magnetic field parameters, e.g.\ the axisymmetric field, so that a solar-like cycle can only be identified by analysing several magnetic field parameters and activity tracers.
Observing slowly rotating sun-like stars is still challenging and magnetic field maps can be misleading, e.g., it can be hard to accurately recover the field distributions for stars that are seen more pole-on, see \cite{Lehmann2019}. We present a more reliable method for detecting solar-like cycles using a parameter-based analysis of the magnetic field topology.

The ZDI technique is limited among others by the stellar rotation rate given by the projected equatorial velocity $v_e \sin i$, which determines the size of the resolution elements, see \citet[Eq.~3]{Morin2010}. Within the resolution elements the magnetic fields of different polarities cancel each other. ZDI can therefore only detect the magnetic field topologies on length scales larger than the resolution elements. 

In \cite{Lehmann2019} we showed that ZDI can recover the large-scale magnetic field topology accurately for sun-like stars with the exception of magnetic field strengths, typically mapping magnetic energies that are one order of magnitude lower than input. ZDI recovers the ratio of the toroidal:poloidal field well, in stars with moderate to high inclination angles. For more pole-on inclination angles the poloidal energy is underestimated and the level of axisymmetry tends to be overestimated. Understanding the properties of the large-scale field recovered in ZDI maps is essential to interpreting the observed magnetic maps correctly. In \cite{Lehmann2019} we showed that ZDI can also recover hints of the solar active latitudes in the toroidal quadrupolar mode.

This particular study focuses on solar-age, slowly rotating solar-type stars which have long rotation periods and therefore low $v_e \sin i$ values. The solar $v_e \sin i$ is only $1-2\,\mrm{km\ s^{-1}}$, depending on the line-of-sight inclination $i$. As noted in \cite{Lehmann2019}, rapidly rotating  younger stars will have correspondingly large $v_e \sin i$ values. The greater number of Doppler-shifted contributions across the stellar profile in a rapid rotator corresponds to an increasing amount of spatial information encoded in the star's spectral line profiles. It is also worth noting that the rotationally broadened absorption line profiles become increasingly shallow in rapid rotators and Doppler imaging campaigns need to be planned so that they can obtain sufficiently high SNR to detect structure robustly, while at the same time minimising the amount of phase blurring due to rapid rotation. This is out of the scope of the current paper, which focuses on our ability to recover solar-type activity cycles in stars that are of similar age and therefore similar rotation and activity levels as the Sun.

To find the best strategy for detecting solar-like cycles, we analyse 118 simulated surface vector magnetic field maps based on solar observations covering the years 1996--2011. The simulations include SC23 and the first part of SC24. The simulations were presented in \cite{Yeates2012}. They modelled the solar magnetic field from 1996--2011 using a non-potential flux transport model, using the observed properties of the emerging flux extracted from solar synoptic maps observed by the National Solar Observatory (NSO), Kitt Peak (KP). The simulated vector magnetic field maps reach from the solar surface up to the corona and were used to analyse the chirality of the high-latitude filaments over SC23. 
We reconstruct the magnetic field for 41 out of the 118 maps using the ZDI technique \citep{Hussain2016}. We compare the magnetic field topology of the simulations and their reconstructions and set them further in context to the observed magnetic field topology presented by \cite{See2016}.

The paper is organised as follows: Section~\ref{Sec:SimAndTech} introduces the simulations and applied techniques, Section~\ref{Sec:Results} presents the results and Section~\ref{Sec:ConAndSum} closes with a conclusion and summary.

\section{Simulations and Techniques}
\label{Sec:SimAndTech}

We apply the Zeeman-Doppler-Imaging technique to synthetic polarimetric spectra modulated from non-potential flux transport simulations of the Sun.

\subsection{Extracting the magnetic field topology for different length scales}
\label{SubSec:MagneticFieldDescription}

The spatial resolution of the maps obtained from ZDI, i.e. the observed maps, is  determined by the   equatorial projected velocity $v_e \sin i$ of the star. This prevents a direct comparison with the much higher-resolution simulated input maps that we use. A fair order-of-magnitude comparison can be achieved by decomposing the magnetic field into its constituent spherical harmonics \citep{Vidotto2016,Lehmann2018,Lehmann2019}. A specific length scale can then be chosen by the corresponding spherical harmonic mode $\ell$, while the angular length scale can be approximately described by $\theta \approx 180^{\circ}/\ell$. For example the large-scale field can be selected by filtering the low-order spherical harmonic modes, e.g. $\ell \leq 5$ or $\ell \leq 10$  (see e.g. \citealt{Morin2010}, \citealt{Johnstone2014}, \citealt{Yadav2015}, \citealt{Folsom2016}, \citealt{Lehmann2017}).

The vector magnetic field $\vect{B}$ can be described by the following equations for the radial, azimuthal and meridional field:
\begin{align}
B_{r}(\phi, \theta) &= \sum_{\ell m} \alpha_{\ell m} P_{\ell m} e^{im\phi}, \label{Eq:B_rad} \\
B_{\phi}(\phi, \theta) &= - \sum_{\ell m} \beta_{\ell m} \frac{im P_{\ell m} e^{im\phi}}{(\ell + 1) \sin \theta} \nonumber \\ &+ \sum_{\ell m} \gamma_{\ell m} \frac{1}{\ell+1} \frac{\mathrm{d}P_{\ell m}}{\mathrm{d}\theta} e^{im\phi}, \label{Eq:B_azi}\\
B_{\theta}(\phi, \theta) &= \sum_{\ell m} \beta_{\ell m} \frac{1}{\ell+1} \frac{\mathrm{d}P_{\ell m}}{\mathrm{d}\theta} e^{im\phi} \nonumber \\ &+ \sum_{\ell m} \gamma_{\ell m} \frac{im P_{\ell m} e^{im\phi}}{(\ell + 1) \sin \theta}, \label{Eq:B_mer}
\end{align}
so that $(B_{r}, B_{\phi}, B_{\theta}) = \vect{B}$, where $P_{\ell m} \equiv c_{\ell m}P_{\ell m}(\cos \theta)$ are the associated Legendre polynomial of mode $\ell$ and order $m$ and  $c_{\ell m}$ is a normalization constant:
\begin{equation}
c_{\ell m} = \sqrt{\frac{2\ell+1}{4\pi}\frac{(\ell - m)!}{(\ell + m)!}}.
\end{equation}
The coefficients $\alpha_{\ell m}, \beta_{\ell m}$ and $\gamma_{\ell m}$ can characterise any given magnetic field topology without a priori assumptions \citep[see][]{Donati2006a}.
This description follows \cite{Elsasser1946} and \citet[Appendix III]{Chandrasekhar1961} but applying a left-handed coordinate system rather than the right-handed coordinate system of the original literature.
For the left-handed coordinate system the radial field component $B_r$ points towards outwards, the azimuthal component $B_\phi$ runs in the clockwise direction as viewed from North pole and the meridional field $B_\theta$ runs with colatitude (from north to south)\footnote{For a right-handed coordinate system the azimuthal component $B_\phi$ runs in the anti-clockwise direction as viewed from the North Pole. The radial and meridional component remain the same.} We changed to the left-handed coordinate system to be consistent with the previous publications related to this work \citep{Vidotto2016,Lehmann2019}.

Next to the spherical coordinate system ($r, \phi, \theta$)  the toroidal and poloidal components are also often used:
\begin{align}
B_{r,\mathrm{pol}}(\phi, \theta)  &= \sum_{\ell m} \alpha_{\ell m} P_{\ell m} e^{im\phi}, \nonumber \\
B_{\phi,\mathrm{pol}}(\phi, \theta)  &= - \sum_{\ell m} \beta_{\ell m} \frac{im P_{\ell m} e^{im\phi}}{(\ell + 1) \sin \theta}, \nonumber \\ 
B_{\theta,\mathrm{pol}}(\phi, \theta)  &=\sum_{\ell m} \beta_{\ell m} \frac{1}{\ell+1} \frac{\mathrm{d}P_{\ell m}}{\mathrm{d}\theta} e^{im\phi} , \label{Eq:B_pol}
\end{align}
\begin{align}
B_{r,\mathrm{tor}}(\phi, \theta) &= 0, \nonumber \\
B_{\phi,\mathrm{tor}}(\phi, \theta) &= \sum_{\ell m} \gamma_{\ell m} \frac{1}{\ell+1} \frac{\mathrm{d}P_{\ell m}}{\mathrm{d}\theta} e^{im\phi}, \nonumber \\
B_{\theta,\mathrm{tor}}(\phi, \theta) &= \sum_{\ell m} \gamma_{\ell m} \frac{im P_{\ell m} e^{im\phi}}{(\ell + 1) \sin \theta}, \label{Eq:B_tor}
\end{align}
so that $(B_{r,\rm{pol}}, B_{\phi,\rm{pol}}, B_{\theta,\rm{pol}}) = \vect{B_{\rm{pol}}}$, $(B_{r,\rm{tor}}, B_{\phi,\rm{tor}}, B_{\theta,\rm{tor}}) = \vect{B_{\rm{tor}}}$ and $(\vect{B_{\mrm{pol}}}, \vect{B_{\mrm{tor}}}) = \vect{B}$. The two coefficients $\alpha_{\ell m}$ and $\beta_{\ell m}$ describe the poloidal field, while $\gamma_{\ell m}$ characterises the toroidal field. The summation is carried out  from $1\leq \ell \leq \ell_{\mathrm{max}}$ and from $\vert m \vert \leq \ell$, where $\ell_{\mathrm{max}}$ indicates the selected maximum mode of the spherical harmonic decomposition corresponding to the smallest included length scale. 

For the analysis of the magnetic field topology we often determine the mean-squared flux density over the stellar surface by integrating over solid angle where ($\mrm{d}\Omega = \sin \theta \mrm{d}\phi\mrm{d}\theta$). We do this  for single magnetic field components or the total field,
\begin{align}
\langle B^2_k \rangle &= \frac{1}{4\pi} \int B^2_k(\phi,\theta) \mrm{d}\Omega,\ \ k \in (r,\phi,\theta)\ \mrm{or}\ k \in (\mrm{pol, tor}),\\
\langle B^2_{\mrm{tot}} \rangle &= \frac{1}{4\pi} \int \sum_k B^2_k(\phi,\theta) \mrm{d}\Omega,\ \ k \in (r,\phi,\theta),
\label{Eq:Etot}
\end{align}
The mean-squared flux density is often called magnetic energy, e.g., see review by \cite{Reiners2012}. For the magnetic energy of the simulated maps $\langle B^2 \rangle\left[G^2\right]$ is a very good proxy but not equivalent, as the factor $\tfrac{1}{2\mu}$ is missing to obtain the true magnetic energy density, where $\mu$ is the permeability. It is assumed that the permeability $\mu$ is constant across the stellar surface.
For the observed and ZDI reconstructed maps $\langle B^2 \rangle\left[G^2\right]$ is more accurately described as the net magnetic flux energy per resolution element. Nevertheless, we refer to $\langle B^2 \rangle\left[G^2\right]$ as magnetic energy in the following to be consistent with the literature.
Additionally, we compute the total surface average field,
\begin{align}
\langle B \rangle &= \frac{1}{4\pi} \int \sqrt{\sum_k B^2_k(\phi,\theta)} \mrm{d}\Omega ,\ \ k \in (r,\phi,\theta).
\label{Eq:Bave}
\end{align}

The determination of the magnetic energies or surface average field is either possible for a specific length-scale by selecting the corresponding $\ell$-mode or for a specific sub-structure, e.g. the large-scale field, by choosing the corresponding cumulative $\ellsum\ $-modes. The cumulative modes $\ellsum\ $ include all lower modes. For example the cumulative $\ellsum\ = 3$ includes all $\ell$-modes $\ell = 1$ to $\ell = 3$.

\subsection{Simulations}
\label{SubSec:Simulations}

%
%
%
\begin{figure*}
\centering
\begin{minipage}{0.2\textwidth}
\includegraphics[angle=0,width = \textwidth ,trim = {0 0 0 0} ,clip]{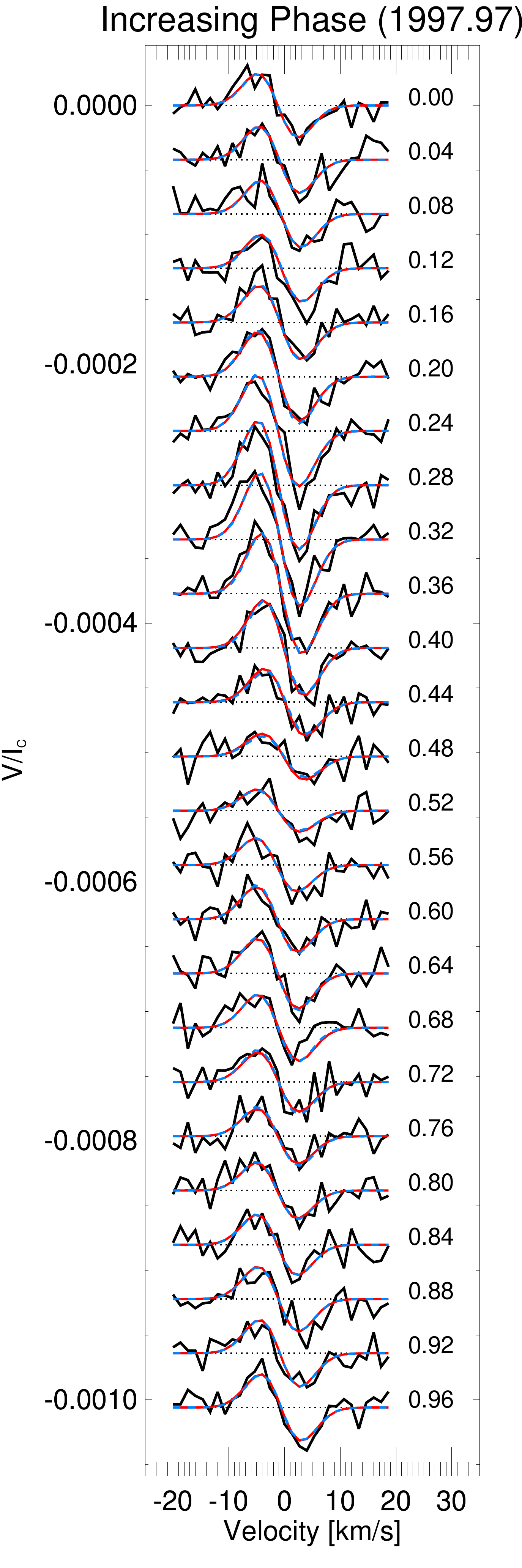}
\end{minipage}
\begin{minipage}{0.79\textwidth}
\raggedright
\large
\hspace{0.8cm}\textbf{Input simulation}\hspace{2.3cm}\textbf{Large-scale field of}\hspace{1.1cm}\textbf{ZDI reconstruction}\\
\hspace{5.5cm}\textbf{the input simulation}\\
\centering
\includegraphics[angle=0,height = 2.3cm ,clip]{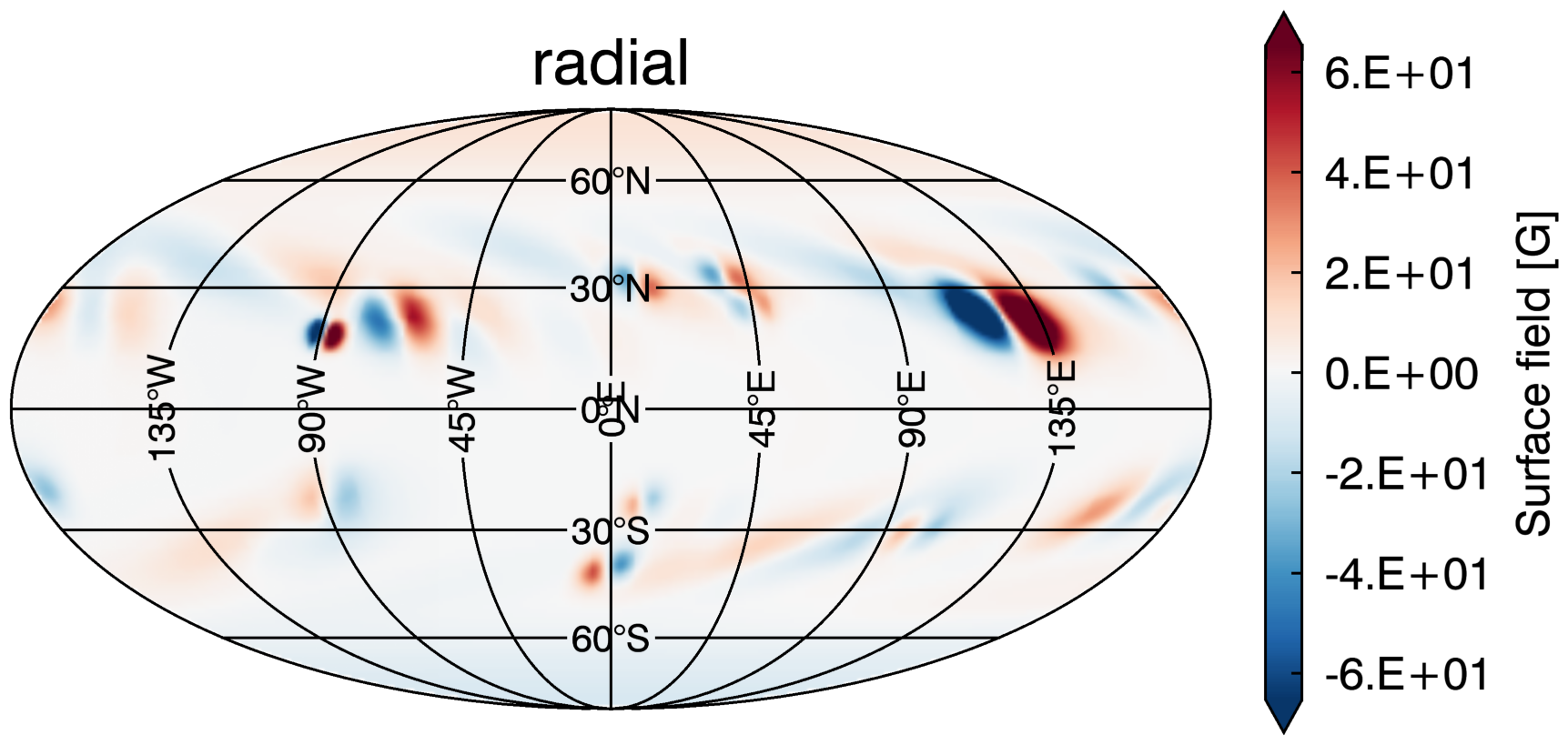}
\includegraphics[angle=0,height = 2.3cm , trim={0 0 3cm  0} ,clip]{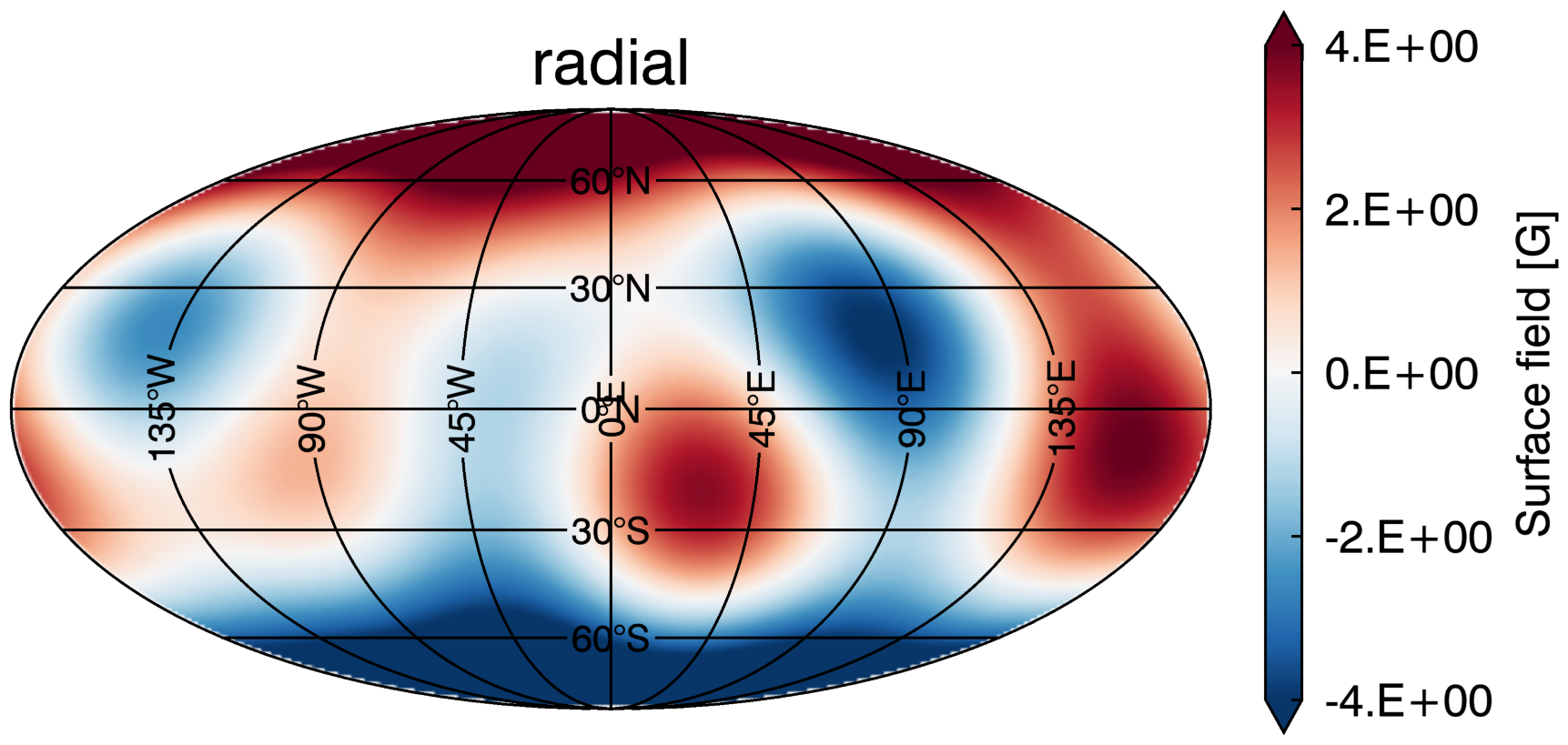}
\includegraphics[angle=0,height = 2.3cm ,clip]{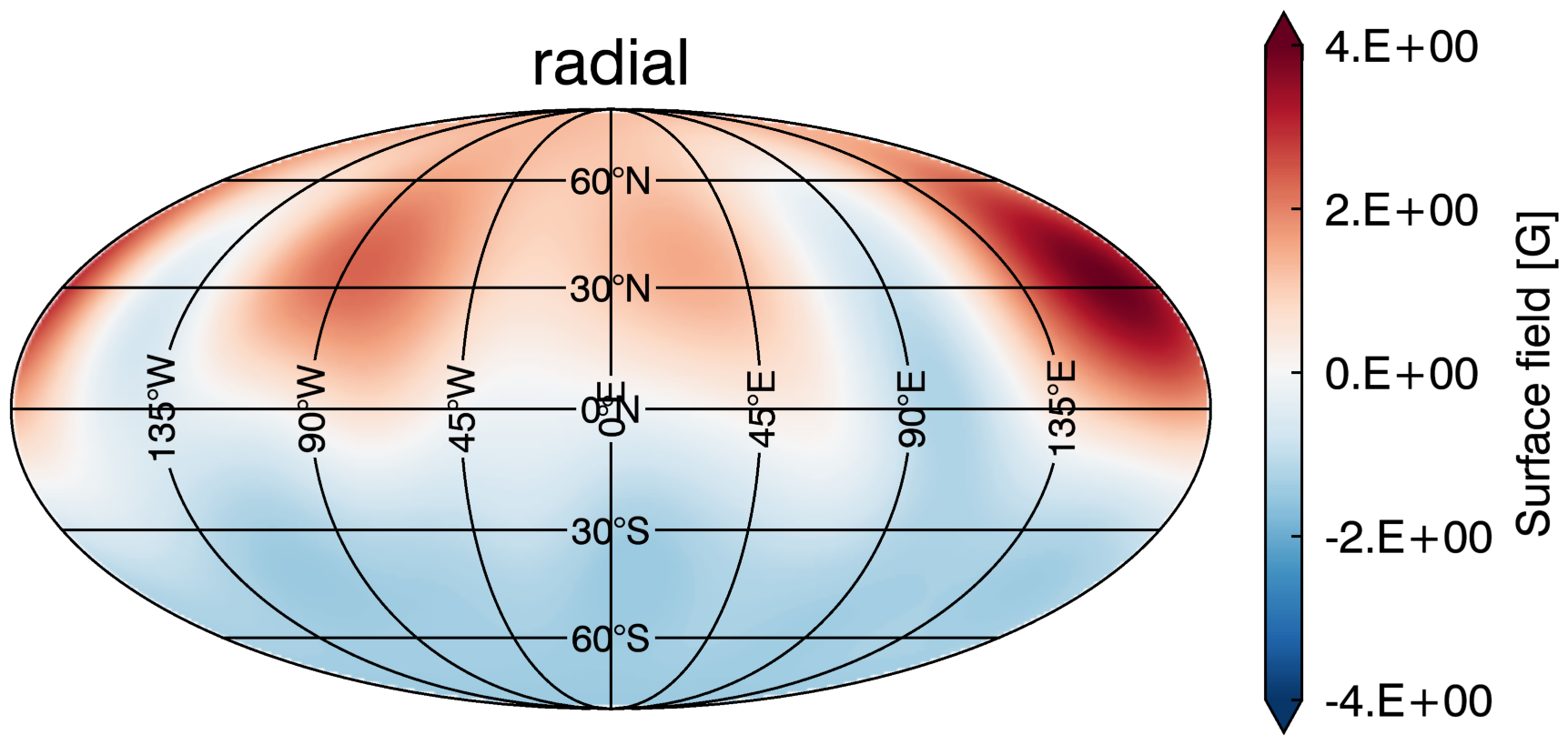} \\
\includegraphics[angle=0,height = 2.3cm ,clip]{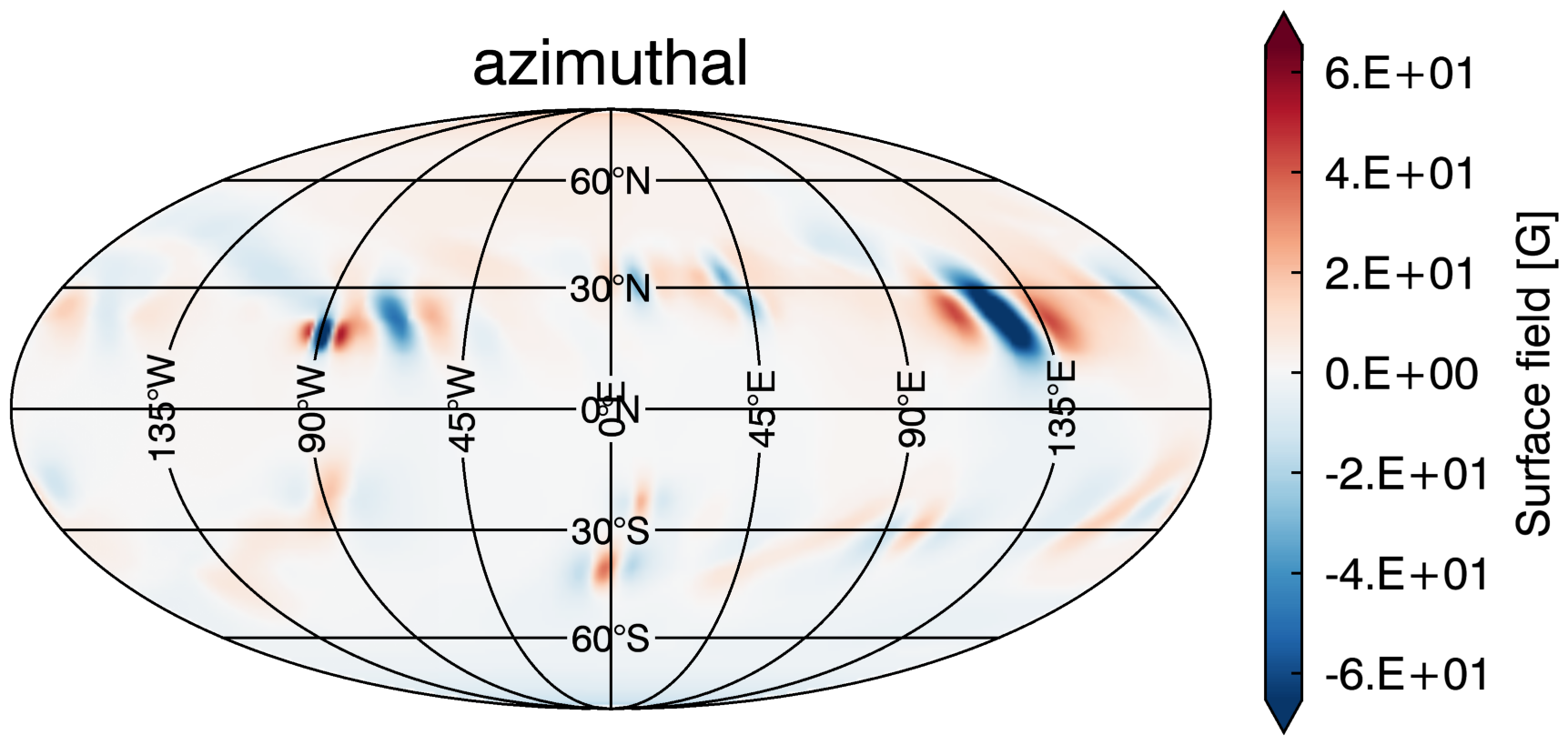}
\includegraphics[angle=0,height = 2.3cm , trim={0 0 3cm  0},clip]{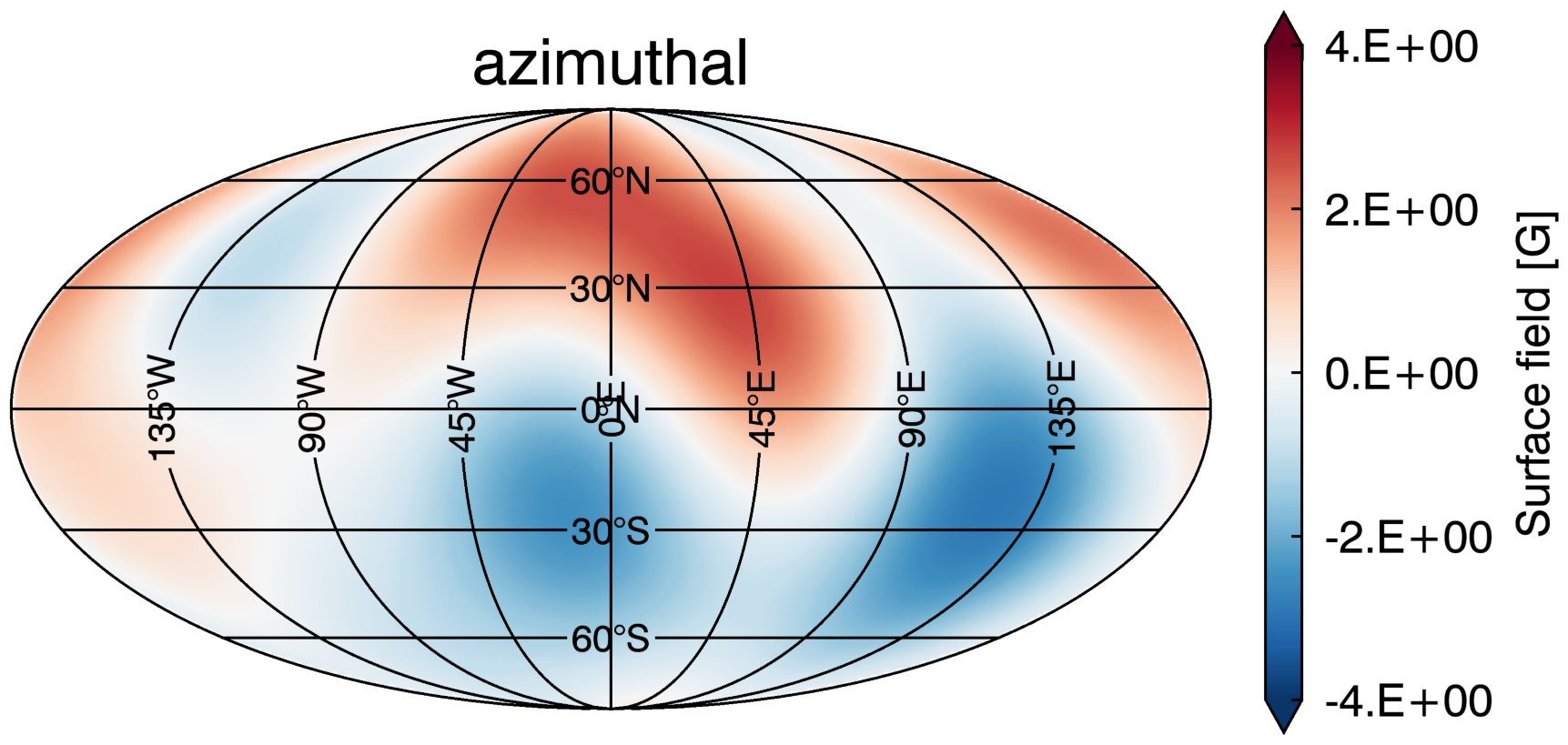}
\includegraphics[angle=0,height = 2.3cm ,clip]{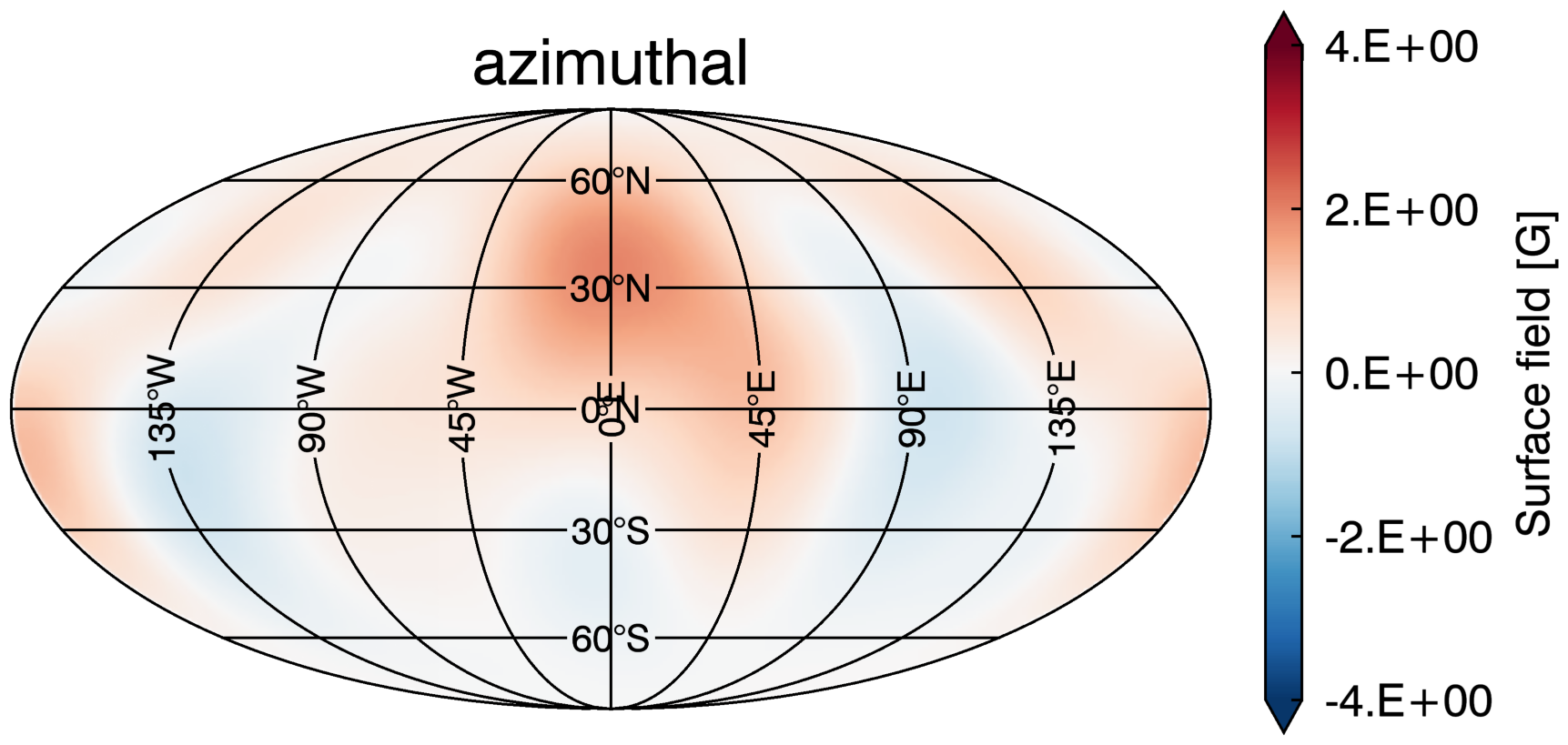} \\
\includegraphics[angle=0,height = 2.3cm ,clip]{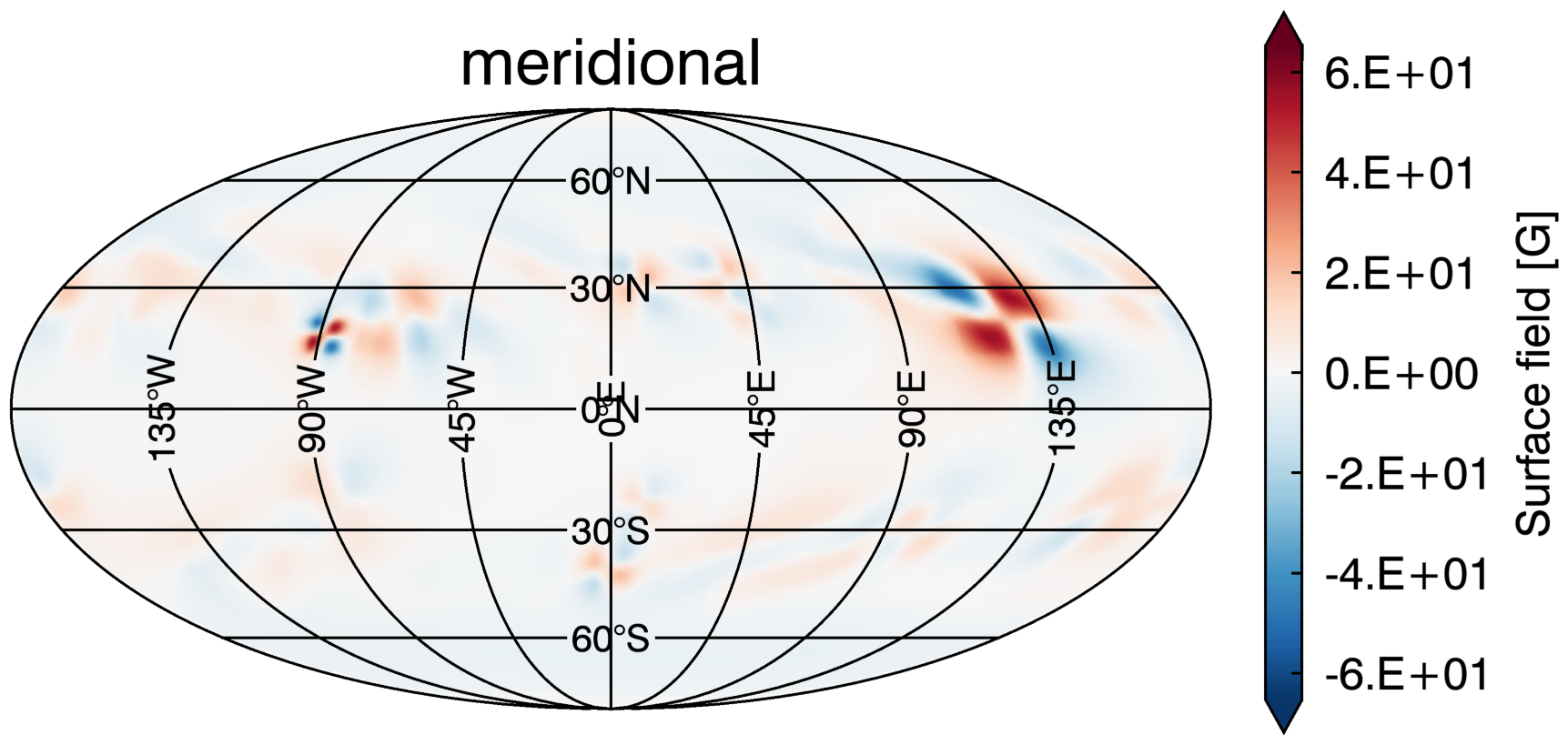}
\includegraphics[angle=0,height = 2.3cm , trim={0 0 3cm  0},clip]{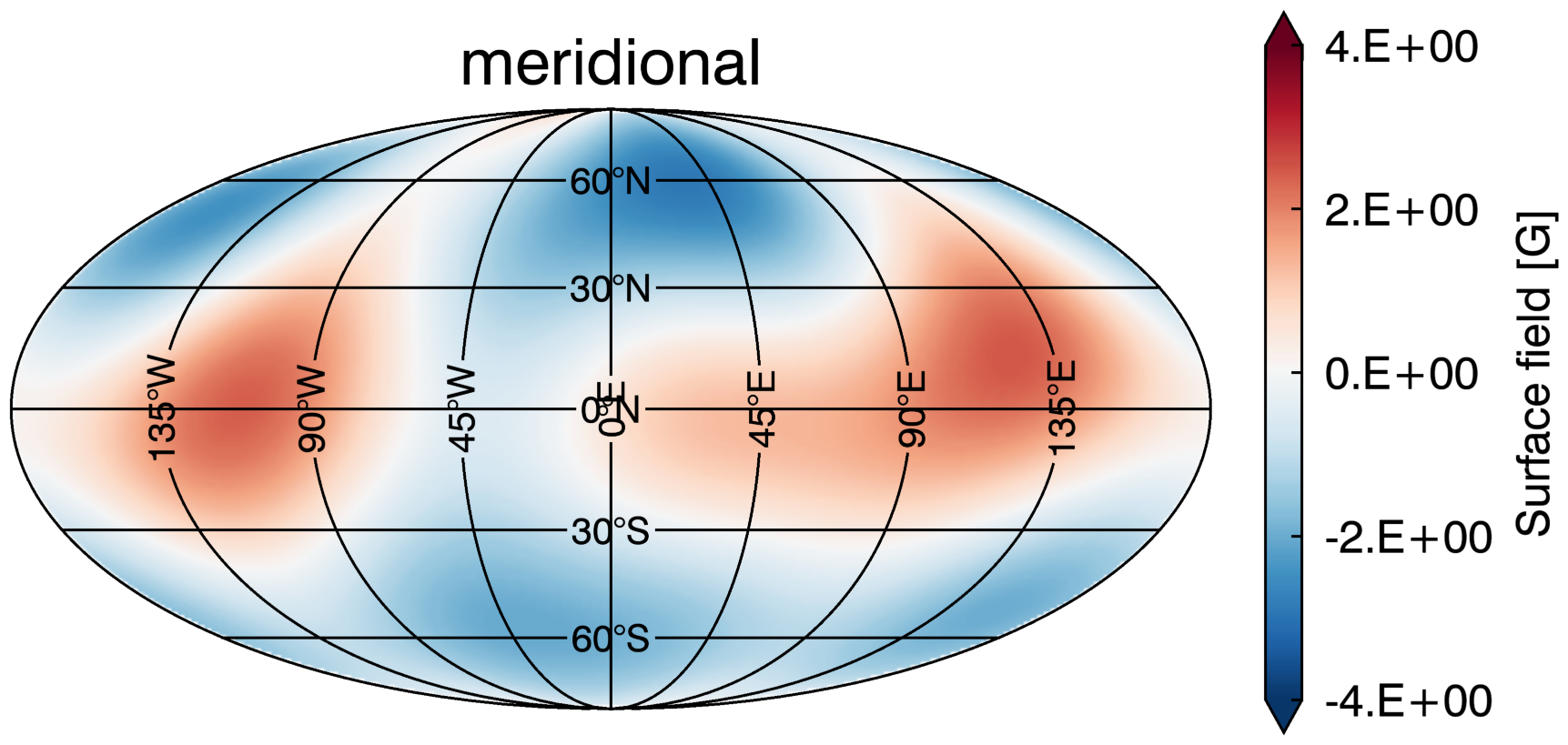}
\includegraphics[angle=0,height = 2.3cm ,clip]{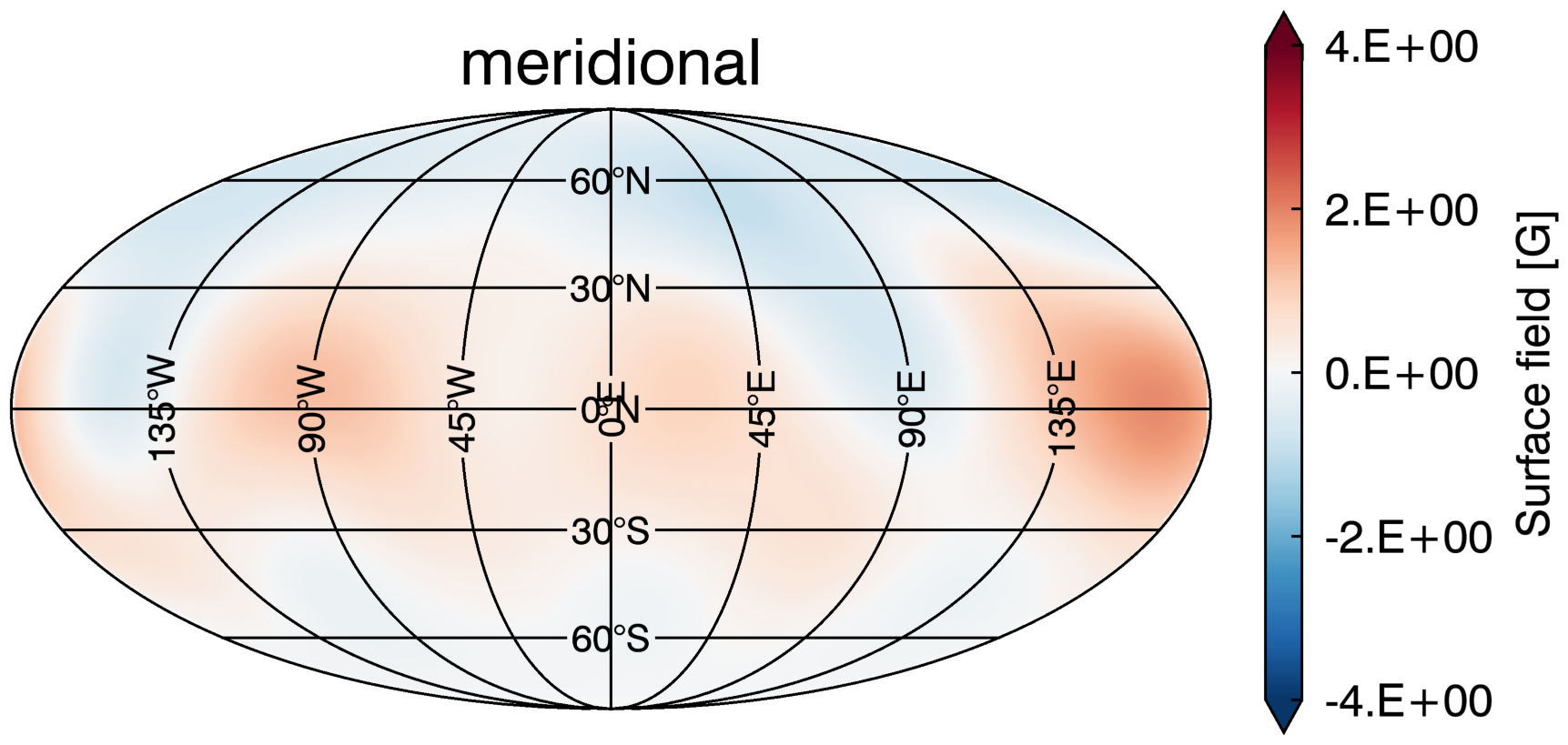} \\
\end{minipage}
\caption{Left panel: The ZDI fit (red solid line) of the noisy Stokes~V profiles (black solid line). For demonstrating the accuracy of the fit we overplot the noise-free Stokes V profiles (dashed blue line) for comparison. The dashed black line is the null line and the observation phases are written to the right. Three panels to the right: The Mollweide projected maps for map number 0720 (year = 1997.97; increasing phase). From left to right: The full resolved simulated input map from which the Stokes profiles were modelled, the large-scale field of the input map for $\ell_{\Sigma} = 3$ and the ZDI reconstructed map. From top to bottom: the radial, azimuthal and meridional surface magnetic field. Be aware of the use of two different colour bars.}
\label{Fig:Map0720}
\end{figure*}

\cite{Yeates2012} simulated the evolution of the photospheric and coronal magnetic field for the Sun, covering 15 years from 1996--2011. They used a non-potential  model \citep{vanBallegooijen2000} connecting a surface flux transport model with the magnetofrictional relaxation in the corona above. Further details are available in Section~2 of \cite{Yeates2012} and \cite{Yeates2008}.
The simulations are initialised using a potential field extrapolation of the synoptic magnetogram for CR1910\footnote{Carrington Rotation (CR) 1910: 1996 June 01 to 1996 June 28.} from the US National Solar Observatory, Kitt Peak (NSO/KP) and were corrected for differential rotation to represent 1996 June 15.
Photospheric motions are modelled using a surface flux transport model \citep{Leighton1964, Wang1989, Sheeley2005, vanBallegooijen2000, Mackay2004}. The supergranular diffusivity is given with $D = 450\,\mrm{km^2\ s^{-1}}$. The meridional flow is described by the profile of \cite{Schuessler2006},
\begin{equation}
u_\theta = C \frac{16}{110}\sin(2\lambda)\exp(\pi-2\vert \lambda \vert),
\end{equation}
where $\lambda = \pi/2 -\theta$ is the latitude and $C = 11\,\mrm{m\ s^{-1}}$ the peak speed. The differential rotation $\Omega(\theta)$ is modelled using the profile of \cite{Snodgrass1983},
\begin{equation}
\Omega(\theta) = 0.18 - 2.30 \cos^2\theta - 1.62\cos^4\theta\,\mrm{deg\,day^{-1}}.
\end{equation}

The emerging active regions drive the evolution of the photospheric and coronal field while the coronal field is simultaneously sheared by photospheric motions. The active regions are emerged as idealised 3D magnetic bipoles (two spots of opposite magnetic polarity) using the model of \cite{Yeates2008}. The properties of the active regions (magnetic flux, location, size, tilt angle) are determined from NSO/KP synoptic maps\footnote{After 2003 with the SOLIS telescope, before with the older vacuum telescope.}. Further details can be found in \cite{Yeates2012}. 
In total, 1838 bipoles with a magnetic flux between $2\cdot10^{20}\,\mrm{Mx}-5.3\cdot10^{22}\,\mrm{Mx}$ were emerged between CR1911 (1996 June) and CR2110 (2011 May), although there are three data periods (CR2015--16, CR2040--41, CR2091) where no bipoles are inserted due to missing observational data.

We are only using the photospheric ($r = \mrm{R_{\odot}}$) vector magnetic field maps of this non-potential 3D simulation which extends up to the corona. In total we analysed all 118 maps. Table~\ref{Tab:MapsofSC23} in the appendix lists the key parameters of the subsample that have a ZDI reconstructed map and their related sunspot number (SSN) and S-index. We used the monthly mean total sunspot number of the World Data Center SILSO, Royal Observatory of Belgium, Brussels and the S-index values published by \cite{Egeland2017} from the Mount Wilson Observatory HK program. Both, SSN and S-index, were averaged over the Carrington rotation dates per map.

The S-index is a measurement of chromospheric emission observed in the Ca II H \& K line profiles. The S-index is defined by the Mount Wilson Observatory HK project as:
\begin{equation}
S = \alpha \frac{N_H + N_K}{N_R + N_V},
\label{Eq:SIndex}
\end{equation}
where $N_H$ and $N_K$ are the counts in the 1.09$\AA$ triangular bands centred on the Ca II H \& K for the HKP2-spectrophotometer and $N_R$ and $N_V$ are the counts in the 20$\AA$ reference bandpasses in the nearby continuum region. The parameter $\alpha$ refers to a calibration constant \citep{Vaughan1978}. For further details see \cite{Egeland2017}.

\subsection{Map Selection and Stokes Profiles Modelling}
\label{SubSec:MapSelectionStokesModelling}

%
%
\begin{figure*}
\centering
\begin{minipage}{0.2\textwidth}
\includegraphics[angle=0,width = \textwidth ,trim = {0 0 0 0} ,clip]{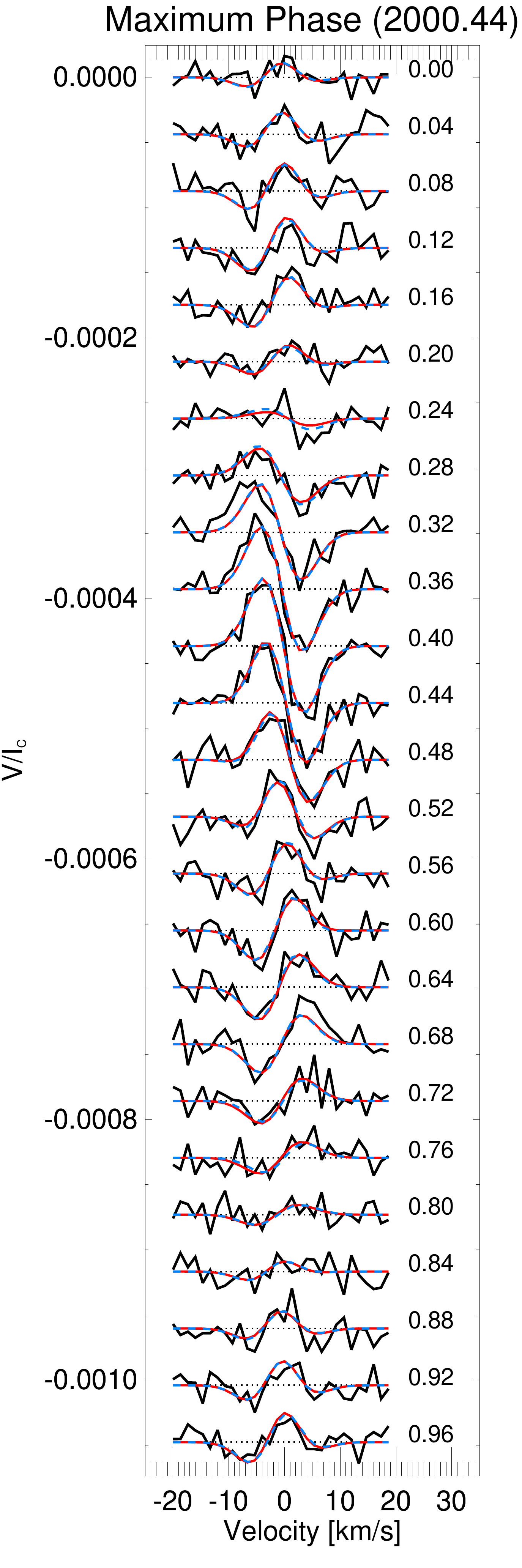}
\end{minipage}
\begin{minipage}{0.79\textwidth}
\raggedright
\large
\hspace{0.8cm}\textbf{Input simulation}\hspace{2.3cm}\textbf{Large-scale field of}\hspace{1.1cm}\textbf{ZDI reconstruction}\\
\hspace{5.5cm}\textbf{the input simulation}\\
\centering
\includegraphics[angle=0,height = 2.3cm ,clip]{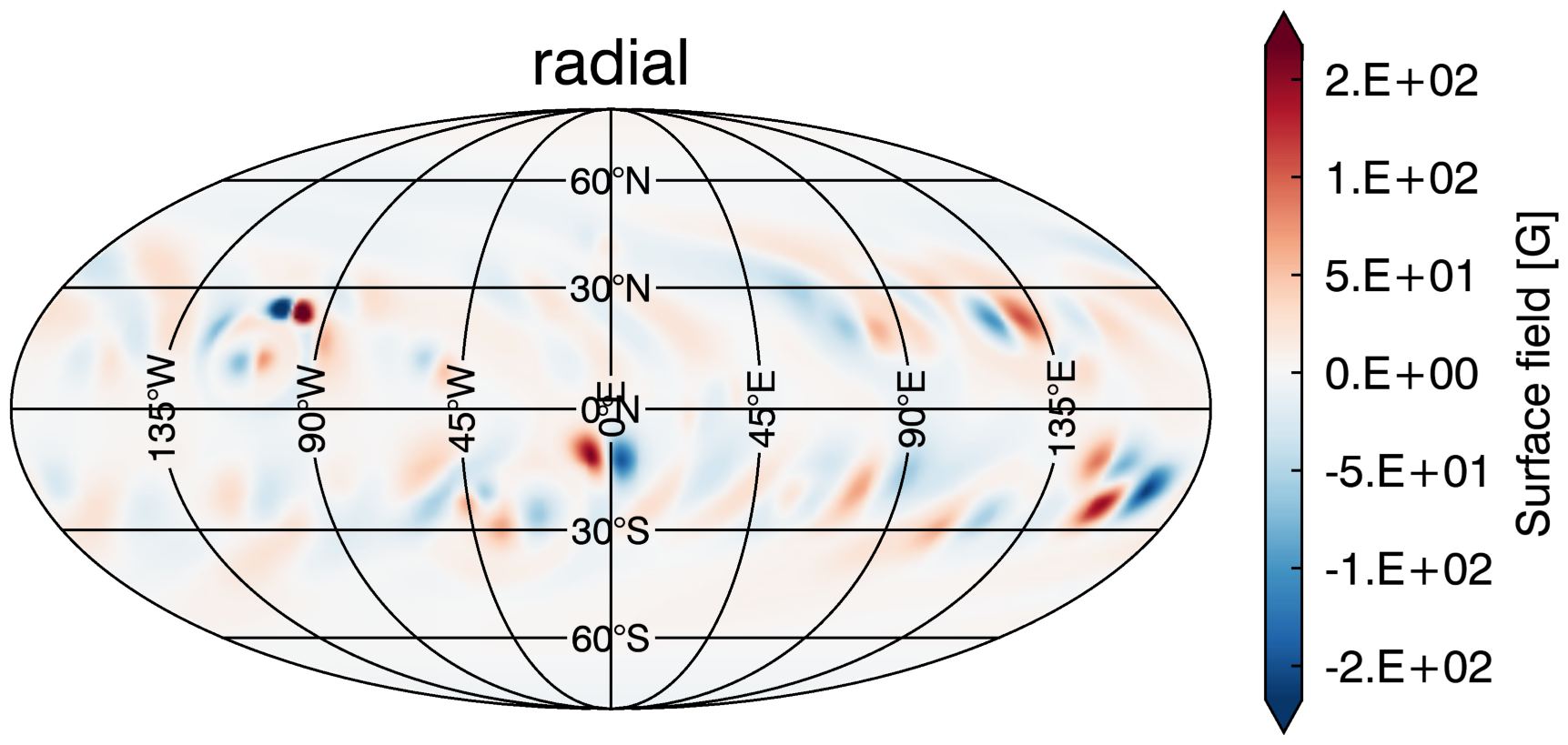}
\includegraphics[angle=0,height = 2.3cm , trim={0 0 3cm  0} ,clip]{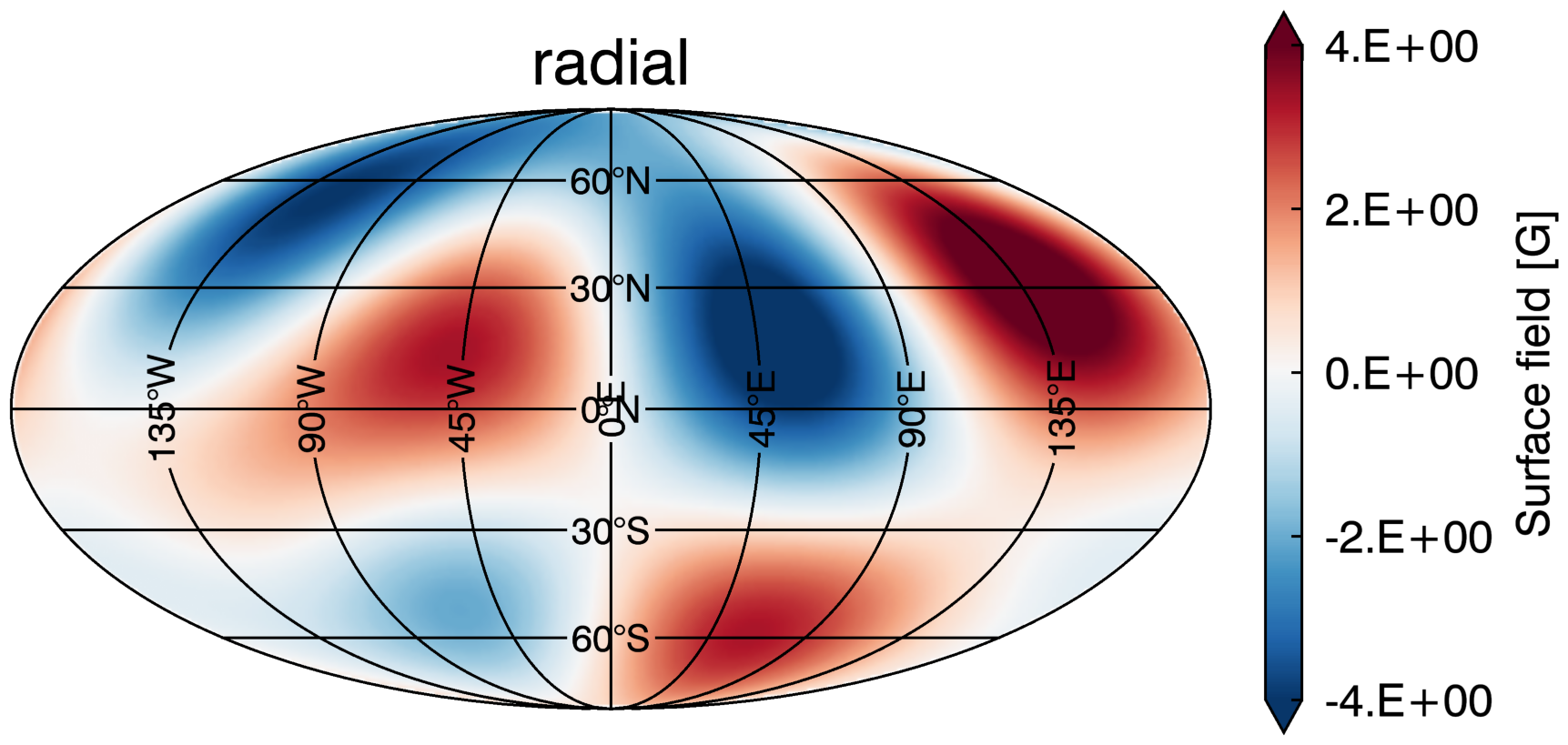}
\includegraphics[angle=0,height = 2.3cm ,clip]{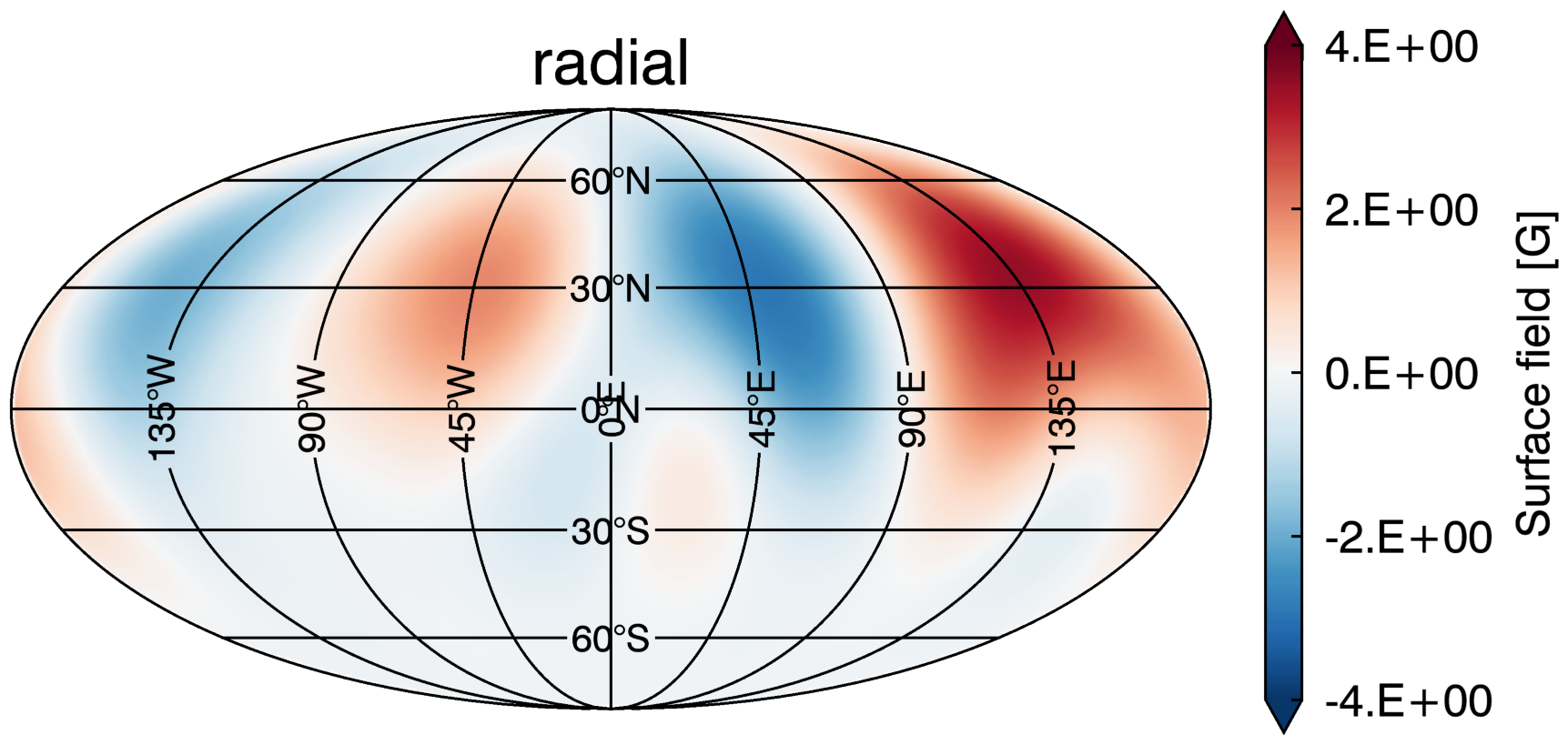} \\
\includegraphics[angle=0,height = 2.3cm ,clip]{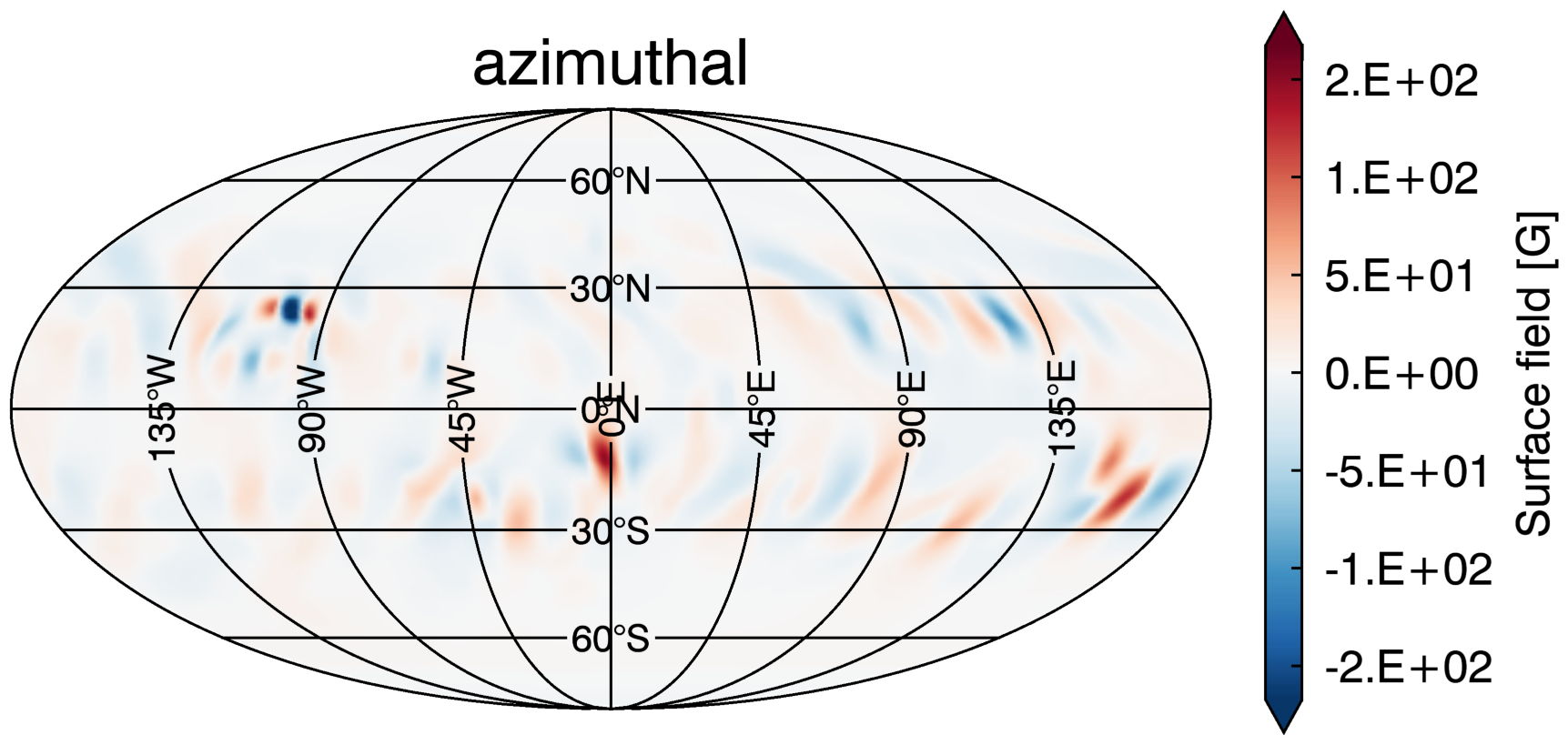}
\includegraphics[angle=0,height = 2.3cm , trim={0 0 3cm  0} ,clip]{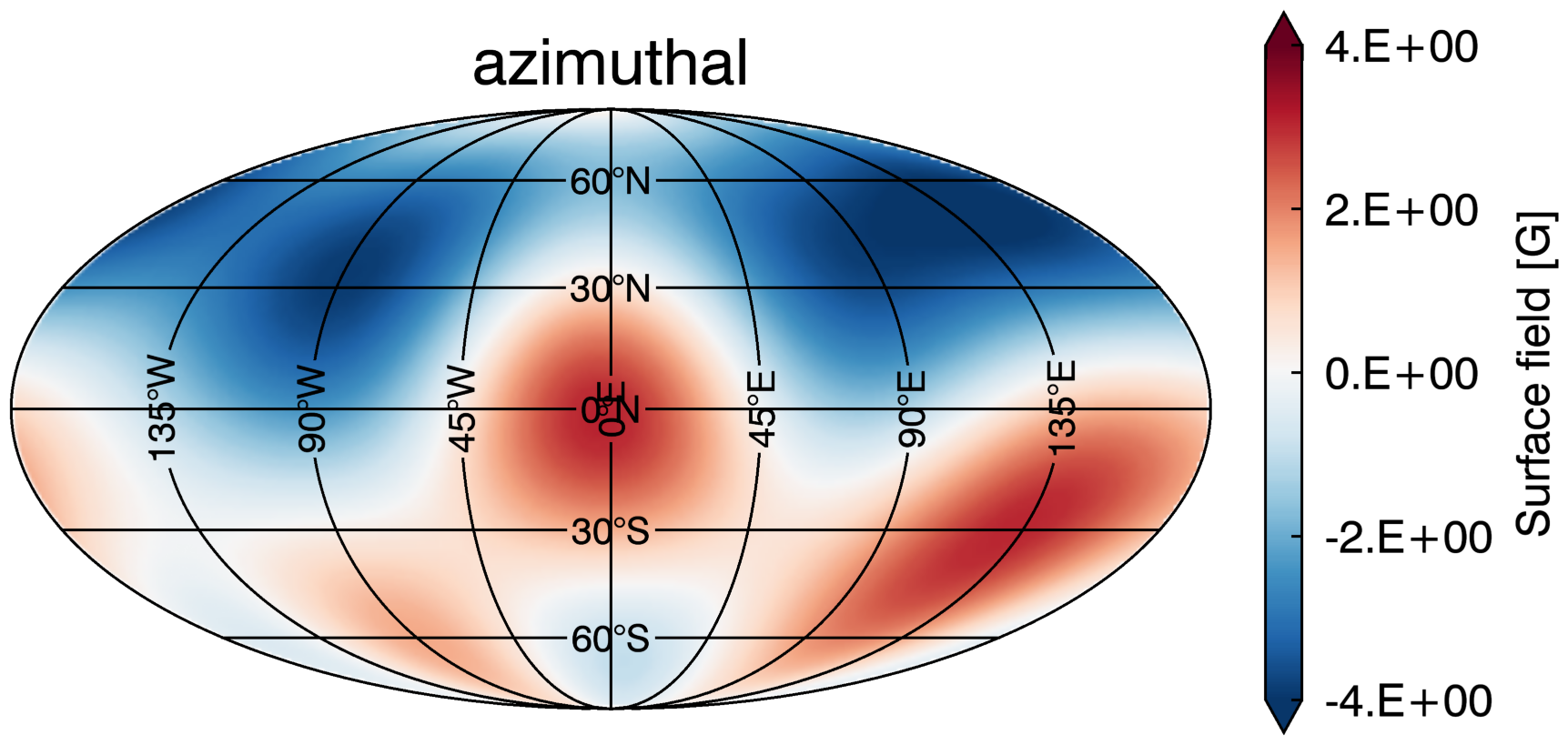}
\includegraphics[angle=0,height = 2.3cm ,clip]{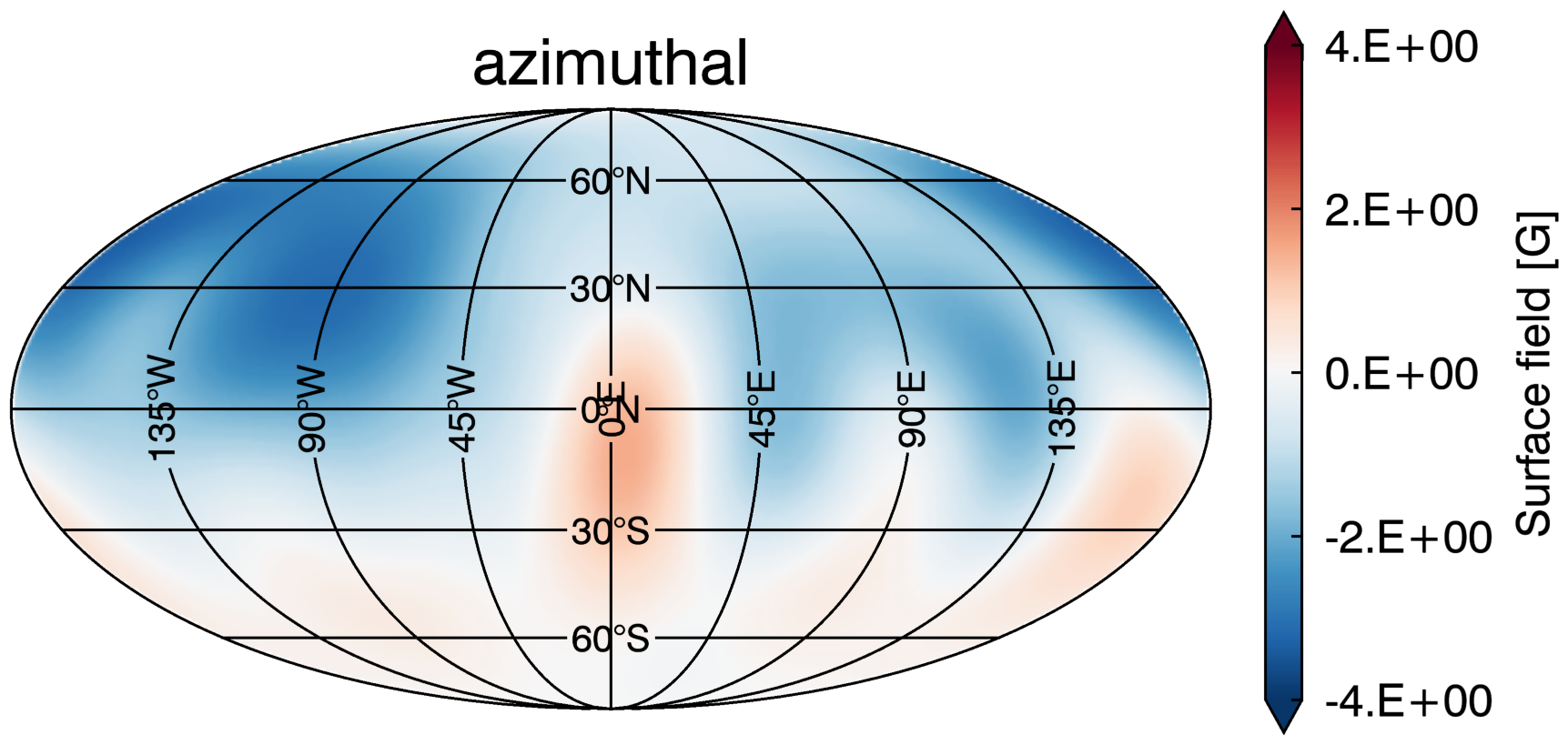} \\
\includegraphics[angle=0,height = 2.3cm ,clip]{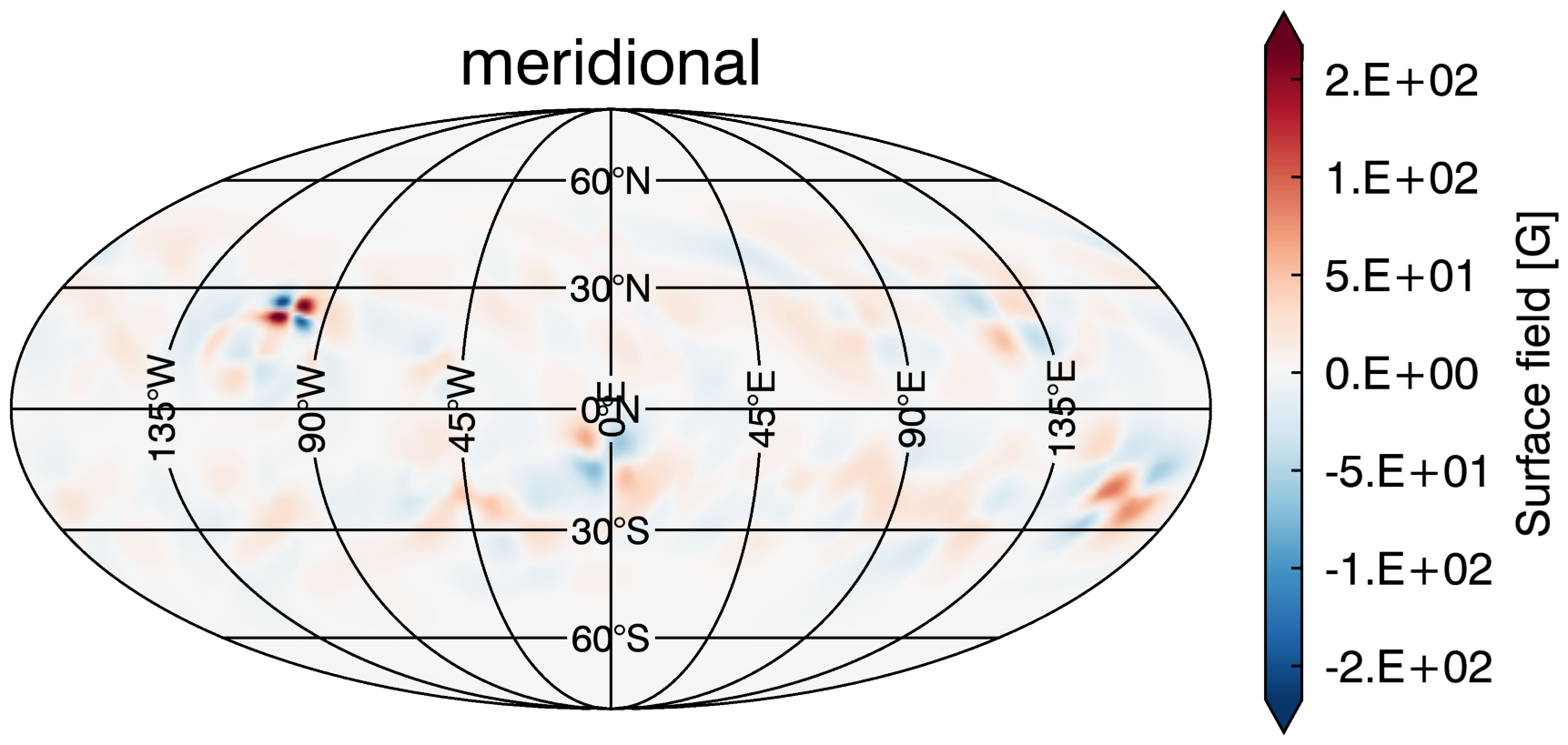}
\includegraphics[angle=0,height = 2.3cm , trim={0 0 3cm  0} ,clip]{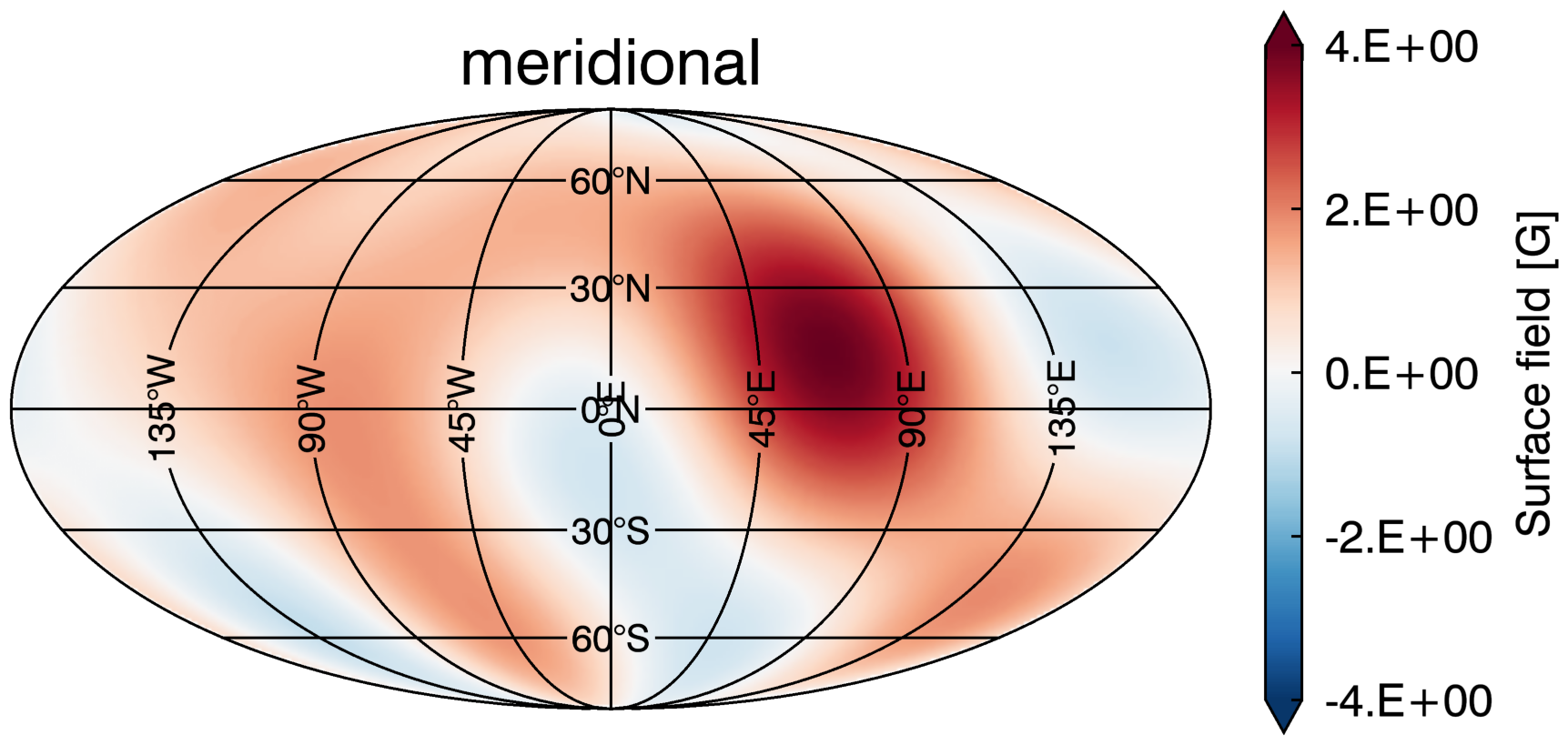}
\includegraphics[angle=0,height = 2.3cm ,clip]{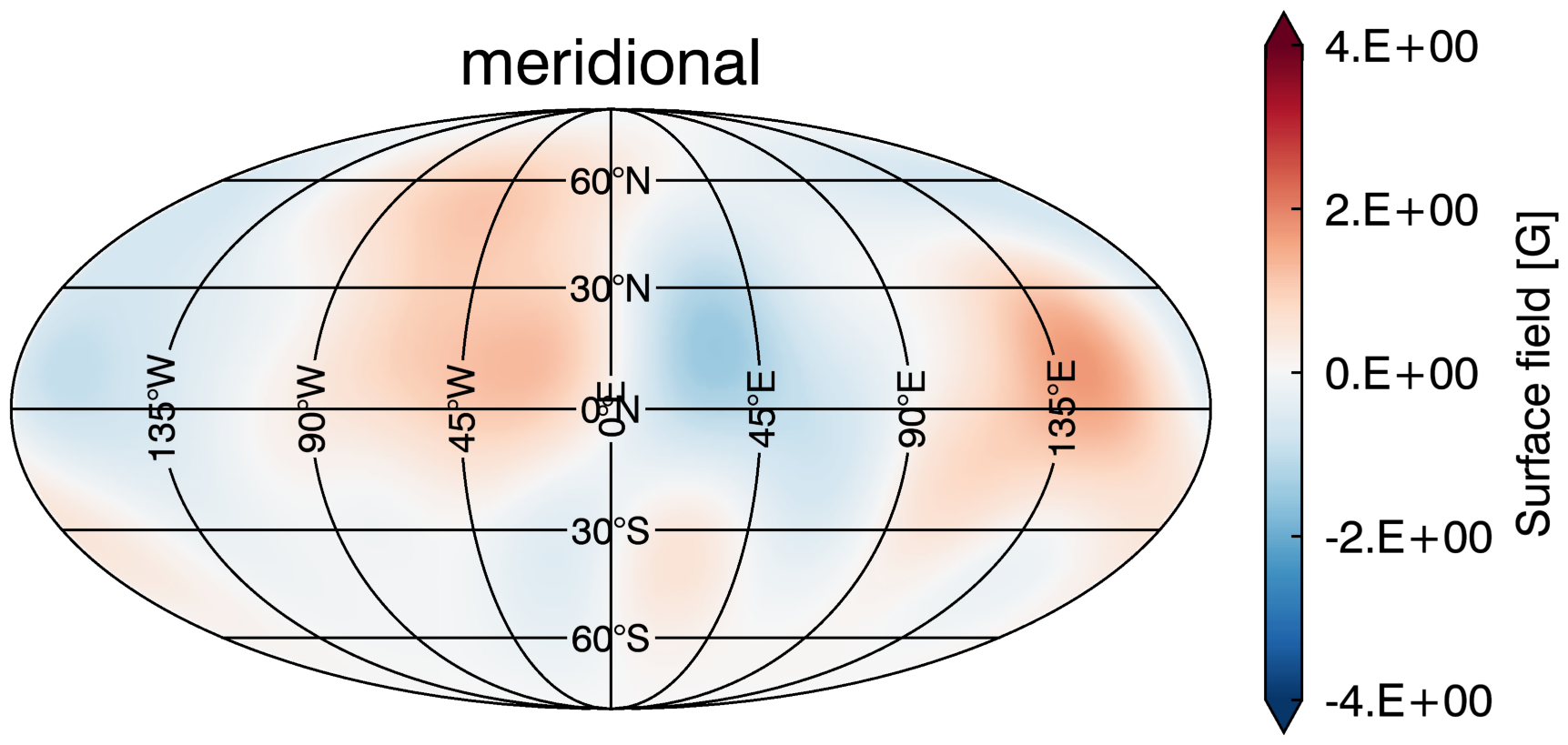} \\
\end{minipage}
\caption{The Stokes~V profiles and Mollweide projected maps for map number 1620 (year = 2000.44, maximum phase). The same format is used as for Fig.~\ref{Fig:Map0720}.}
\label{Fig:Map1620}
\end{figure*}

We analysed the magnetic field topology of all 118 simulated maps covering SC23, looking at the small and large-scale field described by the \lmod\ \ from $1-28$.

From the 118 simulated maps, we selected 41 maps for  ZDI reconstruction. It would be highly unlikely to be awarded sufficient  telescope time to cover a stellar magnetic cycle in greater detail than that for a slowly rotating and low-activity star like the Sun. Half of the selected maps (21 maps) are equally spaced in time from mid-1996 until the beginning of 2011. In addition, 20 maps are selected in between, predominantly around the activity maximum, as the simulated maps show the highest variation during this time, e.g.\ see Fig.~\ref{Fig:EtotvsTime}a. Table~\ref{Tab:MapsofSC23} lists the 41 selected maps and some of their key properties. We modelled Stokes~IV profiles for each epoch at a number of rotational phases and used the resulting 41 time-series of profiles  as inputs for the ZDI code. 

We modelled the Stokes~IV profiles using the same procedure as described in Section~3 of \cite{Lehmann2019}. 
The line profile modelling is based on a Milne-Eddington atmosphere model. The parameters of the line profile were fine-tuned to match the observed spectropolarimetic data of the solar analogue star 18~Sco, as described in \cite{Lehmann2019}. The line profiles of simulated maps of the SC23 are modelled, setting the rotation period of the Sun to 27.0 days and assuming an inclination angle of $60^{\circ}$. This corresponds to an equatorial velocity of $v_e \sin i = 1.62\,\mathrm{km\ s^{-1}}$. We apply 30 velocity bins ranging from $-20\,\mrm{km\ s^{-1}}$ to $20\,\mrm{km\ s^{-1}}$ per Stokes profile, corresponding to a spectral resolution similar to HARPS. 
The local Stokes V profile for each element across the stellar disc is modelled as the derivative of the local Stokes I profile. Each time-series of Stokes V profiles consists of 25 observational phases equally spaced in time, which corresponds to an observation every 0.04 observational phase or every $14.4^{\circ}$ in longitude.
We assumed a fixed S/N corresponding to 10\% of the maximum amplitude of all Stokes~V profiles across the solar cycle, so that each map is differently affected by the noise. 
Naturally when observing distant stars, it is often not practical or even possible  to fine-tune the noise levels of spectropolarimetric datasets acquired over multi-year timescales. 

The Stokes~V amplitudes of the 118 maps range from $2.02\cdot 10^{-5}\,\mrm{V/I_C}$ to $7.83\cdot 10^{-5}\,\mrm{V/I_C}$, where $I_C$ is the Stokes~I continuum level. The corresponding continuum S/N of 10\% of the maximum amplitude is then $S/N \approx 130\,000$.
A fixed S/N allows us to analyse the effect of the noise on the detected large-scale magnetic field topology reconstructed along the magnetic cycle. Further S/N ratios were tested, e.g. S/N = 30\,000, 80\,000, guided by typical and very good signal-to-noise-ratios of current observations, see e.g. \citet{Jeffers2014,BoroSaikia2018}, and a S/N = 425\,000 which corresponds to 3\,\% of the maximum amplitude of all Stokes~V profiles along the cycle. It was not possible to extract reliable maps using a S/N = 30\,000 or 80\,000 as the Stokes~V profiles become too noisy. A S/N = 425\,000, corresponding to 3\,\% of the maximum amplitude, leads of course to less noisy Stokes~V profiles and more detailed maps, but obtaining such a high S/N for a solar-like star is not possible with current telescope facilities. Even a S/N of 130\,000 is challenging today. 
For one example map, we artificially increased the field of the input map for all three components by a factor of 10 dropping the S/N to 12400 for the resulting Stokes~V profiles. The corresponding ZDI map looks the same  as for the higher S/N map from the morphological point of view.

\subsection{Zeeman-Doppler-Imaging}
\label{SubSec:ZDI}

We applied the same ZDI methodology as described in Sections~2 and 3 of \cite{Lehmann2019} using the ZDI code presented in \cite{Hussain2016}.
The $\ell$-modes are equally weighted, so that all $\ell$ and $m$-modes contribute equally to the topology. No length scale and no axi- or non-axisymmentric field configuration is preferred. We allowed a maximum $\ellsum\ $-mode of $\ell_{\Sigma , \mrm{max}} = 7$ following the results of \cite{Lehmann2019}. For the majority of the maps only the \lmod\ \ up to $\ellsum\ \le 5$ show significant magnetic energy. 
ZDI suffers from the ambiguity that an infinite number of magnetic field distributions can fit the Stokes~IV timeseries. To overcome this problem, we apply the maximum entropy approach, see e.g.\ \citet{DonatiBrown1997,Hussain2001,Hebrard2016}. The maximum entropy approach chooses the field configuration containing the least amount of information required to fit the observations to a specific $\chi^2$. The amount of the information is determined via the entropy $S$. We use the form of entropy presented in \citet[eq.~B11]{Folsom2018}.

%
%
\begin{figure*}
\centering
\begin{minipage}{0.2\textwidth}
\includegraphics[angle=0,width = \textwidth ,trim = {0 0 0 0} ,clip]{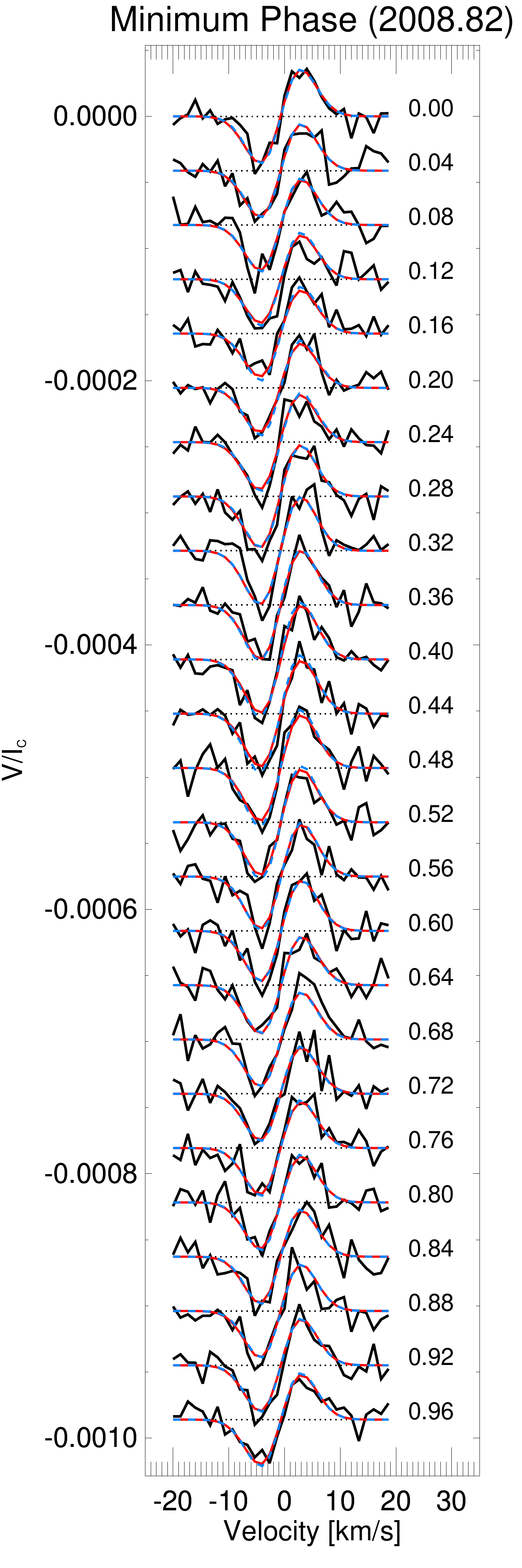}
\end{minipage}
\begin{minipage}{0.79\textwidth}
\raggedright
\large
\hspace{0.8cm}\textbf{Input simulation}\hspace{2.3cm}\textbf{Large-scale field of}\hspace{1.1cm}\textbf{ZDI reconstruction}\\
\hspace{5.5cm}\textbf{the input simulation}\\
\centering
\includegraphics[angle=0,height = 2.3cm ,clip]{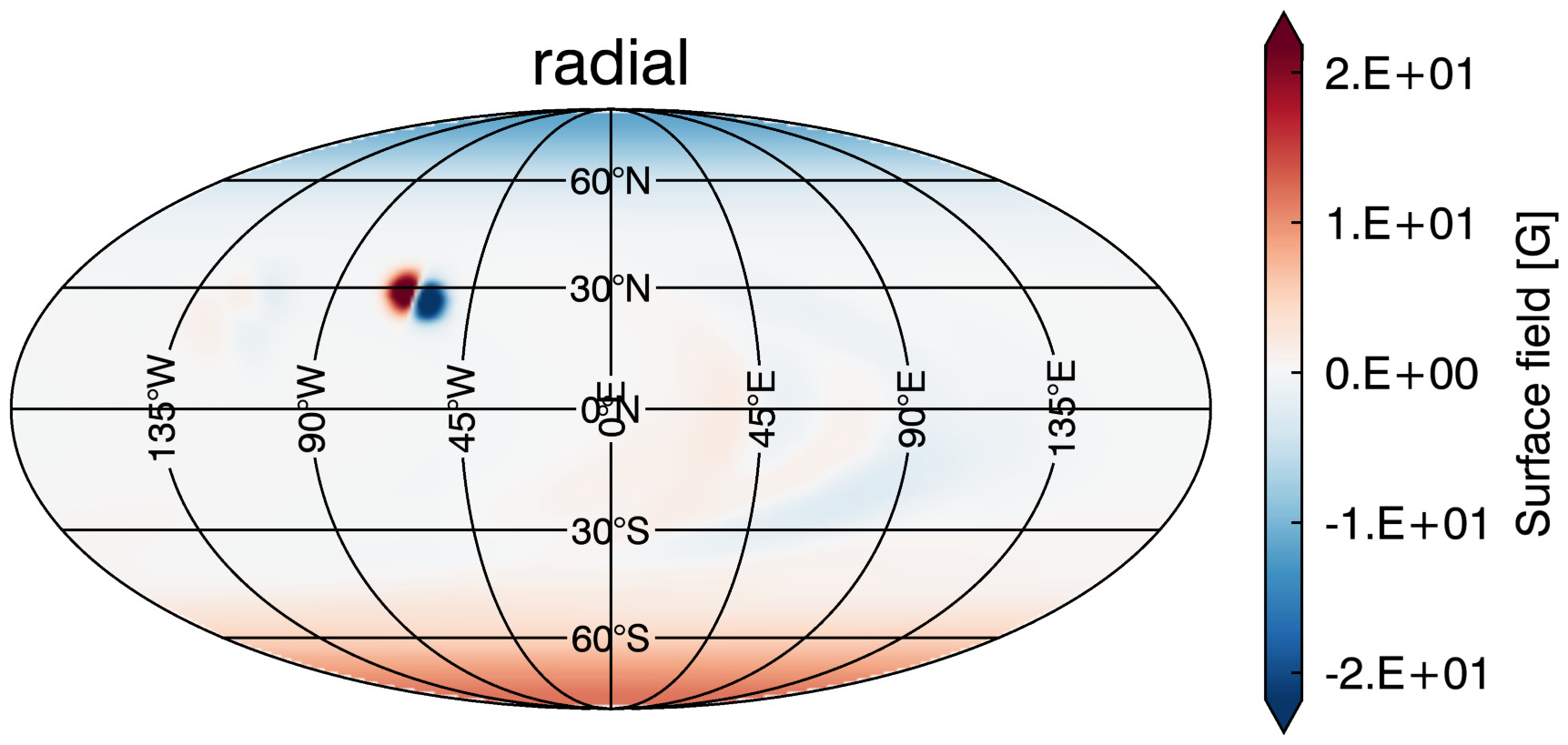}
\includegraphics[angle=0,height = 2.3cm , trim={0 0 3cm  0} ,clip]{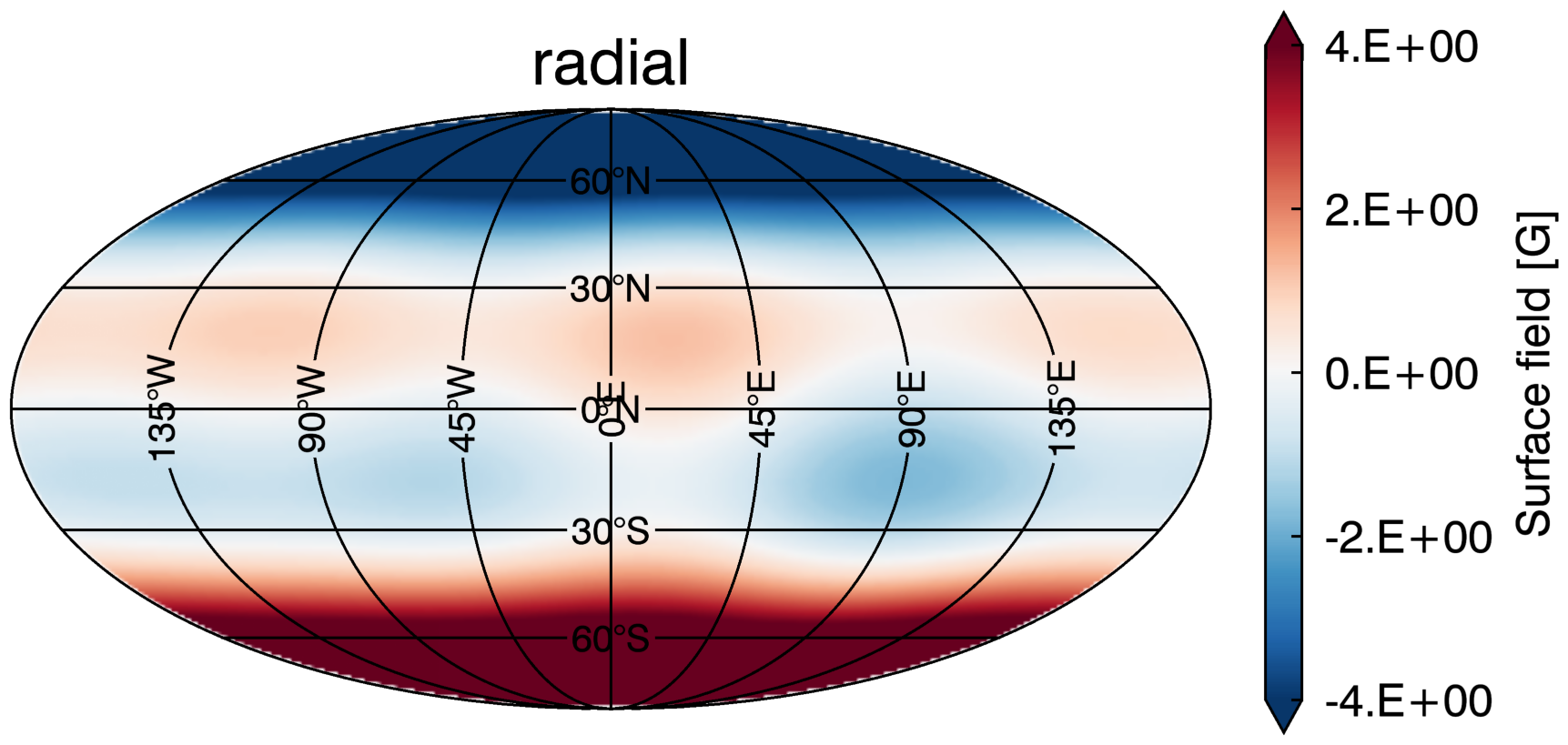}
\includegraphics[angle=0,height = 2.3cm ,clip]{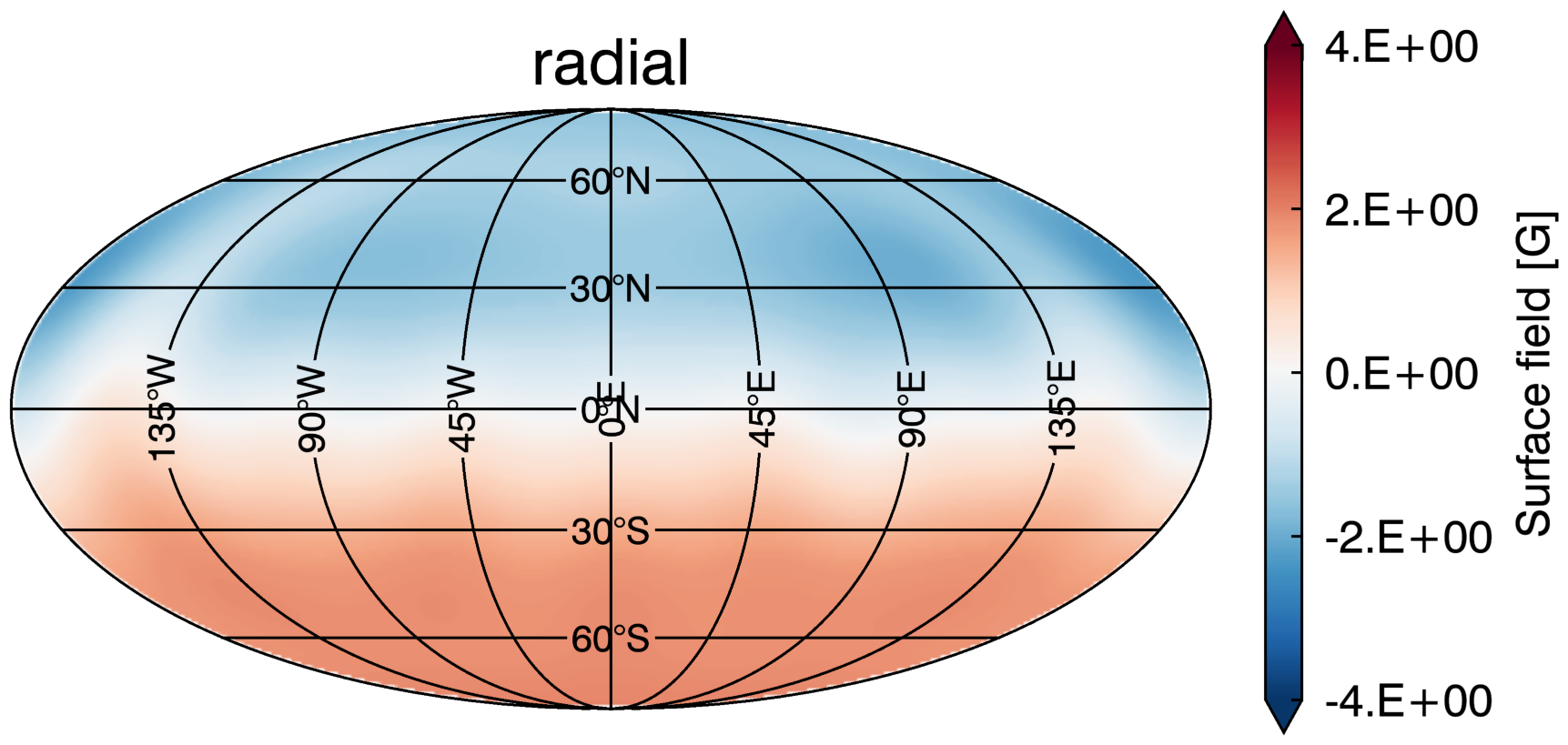} \\
\includegraphics[angle=0,height = 2.3cm ,clip]{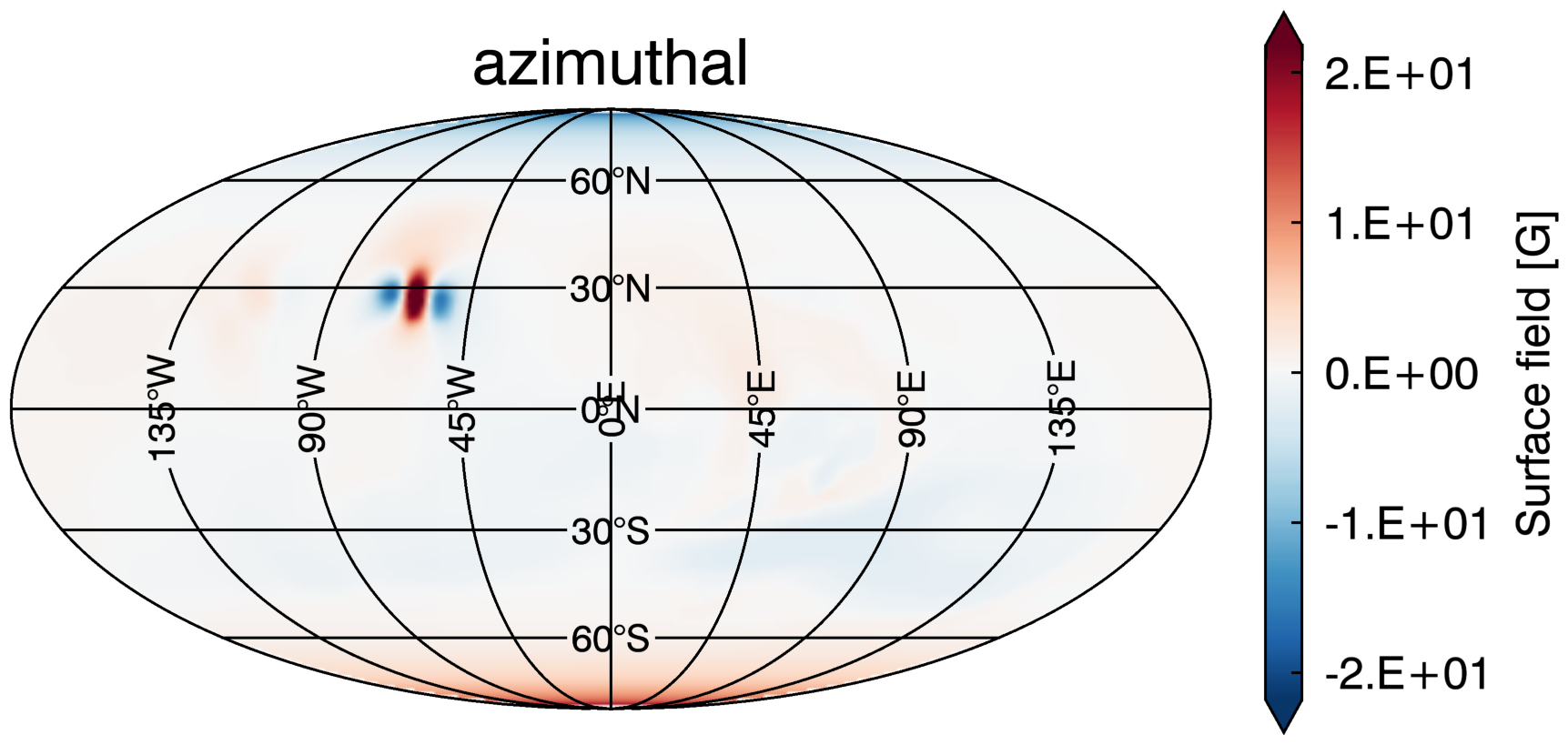}
\includegraphics[angle=0,height = 2.3cm , trim={0 0 3cm  0} ,clip]{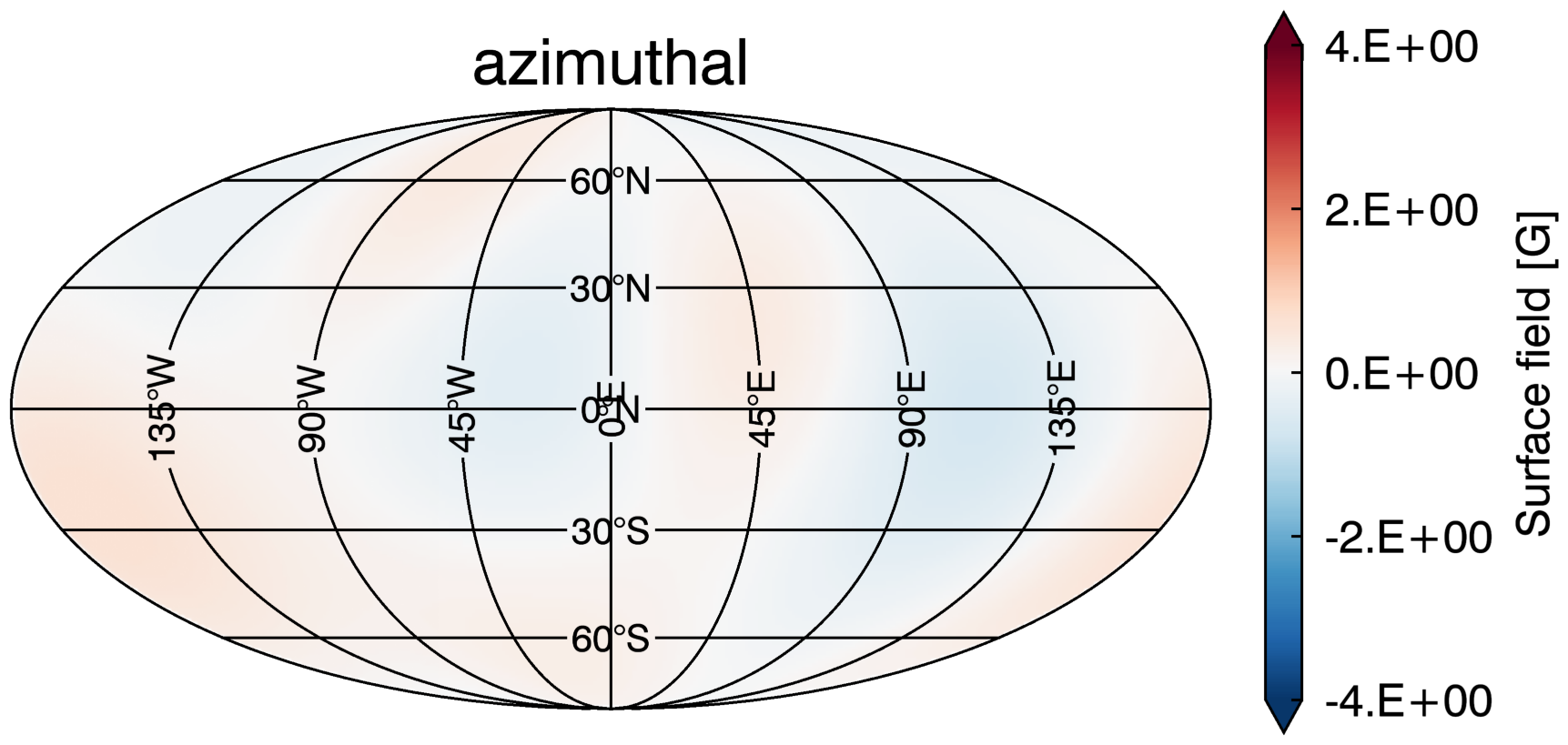}
\includegraphics[angle=0,height = 2.3cm ,clip]{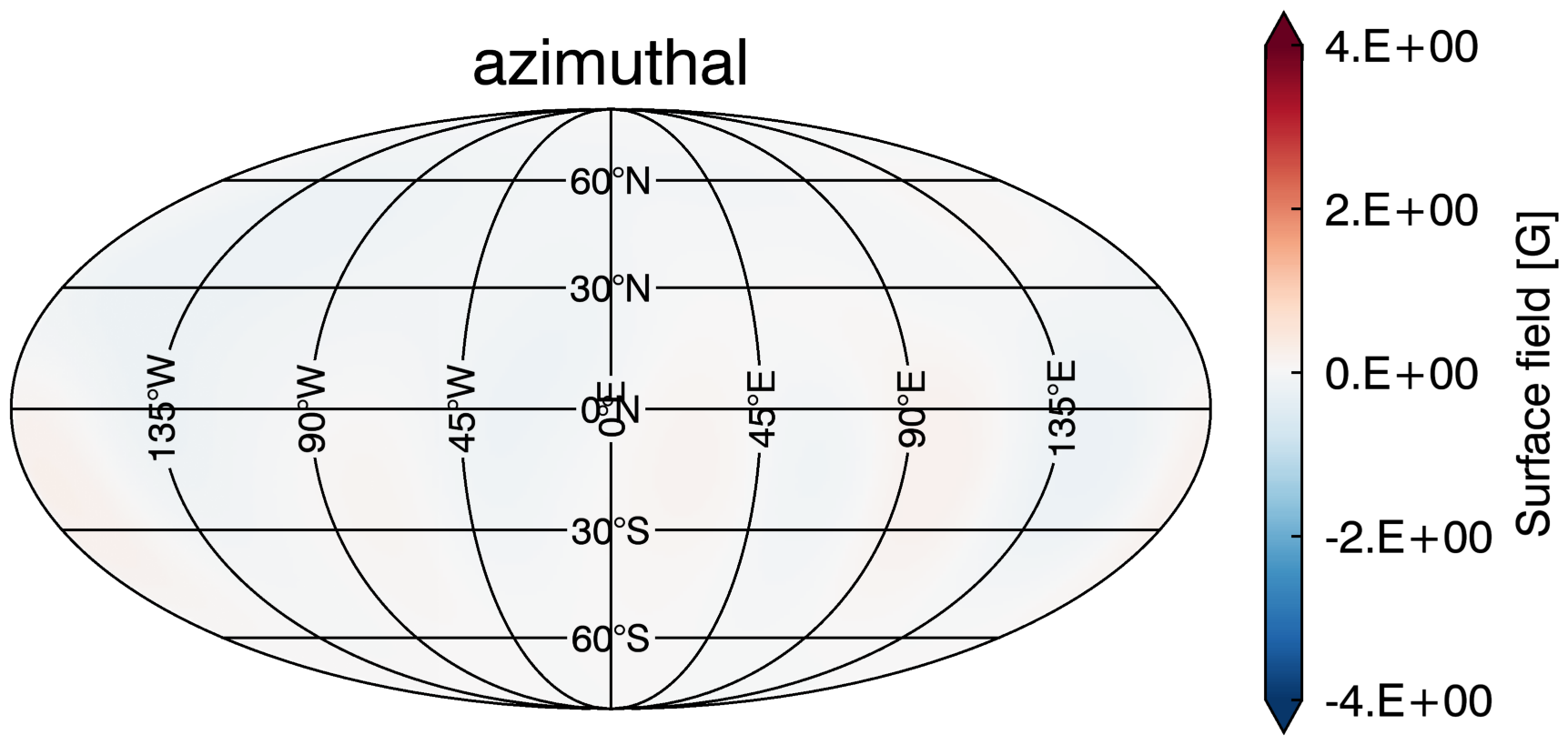} \\
\includegraphics[angle=0,height = 2.3cm ,clip]{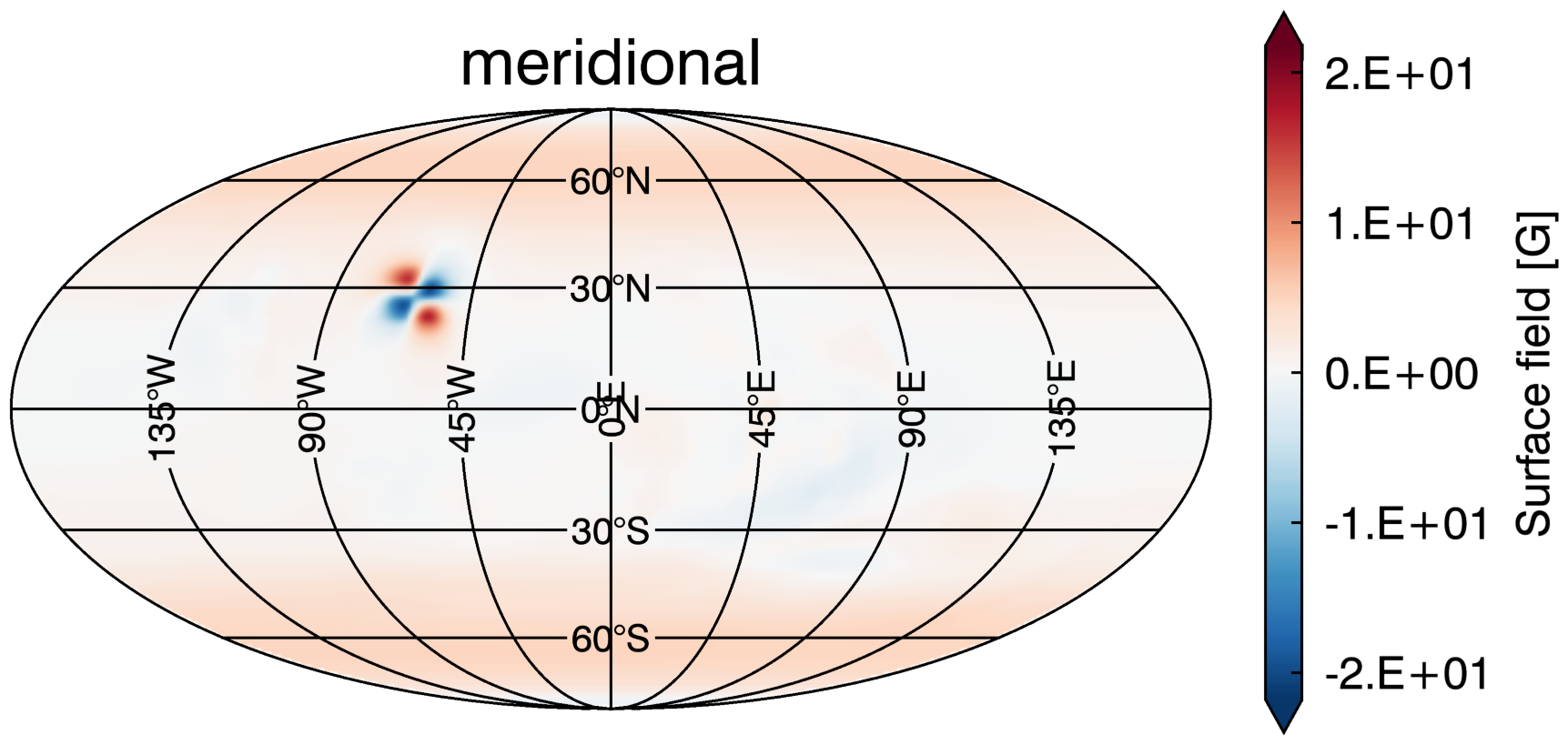}
\includegraphics[angle=0,height = 2.3cm , trim={0 0 3cm  0} ,clip]{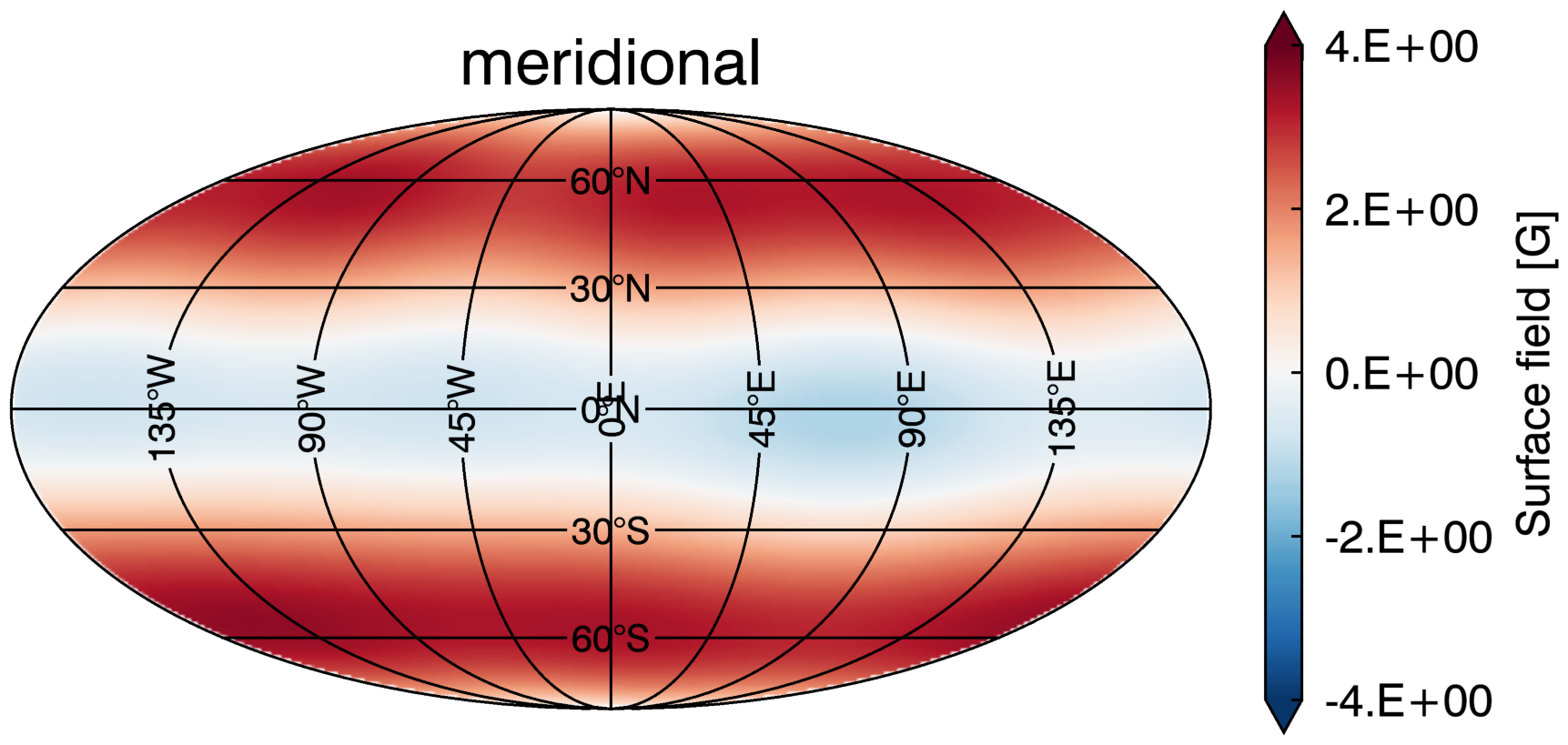}
\includegraphics[angle=0,height = 2.3cm ,clip]{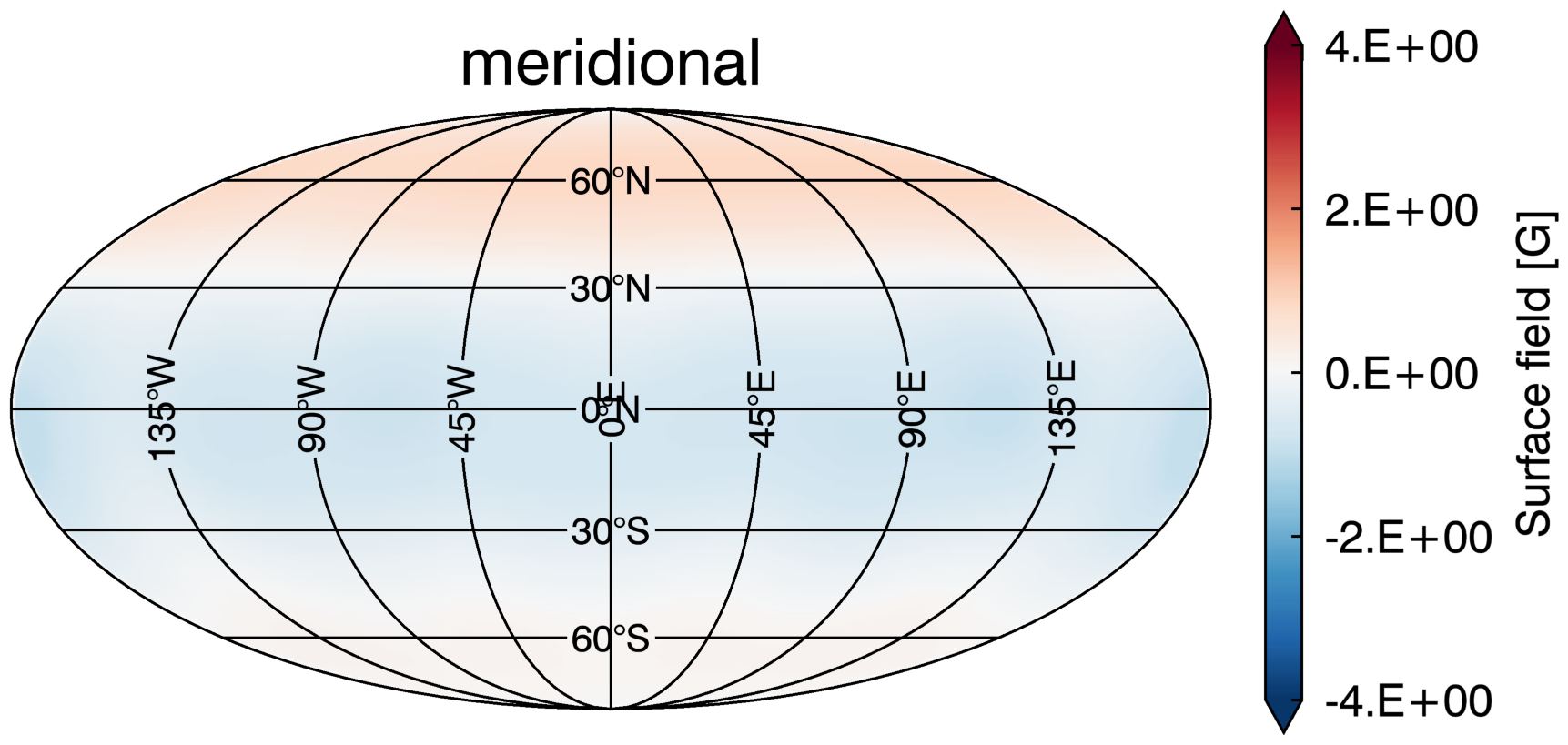} \\
\end{minipage}
\caption{The Stokes~V profiles and Mollweide projected maps for map number 4680 (year = 2008.82, minimum phase). The same format is used as for Fig.~\ref{Fig:Map0720}.}
\label{Fig:Map4680}
\end{figure*}

The left panels of  Figures~\ref{Fig:Map0720}-\ref{Fig:Map4680} show the   Stokes~V spectra fits obtained from ZDI for three maps taken from our dataset. We present one map for the rising phase of the cycle (map 0720 corresponds to the year 1997.97), one for activity maximum (map 1620 = 2000.44) and one from activity minimum (map 4680 = 2008.82). Table~\ref{Tab:MapsofSC23} lists further information about the individual maps. The  Stokes~V profiles with noise added are plotted as black solid lines and their ZDI fits as red solid lines. In addition, we overplot the original noise-free Stokes~V profiles as blue dashed lines to indicate what was initially input. In most cases the ZDI fits and the noise-free input Stokes~V profiles are very similar.

The three example maps presented here were fitted with reduced $\chi^2 = 1.1-1.2$. 
We fit every map to the level of $\chi^2$ where the map and magnetic field values no longer change significantly, when pushing it to lower $\chi^2$ values, but are not overfit. If we push the $\chi^2$ lower by 0.1, the changes in the maps can barely be seen with the eye. The magnetic energy values change at most by $2-4\%$ and the fractions only less than $0.5\%$. Compared to the general spread in the input and ZDI maps this effect is very small. By pushing the fit to even lower $\chi^2$ values, we start to overfit the maps. This can be seen by a regular non axisymmetric pattern starting to appear predominantly at the southern hemisphere and an increase in the non-axisymmetric energy and fractions.
The best fit was achieved for map 0945 with a reduced $\chi^2 = 1.05$. The ZDI maps were successfully fitted with $\chi^2 \le 1.8$ with one exception: map 1665 could only be fitted with a reduced $\chi^2 = 2.7$; this is the map with the highest number of sunspots. We noticed that the Stokes~V amplitude decreases around activity maximum. The maps around the activity maximum are therefore more strongly affected by the noise, which is set to be constant along the cycle, due to their lower Stokes~V amplitude. Small mismatches (barely noticeable by eye) can cause quickly rising $\chi^2$ values for such low-amplitude Stokes~V signatures. 
A low-amplitude Stokes~V signature at activity maximum might appear counter-intuitive at first glance. 
During  solar activity maximum, the solar surface is filled with strong magnetic field regions with sizes that are still far below the corresponding spatial resolution, so the strongest magnetic fields remain undetectable in the disc-integrated Stokes~V profiles and therefore in the ZDI maps. The contributions from the magnetic field regions cancel out within each resolution element across the stellar disc, the size of which is relatively large for the slowly rotating Sun; reaching an angular size of $\theta \approx 35^{\circ}$ in diameter at the equator. 

The panels to the right in Figures~\ref{Fig:Map0720}--\ref{Fig:Map4680} display the Mollweide projected maps for the radial (top), azimuthal (middle) and meridional field (bottom) for the Stokes~V timeseries displayed in the left panel of each figure. 
The maps in the left column are the fully resolved simulated maps from which the  Stokes~V profiles were modelled. The next column shows the large-scale field for $\ellsum\ = 3$ of the simulation and the last column shows the ZDI reconstructed maps. The ZDI reconstructed maps show the best agreement to the magnetic field structures of the $\ellsum\ = 3$ of the simulation. 
ZDI is restricted by the stellar  inclination angle of $60^{\circ}$, which prevents the detection of parts of the southern hemisphere magnetic field. The maps therefore show  very few structures below $30^{\circ}\mrm{S}$ in latitude. 
We see that the magnetic features in the radial map and the azimuthal map are well recovered by ZDI, e.g. for the more structured maps of the increasing and the maximum phase, see Fig.~\ref{Fig:Map0720} and \ref{Fig:Map1620}. The meridional component is sometimes affected by crosstalk with the radial field, especially for low-amplitude Stokes~V observations during the activity maximum, see Fig.~\ref{Fig:Map1620}.
The dipole-dominated field during the activity minimum is also very well reconstructed by ZDI, see Fig.~\ref{Fig:Map4680}. This map also illustrates very well that the  large-scale field of the Sun, and therefore also ZDI, is blind to the presence of small-scale single bipolar sunspots. 

\label{Page:StatAnalysis}
To observe a stellar magnetic cycle with 41 well-sampled ZDI maps is still very unlikely even for longer solar-like cycles. We apply therefore a small statistical analysis to mimic eight observations per cycle. The timescale of the simulated cycle is divided into eight bins equally spaced in time. As the selected 41 maps are not equally spaced in time, it appears that some bins contain a higher number of maps than others. We randomly select one map per bin and repeat the draw of the 8 maps 100 times.

\onecolumn

%
%
%
\begin{longtable}{cc|ccc|ccc|cc}
\caption{Selected data of the 41 simulated input maps and their ZDI reconstructed maps. The table is giving the map index, the time in years and for the cumulative $\ellsum\ = 3$ the total magnetic energy $\langle B^2_{\ellsum\ = 3, \mrm{tot}} \rangle$, the fraction of the poloidal field $f_{\ellsum\ = 3, \mrm{pol}}$, the fraction of the axisymmetric field $f_{\ellsum\ = 3, \mrm{axi}}$ for the simulated input maps and their ZDI reconstructed maps. The last two column refers to the sunspot number (SSN) and S-index averaged over the Carrington rotation per map. The map index in the first column is a simple index number to identify single maps, where a bold map index highlights the maps presented in Fig.~\ref{Fig:Map0720}-\ref{Fig:Map4680}.}
\label{Tab:MapsofSC23}\\
\hline
& & \multicolumn{3}{c}{Simulated Input Map} & \multicolumn{3}{c}{ZDI Reconstructed Map} & & \\
Map Index & Time & $\langle B^2_{\ellsum\ = 3, \mrm{tot}} \rangle$ & $f_{\ellsum\ = 3, \mrm{pol}}$  & $f_{\ellsum\ = 3, \mrm{axi}}$ & $\langle B^2_{\ellsum\ = 3, \mrm{tot}} \rangle$ & $f_{\ellsum\ = 3, \mrm{pol}}$  & $f_{\ellsum\ = 3, \mrm{axi}}$ & SSN & S-index \\
& [yr] & [$\mrm{G^2}$] & [\%] & [\%]& [$\mrm{G^2}$] & [\%] & [\%] &  &  \\
\hline
\endfirsthead
\caption[]{continued}\\
\hline
Map Index & Time & $\langle B^2_{\ellsum\ = 3, \mrm{tot}} \rangle$ & $f_{\ellsum\ = 3, \mrm{pol}}$  & $f_{\ellsum\ = 3, \mrm{axi}}$ & $\langle B^2_{\ellsum\ = 3, \mrm{tot}} \rangle$ & $f_{\ellsum\ = 3, \mrm{pol}}$  & $f_{\ellsum\ = 3, \mrm{axi}}$ & SSN & S-index \\
& [yr] & [$\mrm{G^2}$] & [\%] & [\%] & [$\mrm{G^2}$] & [\%] & [\%] &  & \\
\hline
\endhead
\hline
\endfoot
450 & 1997.23 &  10.19 &   88  &   94  &   1.44 &   86  &   95  &   12.1 & 0.1631 \\
585 & 1997.60 &   8.77 &   90  &   92  &   1.46 &   91  &   96  &   35.7 & 0.1630 \\
\textbf{720} & 1997.97 &   9.83 &   85  &   64  &   1.80 &   89  &   70  &   55.5 & 0.1691 \\
855 & 1998.34 &  10.71 &   81  &   44  &   2.21 &   88  &   63  &   74.0 & 0.1704 \\
945 & 1998.59 &   8.73 &   86  &   48  &   1.39 &   93  &   73  &  121.1 & 0.1687 \\
990 & 1998.71 &  12.26 &   91  &   36  &   1.95 &   91  &   41  &  132.0 & 0.1716 \\
1035 & 1998.84 &  10.20 &   86  &   33  &   2.67 &   87  &   23  &   97.3 & 0.1716 \\
1215 & 1999.33 &   5.94 &   82  &   27  &   1.02 &   79  &   32  &   93.6 & 0.1721 \\
1260 & 1999.45 &   8.21 &   73  &   30  &   0.90 &   71  &   49  &  207.2 & 0.1739 \\
1395 & 1999.82 &  19.52 &   93  &    6  &   4.26 &   81  &    6  &  168.7 & 0.1727 \\
1440 & 1999.95 &  31.17 &   90  &    7  &   7.42 &   84  &    3  &  116.8 & 0.1752 \\
1530 & 2000.19 &  18.34 &   75  &    9  &   5.63 &   82  &    2  &  217.7 & 0.1737 \\
\textbf{1620} & 2000.44 &   9.53 &   63  &   39  &   3.18 &   49  &   46  &  188.0 & 0.1772 \\
1665 & 2000.56 &  10.60 &   83  &   22  &   1.58 &   69  &   23  &  244.3 & 0.1796 \\
1710 & 2000.68 &  10.17 &   81  &   27  &   0.87 &   75  &   20  &  156.0 & 0.1826 \\
1845 & 2001.05 &   9.30 &   77  &   28  &   2.51 &   70  &   10  &  142.6 & 0.1754 \\
1935 & 2001.30 &   8.37 &   93  &   25  &   2.33 &   83  &    5  &  161.7 & 0.1765 \\
1980 & 2001.42 &  10.59 &   91  &   22  &   1.64 &   86  &   10  &  202.9 & 0.1751 \\
2025 & 2001.55 &  12.90 &   74  &   21  &   2.38 &   88  &    3  &  123.0 & 0.1759 \\
2205 & 2002.04 &  35.68 &   81  &   13  &   3.84 &   83  &    8  &  184.6 & 0.1798 \\
2295 & 2002.29 &  21.72 &   82  &   19  &   4.30 &   85  &   10  &  186.9 & 0.1762 \\
2385 & 2002.53 &  15.32 &   92  &   30  &   1.54 &   92  &   22  &  161.0 & 0.1753 \\
2430 & 2002.66 &  13.25 &   89  &   30  &   1.78 &   81  &   42  &  175.6 & 0.1758 \\
2565 & 2003.03 &  17.35 &   82  &   34  &   4.25 &   82  &   21  &  133.5 & 0.1742 \\
2700 & 2003.40 &   8.73 &   94  &   69  &   1.79 &   97  &   63  &   86.8 & 0.1706 \\
2790 & 2003.64 &  11.44 &   89  &   66  &   1.77 &   96  &   73  &  115.4 & 0.1705 \\
2925 & 2004.01 &  18.36 &   85  &   54  &   1.83 &   96  &   52  &   72.2 & 0.1735 \\
3060 & 2004.38 &  14.97 &   91  &   66  &   1.78 &   98  &   74  &   72.8 & 0.1668 \\
3195 & 2004.75 &  19.77 &   89  &   55  &   3.17 &   91  &   40  &   48.8 & 0.1656 \\
3285 & 2005.00 &  17.36 &   93  &   61  &   5.35 &   91  &   44  &   28.9 & 0.1664 \\
3420 & 2005.37 &  13.73 &   94  &   83  &   3.25 &   97  &   88  &   61.9 & 0.1653 \\
3555 & 2005.74 &  15.64 &   97  &   82  &   2.85 &   97  &   83  &   37.2 & 0.1635 \\
3690 & 2006.11 &  14.10 &   96  &   89  &   1.94 &   99  &   91  &    5.7 & 0.1651 \\
3915 & 2006.73 &  14.16 &   97  &   86  &   3.12 &   99  &   88  &   23.7 & 0.1614 \\
4185 & 2007.47 &  13.20 &   98  &   94  &   2.68 &   99  &   93  &   21.3 & 0.1639 \\
4410 & 2008.08 &  12.85 &   99  &   95  &   2.61 &  100  &   97  &    4.1 & 0.1637 \\
\textbf{4680} & 2008.82 &  12.64 &  100  &   99  &   2.01 &  100  &   99  &    4.2 & 0.1631 \\
4905 & 2009.44 &  11.55 &   99  &   99  &   2.02 &   99  &   96  &    2.9 & 0.1648 \\
5175 & 2010.18 &  13.17 &   91  &   65  &   2.70 &   93  &   73  &   28.5 & 0.1642 \\
5265 & 2010.42 &   9.54 &   94  &   78  &   2.47 &   94  &   73  &   13.9 & 0.1652 \\
5400 & 2010.79 &  10.62 &   84  &   62  &   2.36 &   93  &   67  &   33.6 & 0.1649 \\
\end{longtable}

\twocolumn

\section{Discussion of the changing magnetic field topology along the activity cycle}
\label{Sec:Results}

We present an analysis of the magnetic field topologies of 118 surface magnetic field maps simulated with a 3D non-potential flux transport model based on the observations of the solar cycle 23 (SC23) and of their 41 ZDI-reconstructed maps. The simulated maps of SC23 and their ZDI reconstructions are set into the context of other cool star observations in Section~\ref{SubSec:SCvsObs}. We find that the correlations of certain magnetic field parameters with S-Index are detectable with ZDI, see Section~\ref{SubSec:ZDIobsTrends}. We also present how the magnetic field parameters evolve along the magnetic cycle in Section~\ref{SubSec:BAlongCycle} and examine whether ZDI is able to recover hints of the operating dynamo driving the magnetic cycle in Section~\ref{SubSec:DynamoHints}.

\subsection{Placing solar cycle 23 in context of cool stars observations}
\label{SubSec:SCvsObs}

%
%
%
\begin{figure}
\centering
\includegraphics[height=4.9cm, trim = {0 20 0 0}, clip]{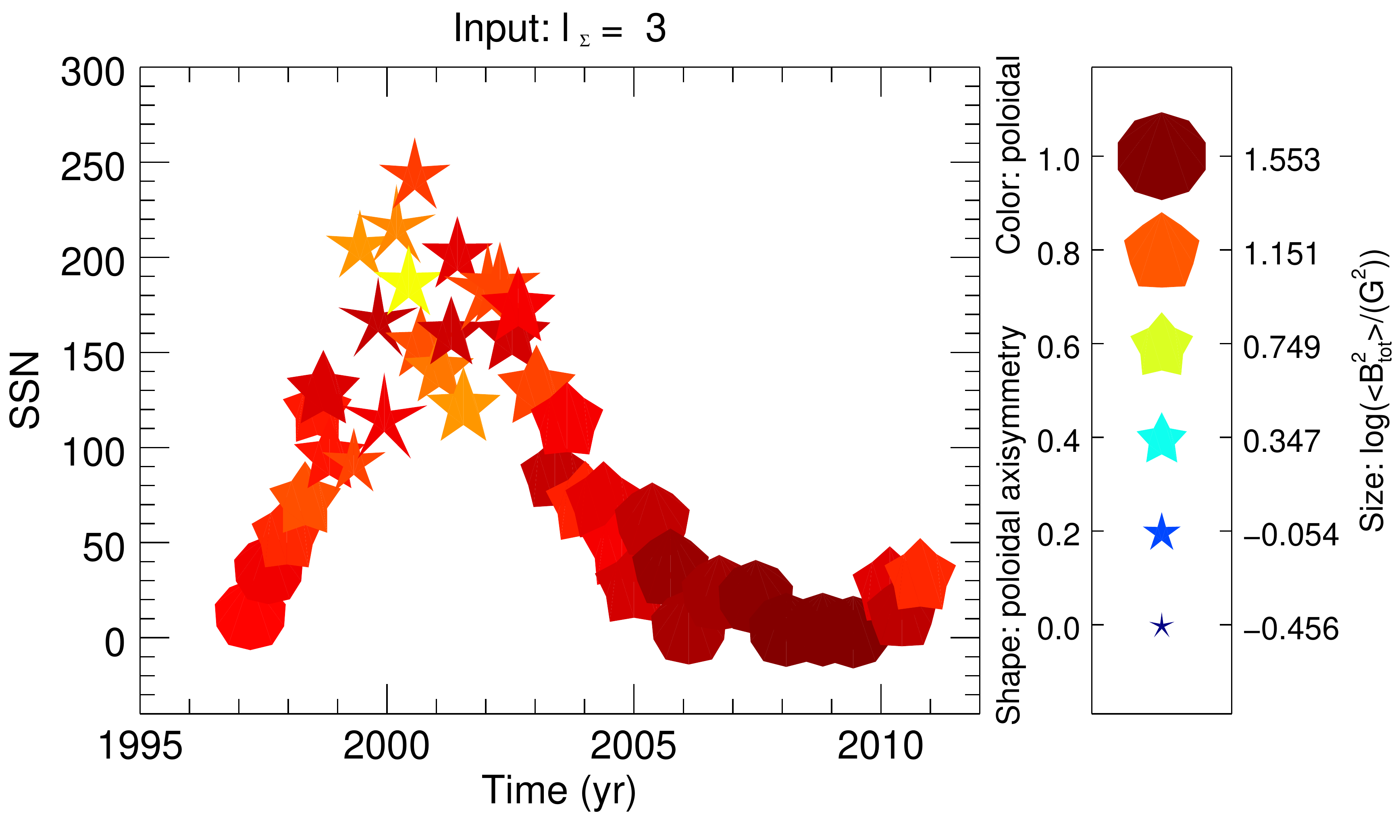}\\
\raggedright
\textbf{a.}\\
\centering
\includegraphics[height=4.9cm, trim = {0 20 0 0}, clip]{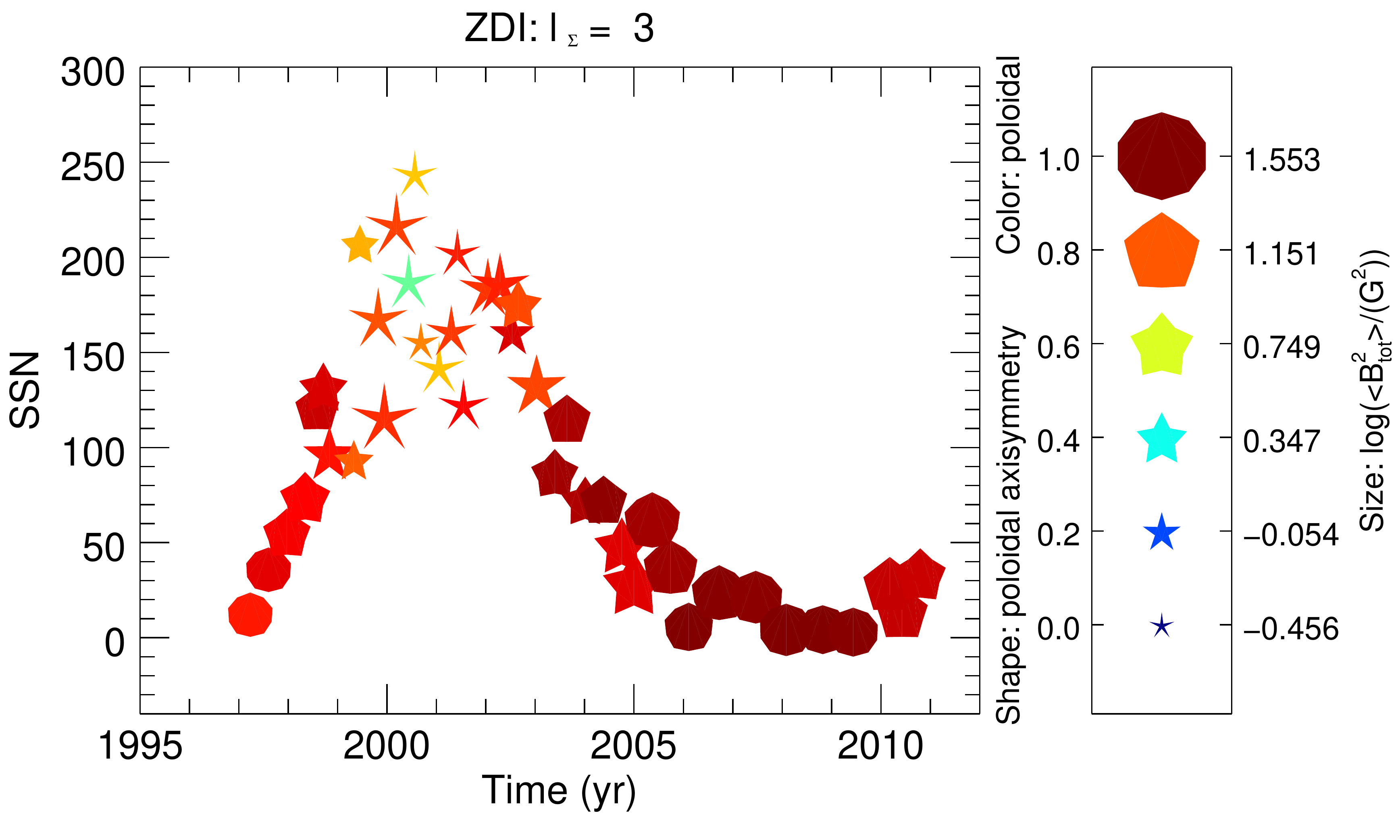}\\
\raggedright
\textbf{b.}
\caption{The large-scale magnetic field topology for the 41 simulated maps (\textbf{a.}) and their ZDI-reconstructed maps (\textbf{b.}) presenting the cumulative $\ellsum\ = 3$ mode. The y-axis indicates the sunspot number (SSN) and the x-axis the time in years. The symbol size displays the logarithmic total magnetic energy $\langle B^2_{\mrm{tot}} \rangle$. The symbol colour represents the fraction of the poloidal field $f_{\mrm{pol}}$ and the symbol shape shows the fraction of the axisymmetric poloidal field $f_{\mrm{axi, pol}}$. See the legend on the right for the numbers.}
\label{Fig:ConfusogramInpZDI}
\end{figure}

First, we compare the large-scale magnetic field properties of the simulations and their ZDI reconstructions using an often used format that was first introduced by \cite[fig.~14]{Donati2008}. Figure~\ref{Fig:ConfusogramInpZDI} presents the total magnetic energy $\langle B^2_{\mrm{tot}} \rangle$, see Eq.~\ref{Eq:Etot}, indicated by symbol size, the fraction of the poloidal field by symbol colour,
\begin{equation}
f_{\mrm{pol},\ellsum\ } = \frac{\langle B^2_{\mrm{pol},\ellsum\ } \rangle}{\langle B^2_{\mrm{tot}, \ellsum\ } \rangle},
\label{Eq:fpol}
\end{equation}
and the axisymmetric poloidal fraction by symbol shape, 
\begin{equation}
f_{\mrm{axi, pol},\ellsum\ } = \frac{\langle B^2_{m=0,\mrm{pol}, \ellsum\ } \rangle}{\sum\limits_m{\langle B^2_{m, \mrm{pol}, \ellsum\ } \rangle}},
\label{Eq:fpolaxi}
\end{equation}
for the subset of the 41 simulations (Fig.~\ref{Fig:ConfusogramInpZDI}a) and their ZDI reconstructions (Fig.~\ref{Fig:ConfusogramInpZDI}b). We present the results including only $\ell$-modes up to $\ellsum\ = 3$. We find that that the large-scale magnetic field properties as well as the maps are most similar for the cumulative $\ellsum\ = 3$ mode. In  Appendix~\ref{App:lmode} we discuss the inclusion of different cumulative $\ellsum\ $-modes and show the corresponding result of Figure~\ref{Fig:ConfusogramInpZDI} by including further $\ellsum\ $-modes. 
The simulations (Fig.~\ref{Fig:ConfusogramInpZDI}a) reproduce the observational results of solar cycle 23 very well. The magnetic topology changes rapidly around the activity maximum in 2001 and shows less poloidal and less axisymmetric poloidal topologies while huge changes appear on short time scales. At activity minimum the topology is strongly poloidal and axisymmetric poloidal, showing only small variations. Unique features of the SC23, e.g. the extended decreasing phase and the long activity minimum, are well captured by the simulations. 
The large-scale magnetic energy $\langle B^2_{\mrm{tot}} \rangle$ recovered by ZDI is smaller for every map  in all $\ellsum\ $-modes. The ZDI maps show lower poloidal and axisymmetric poloidal fractions than the input maps at the activity maximum while the topology is well recovered at the activity minimum. ZDI seems to recover the large-scale field topology very well for SSN $<100$ despite the lower magnetic energy. The simpler the field topologies become for low SSN, the easily ZDI recovers the large-scale topologies.

%
%
%
\begin{figure}
\centering
\includegraphics[angle=0,width = 0.49\textwidth ,trim = {20 10 20 20},clip]{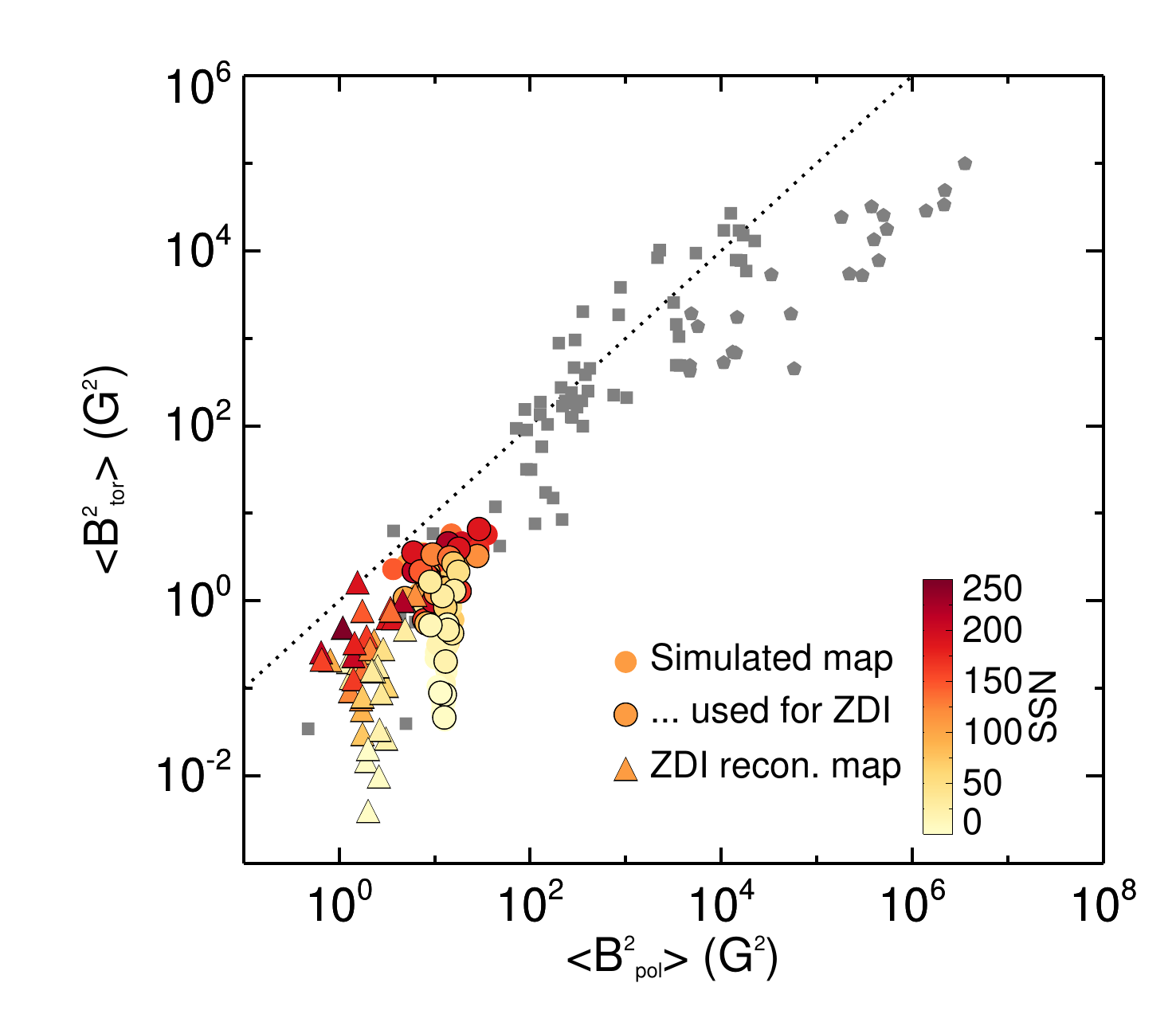}
\caption{The magnetic energy of the toroidal field $\langle B^2_{\mrm{tor}} \rangle$ over the poloidal field $\langle B^2_{\mrm{pol}} \rangle$. The observed stars are displayed by grey symbols, where stars with masses equal or higher than $0.5\,\mrm{M_{\odot}}$ are plotted as squares otherwise as pentagons. The simulated maps of solar cycle 23 are plotted as circles and ringed with a black border if they have a corresponding ZDI map and the ZDI reconstructed maps are displayed as triangles. All symbols are colour-coded by sunspot number (SSN) and the results are shown for $\ellsum\ = 3$, see legend on the right.
}
\label{Fig:EpolvsEtor}
\end{figure}

We compare the large-scale field from the solar cycle simulations with the analysis of the large scale fields recovered from ZDI maps of cool stars from \cite[Fig.~2]{See2015}; showing the toroidal $\langle B^2_{\mrm{tor}} \rangle$ against the poloidal magnetic energy $\langle B^2_{\mrm{pol}} \rangle$ for $\ellsum\ = 3$, see Fig.~\ref{Fig:EpolvsEtor}. The solar cycle simulations (circles) for high SSN occupy the same region of the plot as $\mrm{\epsilon~Eri}$ or $\mrm{HN~Peg}$. The ZDI reconstructions (triangles) show an offset of around one order of magnitude in the toroidal and poloidal energy, similar to our previous results in \cite{Lehmann2019}. There is an even higher offset in $\langle B^2_{\mrm{tor}} \rangle$ for low SSN maps. For SSN below $\approx\,50$ the toroidal magnetic energy in the simulated maps drops suddenly, while the poloidal energy stays constant, see Fig.~\ref{Fig:EpolvsEtor}. This trend is recovered well by ZDI. Or in other words, for $\mrm{SSN} > 50$ $\langle B^2_{\mrm{tor}} \rangle$ and $\langle B^2_{\mrm{pol}} \rangle$ follow the power law relation $\langle B^2_{\mathrm{tor}}\rangle \propto \langle B^2_{\mathrm{pol}}\rangle^{0.8\pm0.3}$. Below this limit $\langle B^2_{\mrm{pol}} \rangle$ stays constant and only $\langle B^2_{\mrm{tor}} \rangle$ decreases further with decreasing SSN. The higher the SSN, the higher $\langle B^2_{\mrm{tor}} \rangle$, which indicates that the large-scale toroidal field energy is driven to a significant degree by the magnetic energy of small-scale field structures like sunspots. As the dipolar field is almost zero for higher SSN, we assume that the power-law relation is mainly determined by small-scale flux emergence. For the large-scale poloidal field $\langle B^2_{\mrm{pol}} \rangle$ we see that small-scale field structures can contribute at $\mrm{SSN} > 50$, but do not have to. The lower limit of the large-scale poloidal energy $\langle B^2_{\mrm{pol}} \rangle$ ($\approx 10\,\mrm{G^2}$ for the input maps) is most likely determined by the global dipolar field. 
The ZDI-reconstructed maps show a wider spread than the input maps but note the log-log scale. Nevertheless, the trends with SSN are reasonably well recovered by ZDI.

\subsection{ZDI observable trends}
\label{SubSec:ZDIobsTrends}

We demonstrate here that magnetic field parameters show defined trends with sunspot number (SSN) and, more importantly for observers, that they also show trends with S-index that should be recoverable in ZDI maps of solar-like stars. 

%
%
%
\begin{figure}
\centering
\includegraphics[angle=0,width = 0.23\textwidth ,trim = {40 10 6 20},clip]{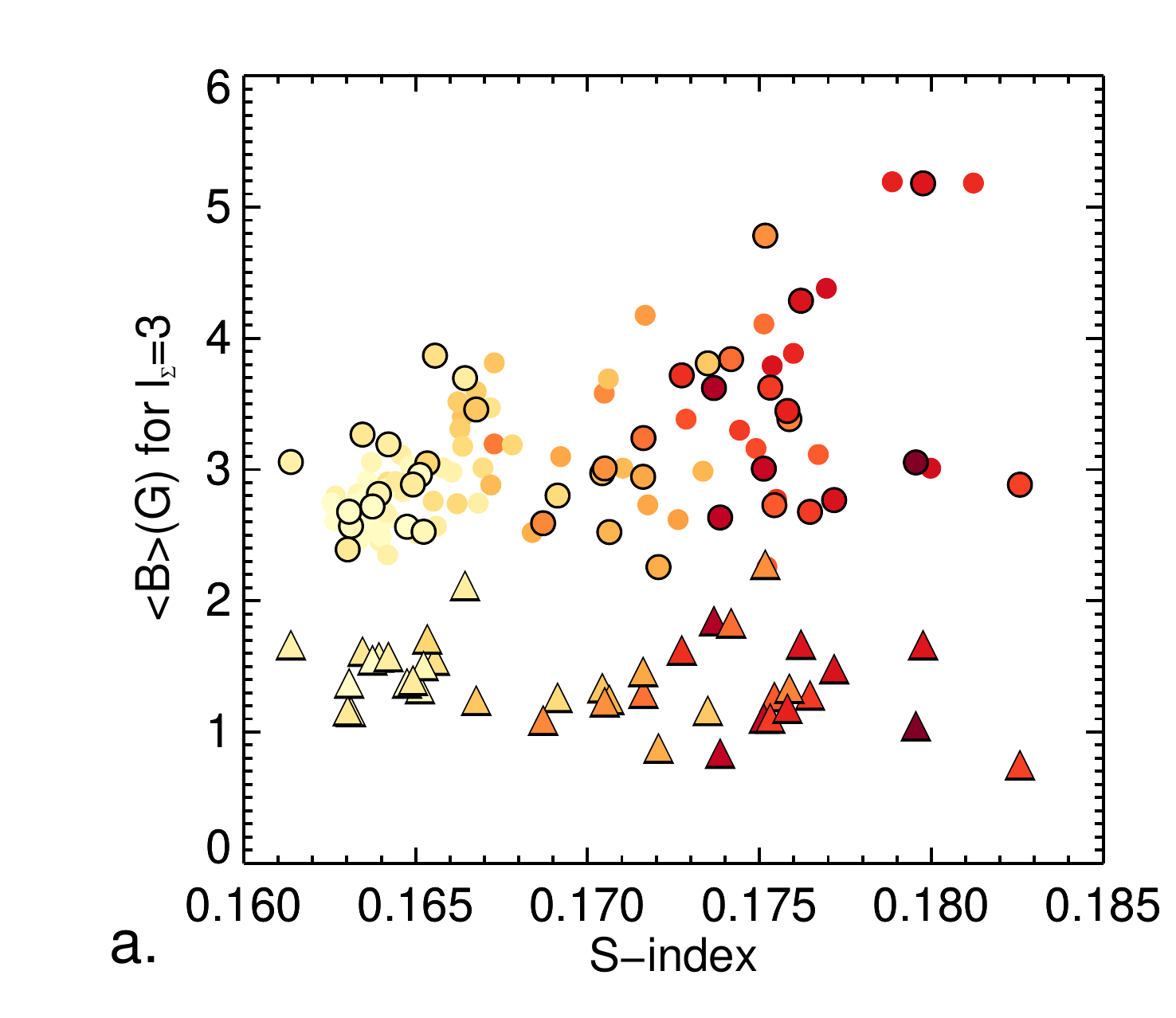}
\includegraphics[angle=0,width = 0.23\textwidth ,trim = {40 10 10 20},clip]{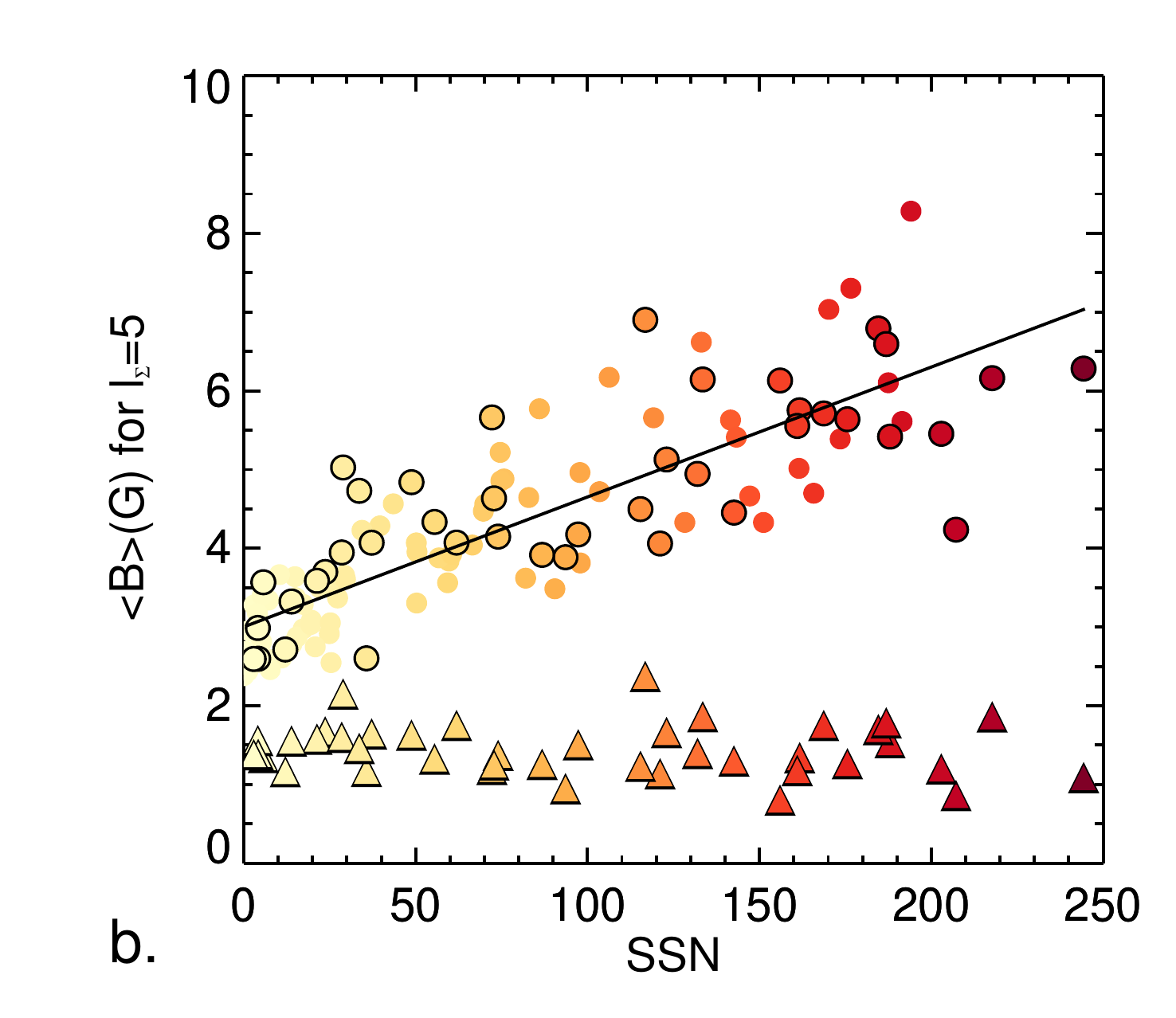}
\caption{The surface averaged large-scale magnetic field $\langle B \rangle$ including $\ellsum\ = 3$ (\textbf{a.}) and $\ellsum\ = 5$ (\textbf{b.}) plotted against S-index  (\textbf{a.}) and sunspot number (SSN) (\textbf{b.}), where the symbols are colour-coded by SSN using the same colour bar as in Fig.~\ref{Fig:EpolvsEtor} for both diagrams. The simulated input maps are displayed as circles, where the 41 maps that have a ZDI reconstructed map are ringed with a black boarder. The ZDI reconstructed maps are plotted as triangles. For \textbf{b.}: The solid black line indicates the slope of the input simulations determined via a least-squares best fit $\langle B_{\ellsum\ =5} \rangle = (3.0\pm0.1)\,\mrm{G} + (0.016\pm0.001)\,\mrm{G} \cdot \mrm{SSN} . $} 
\label{Fig:Bl3vsSIndAndBl5vsSSN}
\end{figure}

It is important to note that the surface-averaged large-scale field $\langle B \rangle$ is unable to trace the cycle. Figure~\ref{Fig:Bl3vsSIndAndBl5vsSSN}a shows  the surface-averaged large-scale field $\langle B_{\ellsum\ =3} \rangle$ against S-index, where the symbols are colour-coded by SSN, using the same format as in Fig.~\ref{Fig:EpolvsEtor}. The simulations (circles) show an upper envelope, which increases with S-index albeit with a large scatter, while the corresponding ZDI reconstructions show a flat distribution with some spread. A slowly rotating solar-like star is likely to display no detectable variation in the averaged large-scale field recovered with ZDI. 

This becomes even more striking if we go to the averaged large-scale field $\langle B \rangle$ of the ZDI maps including $\ellsum\ $-modes up to $\ellsum\ = 5$ against SSN, see Fig.~\ref{Fig:Bl3vsSIndAndBl5vsSSN}b. The input simulations show a clear increase of $\langle B_{\ellsum\ =5} \rangle$ with SSN, while the ZDI reconstructions show only a flat distribution, and generally lower values, ranging from $0.8-2.4\,\mrm{G}$. That we do not see an increase of $\langle B \rangle$ with activity is a result of the cancellation effects within the resolution elements on the projected stellar disk. For higher activity, we see that the likelihood increases of cancellation of magnetic flux of opposite polarity within each resolution element, as the number of active regions with emerging flux  increases. During less active phases the Sun  usually has only one or two large spot groups at the same time but during the maximum there are several; see also the fully resolved input maps in Figs.~\ref{Fig:Map0720} and \ref{Fig:Map1620}. Higher activity in terms of an increased number of small-scale flux events can therefore not be seen in the large-scale $\langle B \rangle$ for slow rotating solar-like stars.

When comparing simulations and reconstructions it is important to consider the number of $\ellsum\ $-modes included, see detailed discussion in Appendix~\ref{App:lmode}. The simulations and reconstructions are most similar for $\ellsum\ = 3$ but ZDI also recovers magnetic energies at higher $\ellsum\ $-modes, which are included in the full ZDI map. For our simulated Sun the ZDI recovered $\langle B \rangle$ increases only slightly by including $\ellsum\ = 5$ compared to $\ellsum\ = 3$. However, this could be different for other stars. One has to be aware, that with the restriction to $\ellsum\ = 3$, we discard information recovered with ZDI. 
We refer to the discussion in the appendix, where we also provide a version of Fig.~\ref{Fig:Bl3vsSIndAndBl5vsSSN}a displaying $\langle B_{\ellsum\ =5} \rangle$ against S-index including the higher modes $\ell = 4$ and 5, see Fig.~\ref{Fig:Bl5vsSInd}.

The attentive reader will have noticed that the trends are much clearer for SSN than for S-index. Of course, it is impossible to measure SSN  for other stars than the Sun, whereas the S-index can be measured relatively easily. In Appendix~\ref{App:SSNSInd}, we show how S-index and SSN are correlated for our data sample and discuss this further. 

%
%
%
%
\begin{figure}
\centering
\includegraphics[angle=0,width = 0.49\textwidth ,trim = {5 55 5 20},clip]{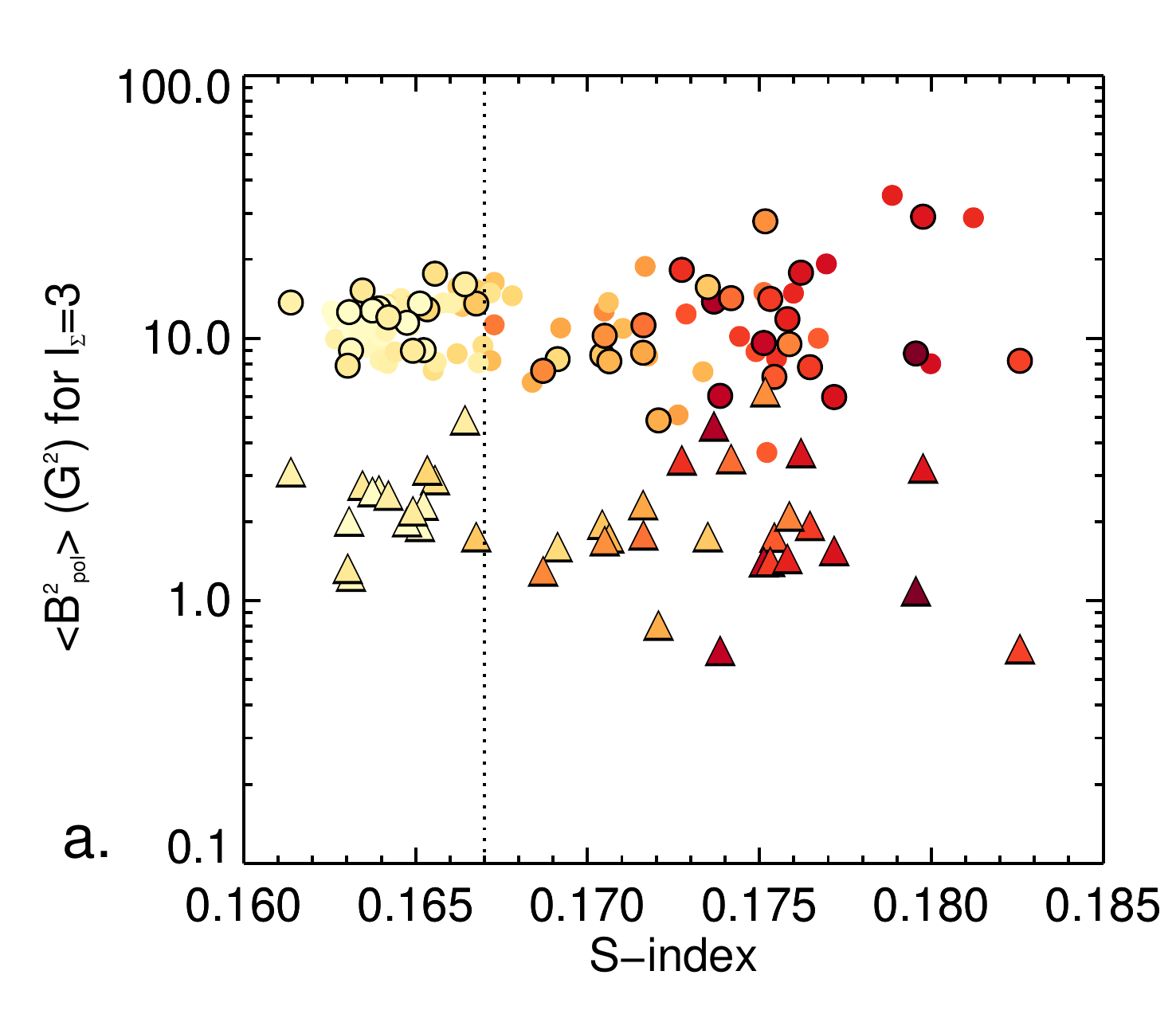}
\includegraphics[angle=0,width = 0.49\textwidth ,trim = {5 10 5 27},clip]{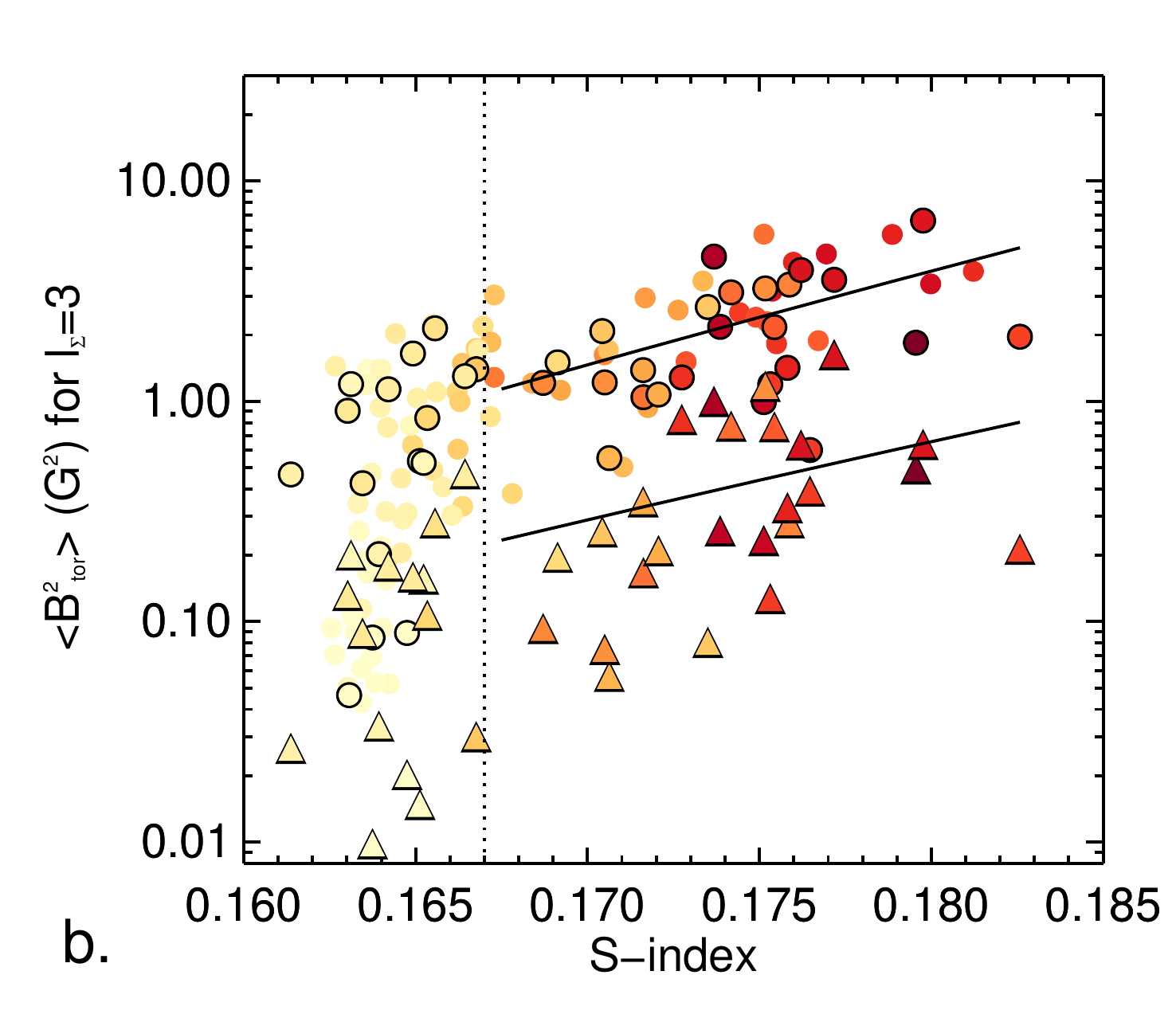}\\
\caption{The poloidal $\langle B^2_{\mrm{pol}} \rangle$(\textbf{a.}) and toroidal $\langle B^2_{\mrm{tor}} \rangle$ magnetic energy  (\textbf{b.}) against S-index for the cumulative $\ellsum\ = 3$ mode using the same format as in Fig.~\ref{Fig:Bl3vsSIndAndBl5vsSSN}a. The dashed line indicates $\mrm{S-index} = 0.167$. The solid lines indicate the least-squared fits for the simulated input maps (circles) and the ZDI maps (triangles) for the toroidal energy using a power law relation.}
\label{Fig:EpolEtorvsSInd}
\end{figure}

The poloidal energy $\langle B^2_{\mrm{pol}} \rangle$ also shows a more or less flat distribution with S-index, especially for S-index $<0.167$, although the spread increases with increasing S-index, see \ref{Fig:EpolEtorvsSInd}a. 
However, the toroidal magnetic energy $\langle B^2_{\mrm{tor}} \rangle$ shows a trend with S-index that seems to be observable by ZDI, see Fig.~\ref{Fig:EpolEtorvsSInd}b. The toroidal energy for $\ellsum\ = 3$ modes of the ZDI reconstructed maps displays an increase with S-index, as shown by the input simulations. We tested different relations between $\langle B^2_{\mrm{tor}} \rangle$ and S-index but the spread was too large to derive secure results. We found that ZDI recovers most relations within the errors, e.g. power or quadratic relations. Only the slope of a linear relation between $\log \langle B^2_{\mrm{tor}} \rangle$ and S-index could not be recovered by ZDI.
We plot a power law relation in Fig.~\ref{Fig:EpolEtorvsSInd}b for S-index $<0.167$ to highlight the increasing trend of $\langle B^2_{\mrm{tor}} \rangle$ with S-index. The S-index value of $=0.167$ correlates to a SSN $\approx 50$. We already saw in Fig.~\ref{Fig:EpolEtorvsSInd} that the behaviour of $\langle B^2_{\mrm{pol}} \rangle$ and $\langle B^2_{\mrm{tor}} \rangle$ changes at this threshold.

The ZDI-recovered increase of the toroidal energy with S-index might be a possible tracer to detect solar-like cyclic behaviour in other cool stars and even to roughly estimate the magnetic activity cycle phase. 
As the scatter is large, this indicator should be only used to indicate higher and lower activity phases and needs to treated with caution in general. The slope of the toroidal energy with S-index is shallow and for sun-like stars we expect only low toroidal energy values, which makes it challenging to use this as a definitive indicator.

The fraction of the toroidal field,
\begin{equation}
f_{\mrm{tor}} = \frac{\langle B^2_{\mrm{tor}} \rangle}{\langle B^2_{\mrm{tot}}\rangle},
\end{equation}
shows a shallow increase with S-index as well, see Fig.~\ref{Fig:ftorvsSIndex} in the Appendix and discussion there. The wide spread and the low values of $f_{\mrm{tor}}$ prevents $f_{\mrm{tor}}$ from being a reliable tracer for solar-like cycle. It can only be used as an additional check if $f_{\mrm{tor}}$ is higher for larger S-index and shows a more extended spread.

The toroidal and poloidal energies show clearer trends if plotted against SSN. The poloidal energies reveal a lower limit for low SSN that is most likely given  by the global dipolar field at activity maximum, while the toroidal energy increases with SSN, see Fig.~\ref{Fig:EpolEtorvsSSN} in the Appendix~\ref{App:SSNSInd}.

%
%
%
\begin{figure}[h]
\centering
\includegraphics[angle=0,width = 0.49\textwidth ,trim = {5 55 5 20},clip]{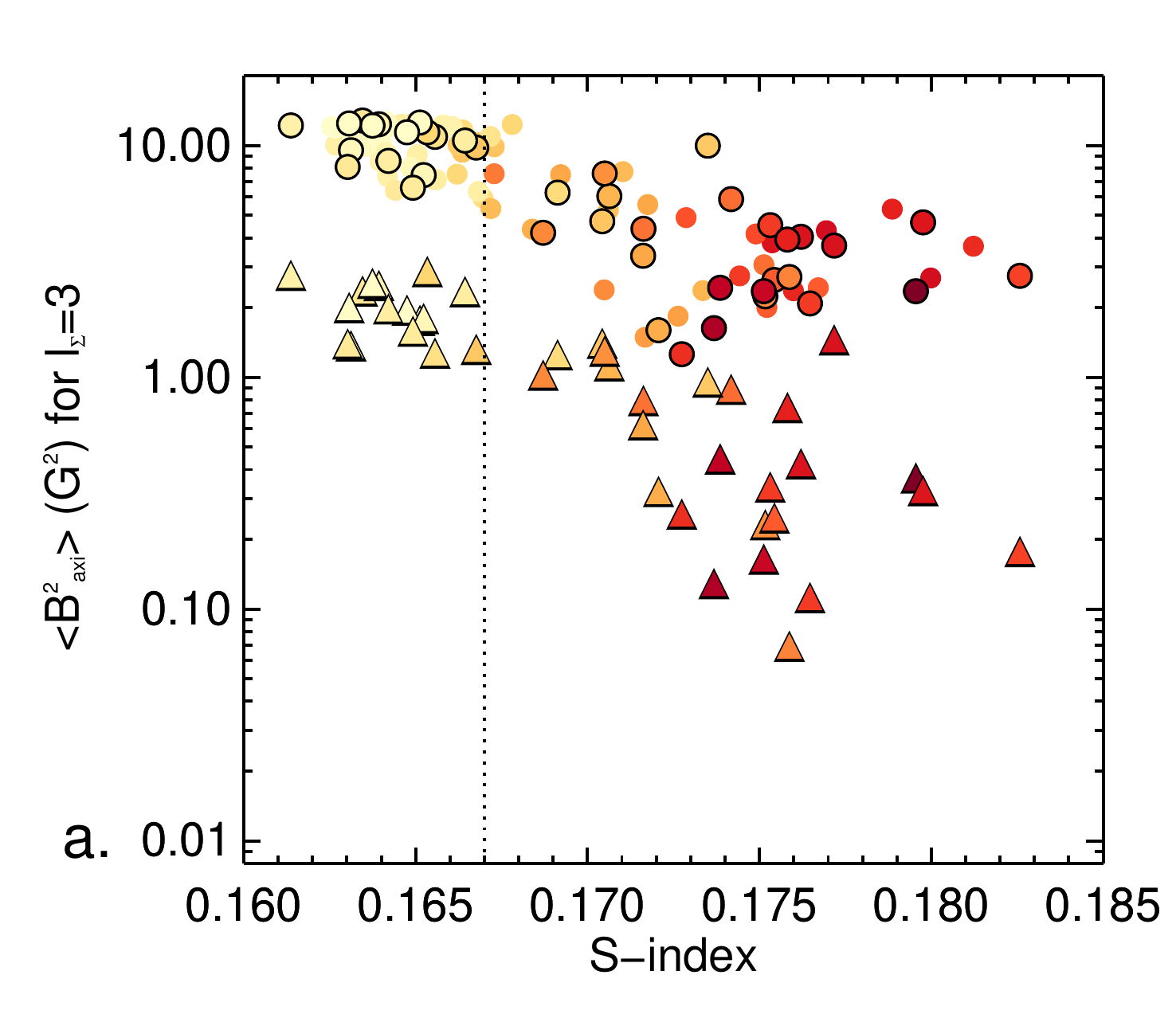}
\includegraphics[angle=0,width = 0.49\textwidth ,trim = {5 55 5 27},clip]{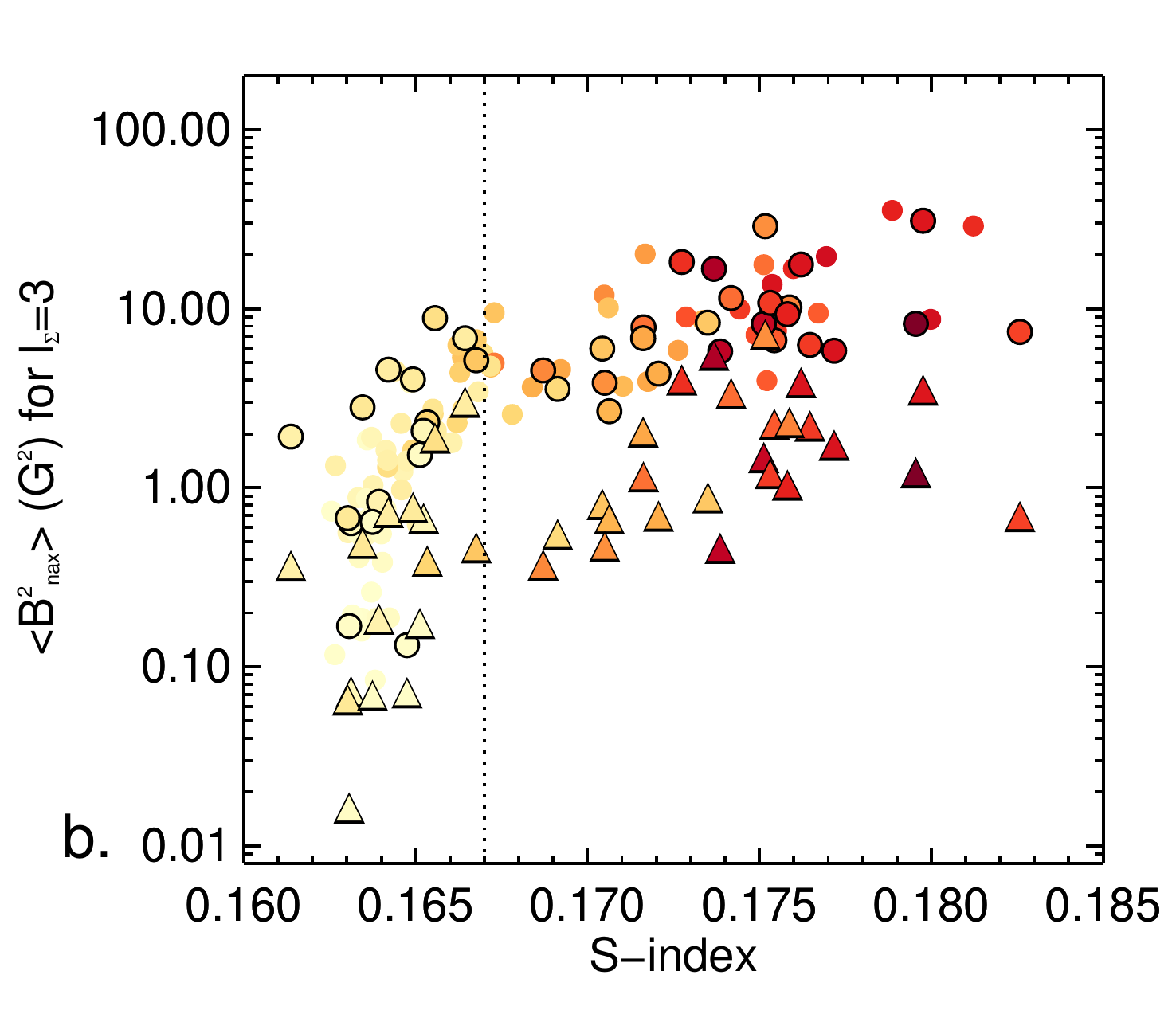}\\
\includegraphics[angle=0,width = 0.49\textwidth ,trim = {5 10 5 25},clip]{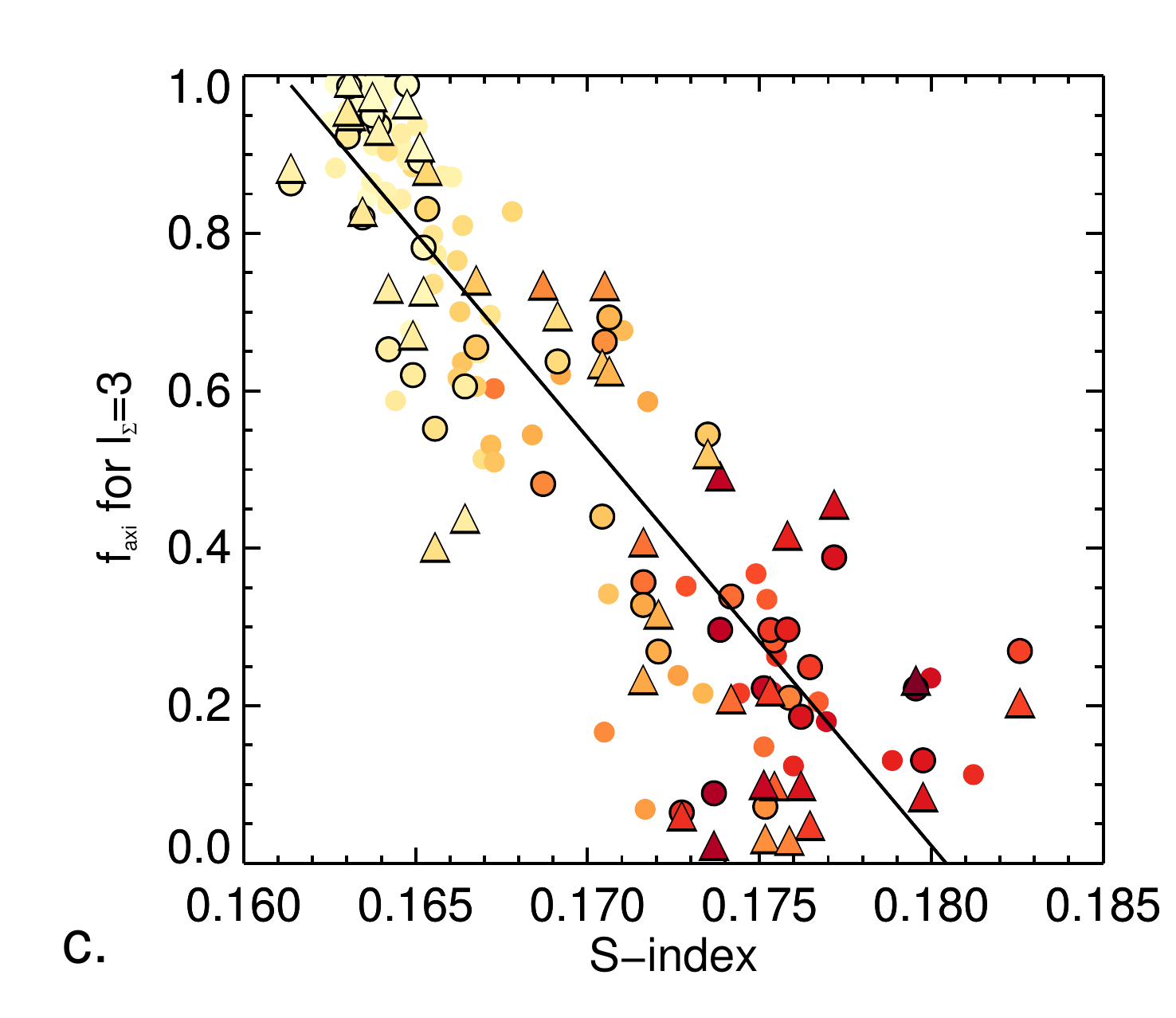}\\
\caption{The axisymmetric $\langle B^2_{\mrm{axi}} \rangle$ magnetic energy (\textbf{a.}), the non-axisymmetric $\langle B^2_{\mrm{nax}} \rangle$ magnetic energy (\textbf{b.}) and the axisymmetric energy fraction $f_{\mrm{axi}}$ (\textbf{c.}) against S-index for the cumulative $\ellsum\ = 3$. The dashed line indicates $\mrm{S-index} = 0.167$ and the solid line the least-square fit through all maps for $f_{\mrm{axi}}$. The same format is used as in Fig.~\ref{Fig:Bl3vsSIndAndBl5vsSSN}a. 
}
\label{Fig:EaxiEnaxFaxivsSInd}
\end{figure}

We find that the S-index variations of the axi- and non-axisymmetric energy are able to uncover solar-like cycles with ZDI. The axisymmetric energy is  defined here as the energy of the $m=0$-mode for the corresponding $\ell$-modes. The non-axisymmetric energy is accordingly the energy stored in all $m$-modes except for $m=0$.  Figure~\ref{Fig:EaxiEnaxFaxivsSInd} displays the axisymmetric $\langle B^2_{\mrm{axi}} \rangle$(Fig.~\ref{Fig:EaxiEnaxFaxivsSInd}a), non-axisymmetric $\langle B^2_{\mrm{nax}} \rangle$ (Fig.~\ref{Fig:EaxiEnaxFaxivsSInd}b) and the axisymmetric fraction
\begin{equation}
f_{\mrm{axi},\ellsum\ } = \frac{\langle B^2_{m=0,\ellsum\ } \rangle}{\sum_m \langle B^2_{m,\ellsum\ } \rangle}.
\label{Eq:Faxi}
\end{equation}
(Fig.~\ref{Fig:EaxiEnaxFaxivsSInd}c) against S-index using the same format as before. 
The axisymmetric energy decreases with S-index, which is recovered by ZDI. For S-index $> 0.167$ the spread increases significantly. For the non-axisymmetric energy we see a strong drop of $\langle B^2_{\mrm{nax}} \rangle$ for S-index $< 0.167$. This drop is well recovered by ZDI and indicates the activity minimum at low SSN (colour-code). 

We find that the axisymmetric fraction $f_{\mrm{axi}}$ is one of the best tracers to detect solar-like activity cycle with ZDI maps. The axisymmetric fraction decreases steeply with S-index following the linear relation
\begin{equation}
f_{\mrm{axi}} \propto (-52 \pm 2) \cdot \text{S-index}
\label{Eq:LinLawfaxiSInd}
\end{equation}
The ZDI reconstructions perfectly recover the trend of the simulations and show the same slope, within the error.
As $f_{\mrm{axi}}$ ranges from $1.0$ to $0.0$ with S-index, a large parameter range is covered over the course of one magnetic cycle, increasing the likelihood to detect this variation for other cool stars. \cite{Lehmann2019} also showed, that $f_{\mrm{axi}}$ is well recovered for stars similar to the Sun observed with a more equator-on inclination, making changes in $f_{\mrm{axi}}$ the most significant marker of solar-like activity cycle with ZDI. The ZDI-detected magnetic cycle of the K~dwarf 61~Cyg~A shows a minimum in $f_{\mrm{axi}}$ for high S-index values \citep{BoroSaikia2016} as well, indicating solar-like cyclic behaviour.

To summarise, we find that ZDI maps can recover the decrease in the axisymmetric fraction $f_{\mrm{axi}}$ with S-index and may be able to recover the increase of the toroidal energy $\langle B^2_{\mrm{tor}} \rangle$ with S-index if $\langle B^2_{\mrm{tor}} \rangle$ is strong enough. We recommend to focus on the axisymmetric fraction to detect solar-like cyclic variations using S-indices. In addition, the non-axisymmetric energy $\langle B^2_{\mrm{nax}} \rangle$ drops suddenly for S-index $< 0.167$, confirming the detection of a magnetic activity minimum. We reiterate that the averaged large-scale field $\langle B \rangle$ or the polodial energy $\langle B^2_{\mrm{pol}} \rangle$ show little variation with S-index in the ZDI-reconstructed maps and are therefore unsuitable tracers of solar-like activity cycles.

\subsection{The variation of the magnetic field properties along solar cycle 23}
\label{SubSec:BAlongCycle}

In this section, we analyse the variation of  magnetic field parameters with time.

%
%
%
\begin{figure}
\centering
\includegraphics[angle=0,width = 0.5\textwidth,clip]{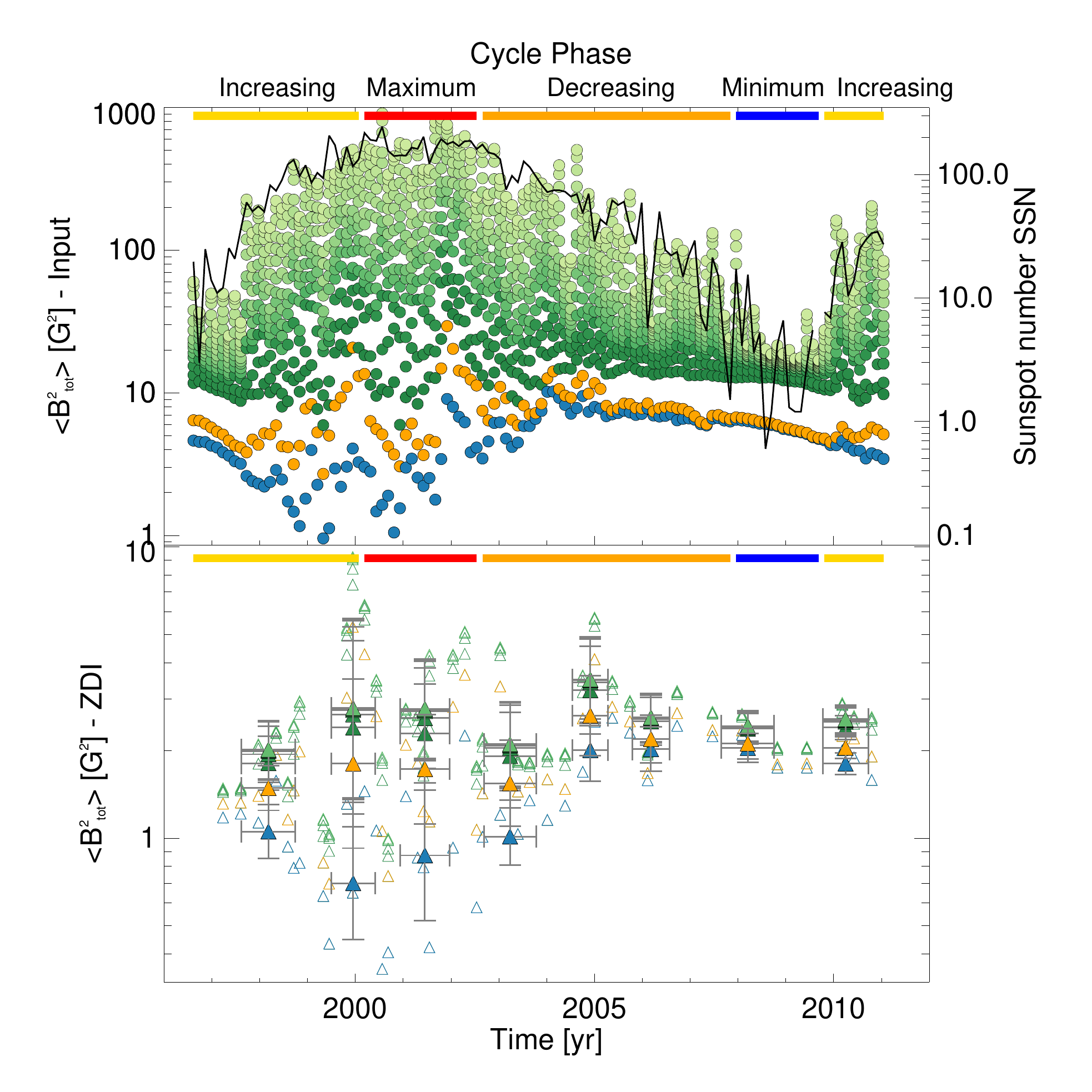}\\
\caption{The total magnetic energy $\langle B^2_{\mrm{tot}} \rangle$ over time for the cumulative $\ellsum\ $-modes. The corresponding cycle phases are indicated by the coloured thick lines at the top of each plot. \textbf{Top:} All 118 simulated maps of solar cycle 23. The dipole mode $\ell = 1$ is blue, the quadrupolar mode $\ellsum\ = 2$ is orange and all higher cumulative modes $\ellsum\ = 3-27$ are plotted as greenish circles, where the colour lightens with $\ellsum\ $--mode. The sunspot number is over-plotted as black line, see y-axis on the right. \textbf{Bottom:} All 41 ZDI reconstructed maps (open triangles) and their binned mean (filled triangles) and standard derivation (error bars) for $\ellsum\ = 1-7$. The same colour schema is used as in the top panel. }
\label{Fig:EtotvsTime}
\end{figure}

First, we show the evolution over time of the total magnetic energy $\langle B^2_{\mrm{tot}} \rangle$. Figure~\ref{Fig:EtotvsTime} top displays $\langle B^2_{\mrm{tot}} \rangle$ against time for the 118 simulations for a range of cumulative $\ellsum\ $--modes $\ellsum\ = 1-28$, indicated by colour. The dipole mode is blue, the quadrupole mode orange and the higher $\ellsum\ $--modes varying green, with the colour getting lighter with higher $\ellsum\ $. The logarithmic SSN is displayed as the black line (see right y-axis). The total magnetic energy of the  cumulative high-\lmod\ \ closely follows the SSN evolution. They trace the magnetic energy of both the large and small-scale fields. The low $\ellsum\ $--modes, which are only sensitive to the large-scale field, show very different trends. The dipole mode $\ell =1$ (blue circles) decreases during the increasing cycle phase, is at a minimum at activity maximum and rises rapidly at the beginning of the decreasing phase, before a slow decline. The cumulative quadrupolar field follows the SSN trend: it increases during the increasing phase, undergoes a maximum at activity maximum and decreases until activity minimum. The simulations reproduce the observed trends in the solar dipole and quadrupole, \cite{DeRosa2012}. The $\ellsum\ = 2$ mode shows a double peak feature during the maximum phase, which is hard to see in the SSN on the log-scale. In general, the large-scale field and the total field including the small-scale structures in the simulations behave as expected from  solar observations, see e.g.\ \cite{DeRosa2012,Kakad2019}. 

The lower plot in Fig.~\ref{Fig:EtotvsTime} shows the 41 ZDI reconstructed maps (empty triangles). The mean and standard derivation of the statistical analysis for the eight bins (introduced at Section~\ref{SubSec:ZDI}) are plotted as filled triangles and corresponding error bars. Note that the range of the $\langle B^2_{\mrm{tot}} \rangle$-axes is different for the top and bottom panels. The ZDI maps are only sensitive to the large-scale field. The first seven \lmod\ \ are displayed using the same colour-scheme as on the top panel. The ZDI-recovered total magnetic energy $\langle B^2_{\mrm{tot}} \rangle$ is approximately one order of magnitude lower compared to the input maps as seen in Fig.~\ref{Fig:EpolvsEtor}. There is a large spread during the activity maximum. We find that, by selecting randomly eight maps across the cycle, one could be lucky and detect a solar-like cycle, but it is also very likely to see no trend or a trend shifted in phase. The large spread prevents a secure and reproducible detection. 
Using $\langle B^2_{\mrm{tot}} \rangle$ from ZDI maps to estimate the cycle phase or even to detect activity cycles on stars similar to our Sun is therefore not recommended. The best tracer among those plotted in Fig.~\ref{Fig:EtotvsTime} is  the dipole mode, which shows the largest variability along time. All higher $\ellsum\ $--modes are, on average over several maps, more or less constant along the cycle.

%
%
%
%
\begin{figure}
\centering
\includegraphics[angle=0,width = 0.5\textwidth]{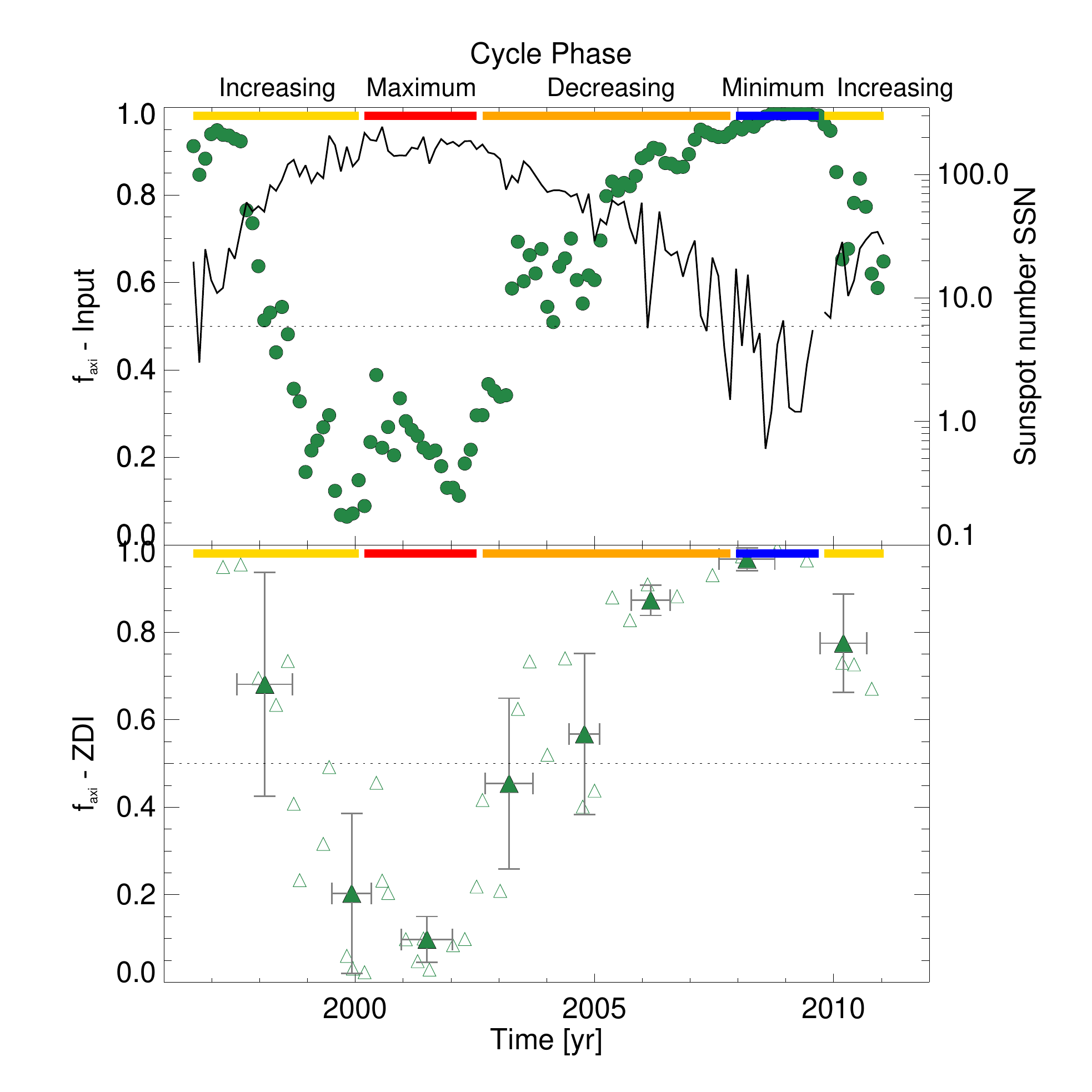}\\
\caption{The fraction of the axisymmetric field $f_{\mrm{axi}}$ for the cumulative $\ellsum\ $-mode $\ellsum\ = 3$. The same format as in Fig.~\ref{Fig:EtotvsTime} is used but only $\ellsum\ = 3$ is plotted.}
\label{Fig:faxivsTime}
\end{figure}

In the previous section we saw, that the fraction of the axisymmetric field  varies significantly over the solar cycle, and that ZDI is able to detect that variation making this magnetic field parameter one of the best cycle tracers.
In the Appendix~\ref{App:AddFigures} we show the  evolution over time of the axisymmetric and non-axisymmetric energy separately, see Fig.~\ref{Fig:EaxiEnaxvsTime} as they show  very distinct trends with time, which are well recovered with ZDI.
Here, we continue with Figure~\ref{Fig:faxivsTime} which shows the  evolution over time of $f_{\mrm{axi}, \ellsum\ = 3}$ combining both quantities, see Eq.~\ref{Eq:Faxi}, for the cumulative $\ellsum\ = 3$ using the same format as Fig.~\ref{Fig:EtotvsTime}. 

The simulations (top panel) show a rapid decrease during the increasing phase in connection with a broad minimum during the maximum activity phase and a slow nearly linear increase during the decreasing phase. ZDI (bottom panel) "sees" fewer details in the increasing and decreasing phase, but clearly catches  the maximum and minimum phases, highlighting how well $f_{\mrm{axi}}$ is able to trace solar-like cycles.

The trend of $f_{\mrm{axi}}$ with time is mainly driven by the axisymmetric poloidal field, see Eq.~\ref{Eq:fpolaxi}. 
\cite{Lehmann2019} showed that $f_{\mrm{axi, pol}}$ is best recovered by ZDI for stars similar to the Sun. The trends with time and S-index look similar to $f_{\mrm{axi}}$ but show even a bit less spread and stronger minima and maxima. 
The fraction $f_{\mrm{axi, pol}}$ is therefore an equally good tracer for solar-like cycles. The axisymmetric toroidal field displays only a large spread independent of activity tracers or time and is not usable for solar cycle detection.

The evolution with time of the toroidal energy fraction $f_{\mrm{tor}}$ is not a reliable tracer for solar-like cycle, see Fig.~\ref{Fig:ftorvsTime} in the appendix and discussion there. A star similar to the Sun is too inactive to use $f_{\mrm{tor}}$ as trustworthy cycle tracer. However, the detection of the $f_{\mrm{tor}}$ maximum can be used as additional check to confirm the detection of the activity maximum for the solar-like cycle. For stars more active than the Sun $f_{\mrm{tor}}$ might be an effective cycle tracer. However, it is unclear whether more active stars will follow solar-like trends.

ZDI recovers very well the poloidal fraction within 5\% on average but the poloidal fraction shows only little variations with time similar to $f_{\mrm{tor}}$ as $f_{\mrm{pol}} = 1-f_{\mrm{tor}}$. This prevents the clear detection of cycle phases or even a cyclic behaviour at all.

\subsection{ZDI uncovers traces of the underlying dynamo}
\label{SubSec:DynamoHints}

%
%
%
\begin{figure}
\centering
\includegraphics[angle=0,width = 0.5\textwidth ,trim = {5 55 5 15}, clip]{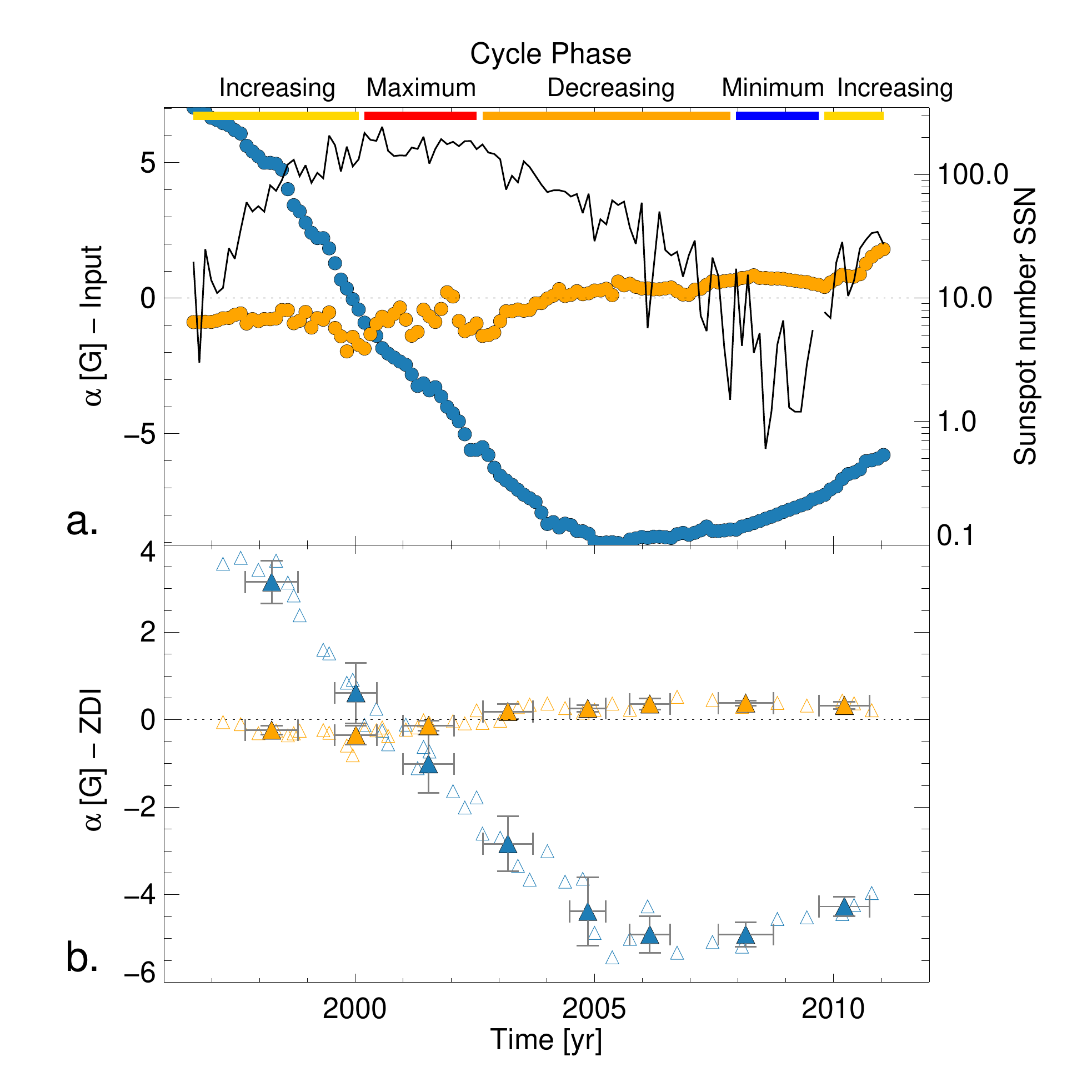}
\includegraphics[angle=0,width = 0.5\textwidth ,trim = {5 10 5 54}, clip]{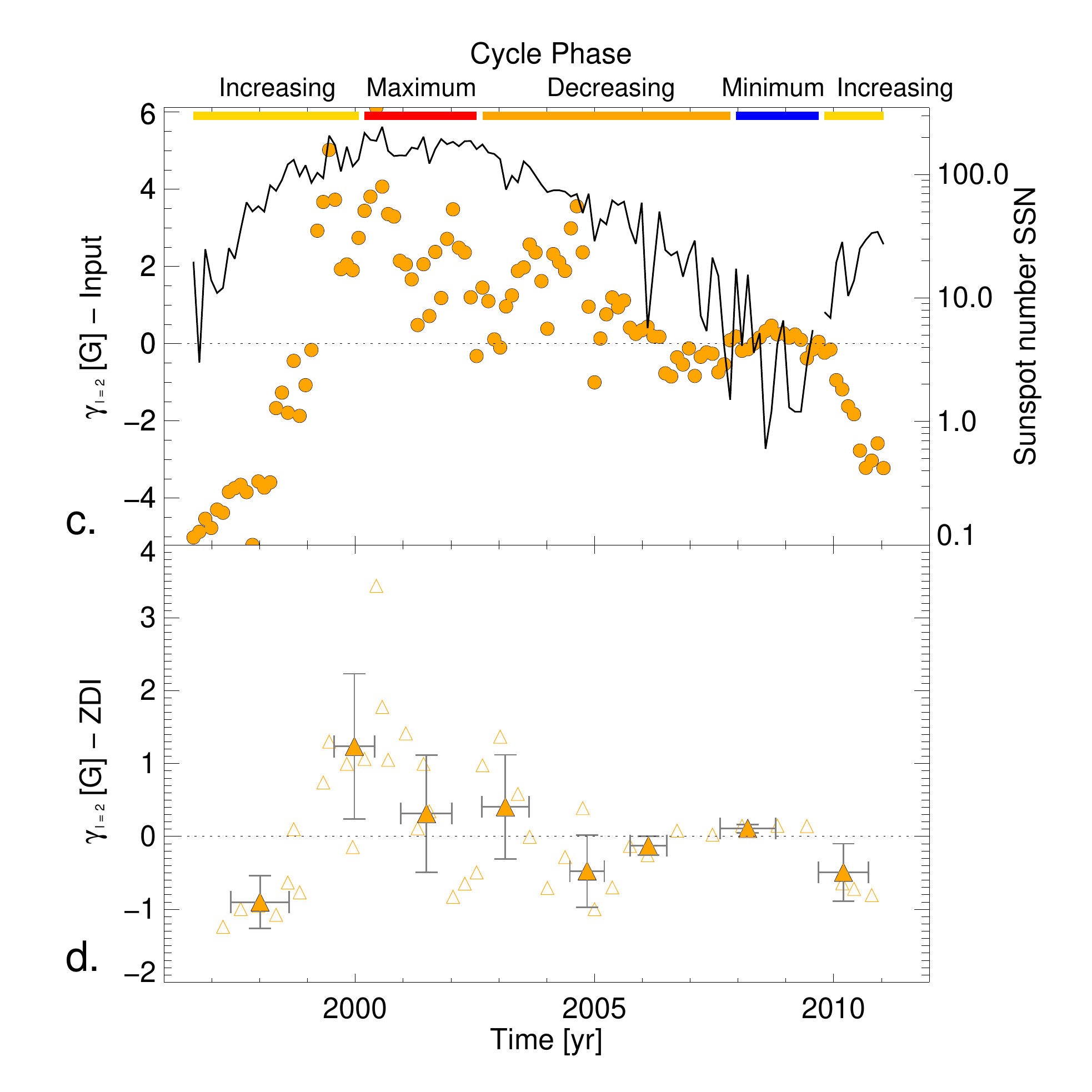}
\caption{The evolution over time of selected magnetic field coefficients for the simulated input maps (\textbf{a.,c.}) and their ZDI reconstructions (\textbf{b.,d.}) using the same format as in Fig.~\ref{Fig:EtotvsTime}. \textbf{a.,b.}: for the $\alpha_{\ell=1,m=0}$ (blue) and $\alpha_{\ell=2,m=0}$ coefficient (orange) tracing the radial poloidal axisymmetric dipole and quadrupole mode; \textbf{c.,d.} for the  $\gamma_{\ell = 2, m=0}$ coefficient tracing the toroidal axisymmetric quadrupole mode.}
\label{Fig:AlphaGammavsTime}
\end{figure}

To uncover traces of the underlying operating dynamo, we recommend the analysis of  specific $\alpha$ and $\gamma$ coefficients of Eq.~\ref{Eq:B_rad}-\ref{Eq:B_mer}. Figure~\ref{Fig:AlphaGammavsTime} shows the  evolution over time of $\alpha_{\ell=1,m=0}$ (blue) and $\alpha_{\ell=2,m=0}$ (orange) (Fig.~\ref{Fig:AlphaGammavsTime}a, b) and of $\gamma_{\ell = 2, m=0}$ (Fig.~\ref{Fig:AlphaGammavsTime}c, d). The $\alpha_{\ell=1,m=0}$ traces the radial poloidal axisymmetric dipole mode of the star. Its evolution over time shows when the global dipole flips its polarity (during the activity maximum) and when the new dipole of opposite polarity is fully established and strongest (during the end of the decreasing and minimum phase). We discover that ZDI is extremely effective in recovering the trend of $\alpha_{\ell=1,m=0}$ obtaining only small errors beside the offset to lower values. ZDI reconstructs less magnetic field for the global dipole in general as seen before. The relative evolution of the global poloidal dipole for the stellar dynamos can be therefore very well studied with the help of ZDI.

The axisymmetric radial poloidal quadrupole $\alpha_{\ell=2,m=0}$ (orange symbols) shows a more complex trend. The poloidal quadrupole mode is expected to behave in anti-phase to the poloidal dipole mode, which is the case for the input maps. We see a wide spread during activity maximum and that $\alpha_{\ell=2,m=0}$ is unusually high during the last increasing phase. However, the polarity changes are still clearly seen. The observed $\alpha_{\ell=2,m=0}$ from ZDI follows the same general trend but the polarity flip appears earlier, presumably due to the larger spread in the input maps. It is challenging to  derive conclusions about the cycle from the poloidal quadrupole based on ZDI observations but stronger trends may be observed in  more active, rapidly rotating stars.

Further information about the non-radial poloidal field can be uncovered by analysing the $\beta$ coefficients, see Eq.~\ref{Eq:B_pol}. As the ZDI-reconstructed meridional field is sometimes affected by crosstalk, the ZDI-reconstructed $\beta$ coefficients are similarly affected, predominantly by the $\alpha$ coefficients. This prevents the use of the $\beta$ coefficient as a tracer. 

The toroidal field can be analysed by focusing on the $\gamma$ coefficients. The best recovered trend is for the $\gamma_{\ell=2,m=0}$, see Fig.~\ref{Fig:AlphaGammavsTime}c, d. This mode, although part of the large-scale field, is affected by  small-scale flux emergence with opposite polarities at mid-latitudes spanning the equator.  The polarity flip of  $\gamma_{\ell=2,m=0}$ between the different cycles is clearly seen but occurs differently between cycles SC22 and SC23 around 1998, and between SC23 and SC24 in 2009. The $\gamma_{\ell=2,m=0}$ coefficient changes rapidly from negative to positive  between 1996-2000. We think this may be due to the fast transition between these two solar cycles. We still see active regions from the old SC22 cycle at low latitudes, while  the first spot group of opposite polarity occurs at high latitudes. At a certain point the new sunspot polarity dominated and quickly changed the sign of the $\gamma_{\ell=2,m=0}$. As the latitudes of flux emergence decrease, the $\gamma_{\ell=2,m=0}$ coefficient gets weaker until it is near zero during the strong minimum between 2008-2010, when nearly no active regions emerged. After 2010, SC24 started with the emergence of active regions at high latitudes, showing the opposite polarity and resulted in the $\gamma_{\ell=2,m=0}$ coefficient becoming strongly negative. We think that the unusually strong minimum causes the different behaviour of $\gamma_{\ell=2,m=0}$ across the border between these two solar cycles.

ZDI recovers the polarity flips but the error bars become pretty large, so  a robust detection of this phenomenon is challenging. The solar toroidal field is relatively weak, so ZDI starts to struggle to reliably recover the trend with time. Nevertheless, it is still worth checking $\gamma_{\ell=2,m=0}$ for  any trends but keeping in mind the effect may be subtle.

The dipolar toroidal mode $\gamma_{\ell=1,m=0}$ displays a wide spread as soon as sunspots emerge for the simulations (not shown). The sign of $\gamma_{\ell=1,m=0}$ swaps on individual maps, depending on which hemisphere  dominates the magnetic flux at the time. The behaviour of $\gamma_{\ell=1,m=0}$ is therefore highly dependent on the actual small scale flux emergence. ZDI is not able to recover this as it  is limited by the inclination. However, more importantly, ZDI is blind to single flux emergence events for slow rotating stars. The $\gamma_{\ell=1,m=0}$ recovered by ZDI is dominated by the behaviour of the $\gamma_{\ell=2,m=0}$, reflecting the global properties of the small scale flux emergence (the active latitudes). An illustration of the toroidal field given by the $\gamma_{\ell=1,m=0}$ and $\gamma_{\ell=2,m=0}$ can be found in \cite[fig.~1]{Lehmann2018}. 

We want to highlight once again, that for low $v_e \sin i$, slowly rotating, low activity solar-like stars, ZDI sees the large-scale field, e.g., the dipolar field and the large-scale reflections or properties of the small-scale flux emergence like the active latitudes reflecting the general latitude of flux emergence. ZDI is able to recover most features of the $\ellsum\ = 3$ restricted stellar magnetic field topology. Only features in the large-scale field that are reflections of single small-scale events, e.g. the sign of $\gamma_{\ell=1,m=0}$, are not recovered by ZDI.  As noted earlier, ZDI maps consistently recover lower magnetic field energies in comparison to  the simulated input $\ellsum\ = 3$  stellar magnetic field topology.

For the lowest order $\ell$-modes the non-axisymmetric modes $m\neq 0$ can also be  recovered by ZDI sometimes, e.g., $\alpha_{\ell = 1,m=1}, \gamma_{\ell = 1,m=1}, \alpha_{\ell = 2,m=1}, \alpha_{\ell = 2,m=2}, \beta_{\ell=2,m=2}, \beta_{\ell=3,m=1}$. However, the trends over time do not show cyclic behaviour at the first glance. They tend to show more variability during the activity maximum than during the activity minimum. For slowly rotating low activity stars like the Sun, ZDI is unlikely to recover strong trends in the $m \neq 0$ components. 

%
%
%
\begin{figure}
\centering
\includegraphics[angle=0,width = 0.5\textwidth]{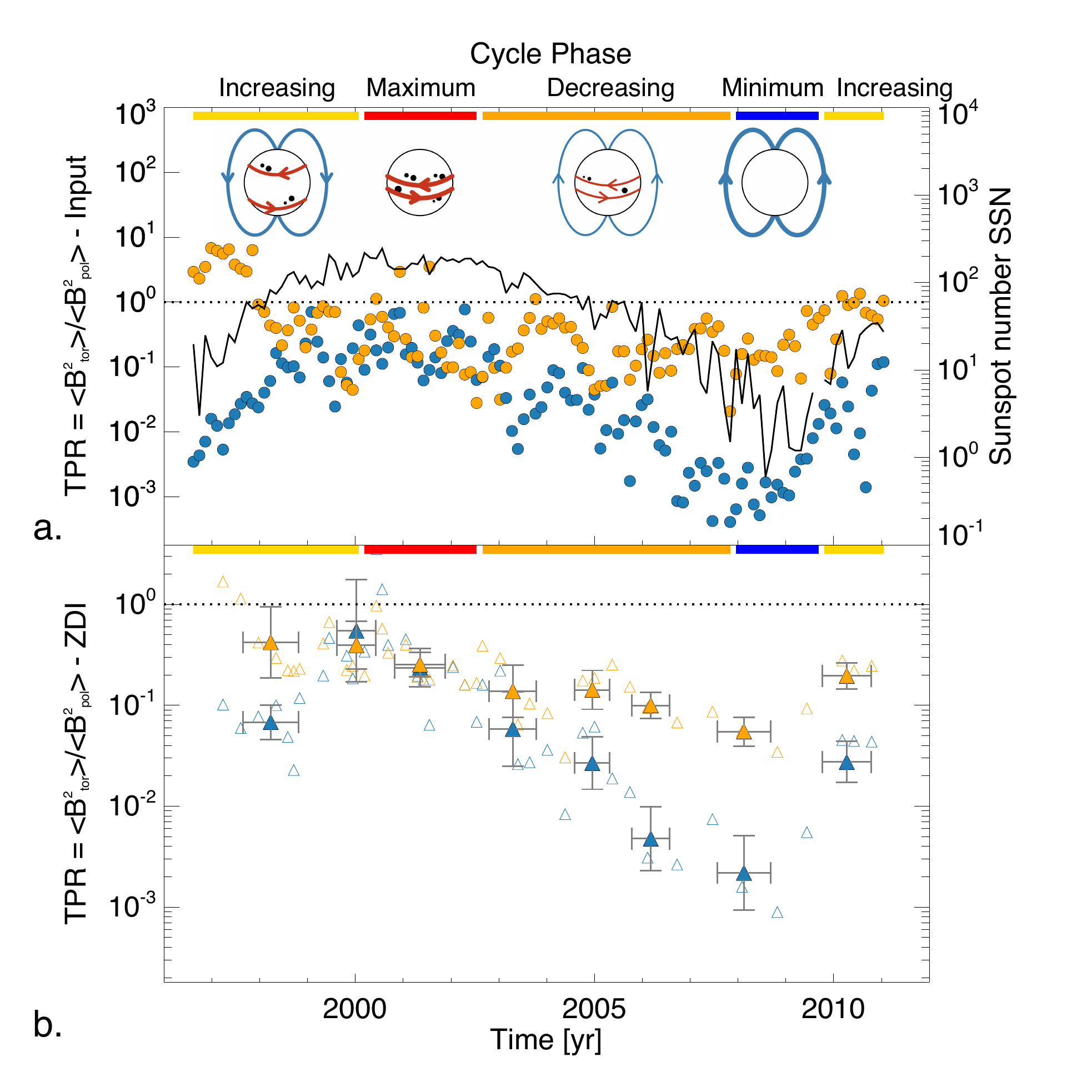}
\caption{The ratio of the toroidal and poloidal magnetic energy $ \mrm{TPR} = \langle B^2_{\mrm{tor}}\rangle / \langle B^2_{\mrm{pol}}\rangle$ for the dipolar $\ell = 1$ and quadrupolar $\ell = 2$ mode. The four little inserts at the top illustrate how the poloidal dipole (blue) and toroidal quadrupole modes (red) vary along the solar cycle. The quadrupole mode responds to the global properties of the small-scale flux emergence (black dots). The same format as in Fig.~\ref{Fig:EtotvsTime} is used but presenting only the first two single $\ell$-modes.}
\label{Fig:EtorSingEpolSingvsTime}
\end{figure}

Tracing the evolution of the toroidal field is challenging in solar-like stars. The ratio between the toroidal and poloidal magnetic energy, 
\begin{equation}
\mrm{TPR} = \frac{\langle B^2_{\mrm{tor}}\rangle}{\langle B^2_{\mrm{pol}}\rangle},
\label{Eq:TPR}
\end{equation} 
is a good alternative to the $\gamma$ coefficient and gives us important information about the interaction between the large-scale and small-scale field distributions for the specific case of our Sun and similarly active stars. Figure~\ref{Fig:EtorSingEpolSingvsTime} shows the toroidal : poloidal magnetic energy ratio (TPR) for the single $\ell$-modes $\ell = 1$ and 2 using the same format as Fig.~\ref{Fig:EtotvsTime}. The dashed line indicates when the toroidal energy dominates over the poloidal. The four little inserts at the top of Fig.~\ref{Fig:EtorSingEpolSingvsTime} sketch the behaviour of the dipole (blue arrows) and of the global properties of the small-scale flux emergence (red arrows\footnote{The solar small-scale flux (sunspots or active regions in general) emerges at mid latitudes with opposite polarity across equator and is therefore captured by the toroidal quadrupolar mode in the large-scale field \citep{Lehmann2019}.}) during the four phases of the solar cycle. 

The $\mrm{TPR}_{\ell=2}$ (orange symbols) becomes strongly, even dominantly, toroidal during the increasing phase, as the sunspots emerge at mid latitude range strengthening the toroidal $\ell = 2$ mode, see Fig.~\ref{Fig:EtorSingEpolSingvsTime}a. The latitude of spot emergence decreases around the maximum activity phase and hence $\mrm{TPR}_{\ell=2}$ decreases as well staying more or less constant for the rest of the cycle. ZDI can reproduce some of the toroidal dominated $\mrm{TPR}_{\ell=2}$ maps but the statistical analysis shows that it will be hard to catch the maximum for $\mrm{TPR}_{\ell=2}$ during the increasing phase, see Fig.~\ref{Fig:EtorSingEpolSingvsTime}b. Nevertheless, a strong $\mrm{TPR}_{\ell=2}$ can be a hint for mid latitudinal spot emergence for other cool stars that have a solar flux emergence pattern.

The $\mrm{TPR}_{\ell=1}$ (blue symbols) follows beautifully the trend of SSN but is driven from the poloidal dipolar field evolution, see Fig.~\ref{Fig:EtorSingEpolSingvsTime}a. The $\mrm{TPR}_{\ell=1}$ is maximum during the activity maximum, as the poloidal $\ell=1$ is weakest, and minimum during the activity minimum, while the poloidal dipole is strongest. ZDI is well recovering this trend, see Fig.~\ref{Fig:EtorSingEpolSingvsTime}b.
The unusual early and rapid increase of the $\mrm{TPR}_{\ell=1}$ for the beginning of SC24 could be related to the long activity minimum 2008--2010. \cite{Janardhan2018} reported an asymmetric polar field reversal that could be the reason for an unusually fast decreasing dipolar poloidal mode. To draw more reliable conclusions further simulations of the increasing phase until the maximum phase and beyond for SC24 are needed.

Summing up: $\mrm{TPR}_{\ell=1}$ is dominated by the poloidal dipolar energy but follows the SSN trend being clearly recoverable by ZDI. A strong $\mrm{TPR}_{\ell=2}$ detected by ZDI can indicate a mid latitude spot emergence for stars with solar flux emergence profile.

\section{Conclusion and Summary}
\label{Sec:ConAndSum}

To detect a solar-like cycle using ZDI is challenging especially for slowly rotating, low activity stars similar to the Sun but likely to be possible. One of the biggest challenges, next to the low $v_e \sin i$ and low activity of the targets, would be to get a sufficient number of observations covering the long time range of a solar-like activity cycle with high enough S/N and good phase sampling. Our study focusses on very low $v_e \sin i$, low activity, mature G~dwarf targets similar to our Sun. In most cases, stars observed with ZDI, show higher $v_e \sin i$ values and/or are more active stars. For these stars the situation is likely to be different, so that some of the limitations we see for solar twins, e.g. the limitation in spatial resolution and the offset in magnetic energy, might not hold for higher $v_e \sin i$ stars.

As noted in the introduction, applying ZDI to younger more active rapidly rotating stars has the effect of increasing spatial resolution in the resulting maps. In addition, it appears that these younger more active systems tend not to show activity cycles that exactly correspond to the type seen on the Sun \citep{Olah2016}. Our simulations do not currently reach the activity levels corresponding to these stars, see \cite{Lehmann2019}, but a future study will investigate the impact of broader $v_e \sin i$ values on the ability to reconstruct more small scale field with similar simulations to those presented here.

Our study finds that the total magnetic energy and the surface-averaged large-scale field are not able to trace solar-like activity cycles for slow rotating mature G~dwarfs. There are two main reasons that a higher magnetic activity level can not be seen in these two parameters. First, the likelihood of cancellation increases for magnetic flux regions of opposite polarity within each resolution element, as the number of emerging flux events increases with an increasing activity. Second, the large-scale field of the Sun is also significantly driven by the dipole component, which is at its lowest at activity maximum. A weak dipole would therefore decrease the detectable large-scale solar field and magnetic energy during the activity maximum.

To identify solar-like cycle with ZDI, the analysis of the fraction of axisymmetric energy is the most promising tracer. This parameter is well recovered by ZDI for slowly rotating solar-like stars and shows a clear trend over time ranging from nearly 0\% to 100\%. Furthermore, the analysis of the axisymmetric and non-axisymmetric magnetic energy over time is also promising, as both parameters show well distinguished and anti-phase trends. We recommend to focus more on the fractions than on the actual energy values, as they are likely to be more robust, and are affected less or not at all by the offset to lower magnetic energies reported for ZDI detections of solar-like stars.

To additionally confirm the detection of an activity maximum, we advise checking if the maximum of the toroidal field fraction $f_{\mrm{tor}}$ appears at the same time as a maximum in non-axisymmetry, although it can be challenging to detect reliable trends in $f_{\mrm{tor}}$ for solar-like stars. A solar-like activity minimum can be additionally confirmed by seeing a sharp decrease in the toroidal energy, while the poloidal energy remains constant.

As S-indices can be derived from the same data that are used to reconstruct ZDI maps, we recommend checking if the axisymmetric fraction decreases with S-index while the toroidal energy increases. It would be particularly interesting to compare the correlations with S-index for several stellar cycles to uncover trends with different stellar parameters like age or rotation period.

The comparison of the dominant coefficients $\alpha_{\ell,m}, \beta_{\ell,m}, \gamma_{\ell,m}$, see Eq.~\ref{Eq:B_rad}-\ref{Eq:B_mer}, for different stars with magnetic cycles has the potential to become important for understanding the underlying dynamo. For the solar case, ZDI is able to recover the important role of the $\alpha_{\ell=1,m=0}$ and $\gamma_{\ell=2,m=0}$ coefficients, corresponding to the global poloidal dipole and the global properties of the emerging small-scale flux dominating the observable toroidal large-scale field.

We are at the threshold of a very interesting time for understanding stellar dynamos. The continued monitoring of cool stars with high-resolution spectropolarimetric instrumentation will enable us to find further stars with magnetic activity cycles, and by building up our understanding of stellar dynamos in general, to search for a second Sun.

In the following we summarise our main findings:
\begin{itemize}
\item The 3D non-potential surface magnetic field simulations of \cite{Yeates2012} reproduce the solar cycle 23 (SC23) very well and are in agreement with other observations of cool star magnetic topologies reconstructed using ZDI.
\item The amplitude of the Stokes~V profiles varies along the cycle, and is lowest around activity maximum due to flux cancellation. The ZDI-reconstructed maps are therefore most affected by noise during the phases with highest sunspot numbers (SSN) and S-index.
\item ZDI recovers approximately one order of magnitude less toroidal and poloidal magnetic energy in the large-scale field,  but does recover the same trends as the input simulations.
\item ZDI reconstructions are most accurate at $\mrm{SSN} < 100$. At higher SSN, ZDI tends to underestimate the poloidal and axisymmetric fraction up to 20\% on average.
\item With decreasing SSN, the large-scale toroidal magnetic energy decreases, while the large-scale poloidal energy shows a lower limit at $\mrm{SSN}<50$; most likely given by the global dipolar field. The large-scale toroidal field seems to be driven to a significant extent by small-scale magnetic field structures like active regions, as we find that the toroidal energy increases with SSN. We caution that in practice changes in the toroidal energy may show an even larger scatter. Differing phase coverage and data quality in maps derived over different epochs will affect the detection of toroidal energy especially for low activity, slowly rotating stars. However, the fractional energies are likely to be more robustly recovered over the full timescale of an activity cycle monitoring campaign. 
\item The surface averaged magnetic field $\langle B_{\ellsum\ = 5} \rangle$ recovered by ZDI is constant ($\langle B_{\ellsum\ = 5} \rangle = 0.8-2.4\,\mrm{G}$) and blind to any  trend in activity.
\item The toroidal energy $\langle B^2_{\mrm{tor}} \rangle$ increases with S-index and SSN, which might be detectable for solar-type stars.
\item The axisymmetric fraction $f_{\mrm{axi}}$ shows a linear decrease with S-index, see Eq.~\ref{Eq:LinLawfaxiSInd}. The non-axisymmetric energy $\langle B^2_{\mrm{nax}} \rangle$ displays a strong drop at S-index $<0.167$.
\item The best tracer for solar-like cycles is the fraction of the axisymmetric field or the evolution of the axisymmetric and non-axisymmetric magnetic energy over time. The total magnetic energy $\langle B^2_{\mrm{tot}} \rangle$ or the surface average field $\langle B \rangle$ is inappropriate to detect solar-like cycles or to estimate cycle phases. 
\item ZDI recovers well the trends in $\alpha_{\ell=1,m=0}$ and $\gamma_{\ell=2,m=0}$ coefficients, which trace the evolution of the global poloidal dipole over time as well as  the global properties of the emerging small-scale flux (active latitudes). These therefore provide hints about the acting solar dynamo.
\end{itemize}

\section*{Acknowledgements}
We would like to thank Jean-Fran\c{c}ois Donati and Victor See for fruitful discussions that improved this paper and helped shape its final form. 
We also thank the referee Colin Folsom for the very constructive comments which have improved this paper.
LTL acknowledges support from the Scottish Universities Physics Alliance (SUPA) prize studentship and the University of St Andrews Higgs studentship. LTL and GH would like to thank the ESO Directorate for Science, which provided funds to support this collaboration. GH also warmly thanks the IDEX initiative at Universit\'e Federale Toulouse Midi-Pyrenees (UFTMiP) for funding the STEPS collaboration program between IRAP/OMP and ESO.
AAV acknowledges funding from the European Research Council (ERC) under the European Union's Horizon 2020 research and innovation programme (grant agreement No 817540, ASTROFLOW).
MMJ acknowledges the support of the Science \& Technology Facilities Council (STFC) (ST/M001296/1).
DHM would like to thank both the UK STFC and the ERC (Synergy Grant: WHOLE SUN, Grant Agreement No. 810218) for financial support.

\section*{Data availability}
The data underlying this article are available under ``Chapter5.zip'', at https://doi.org/10.17630/ae078167-03ea-4af6-9055-c3147e13c286 as part of LTL's PhD thesis data.




\bibliographystyle{mnras}
\bibliography{Lehmann_2020_ZDISolarCycle_Arxiv} 



\appendix

\section{The $\ell$-mode discussion}
\label{App:lmode}

%
%
%
\begin{figure*}
\centering
\includegraphics[height=4.9cm, trim = {0 20 0 0}, clip]{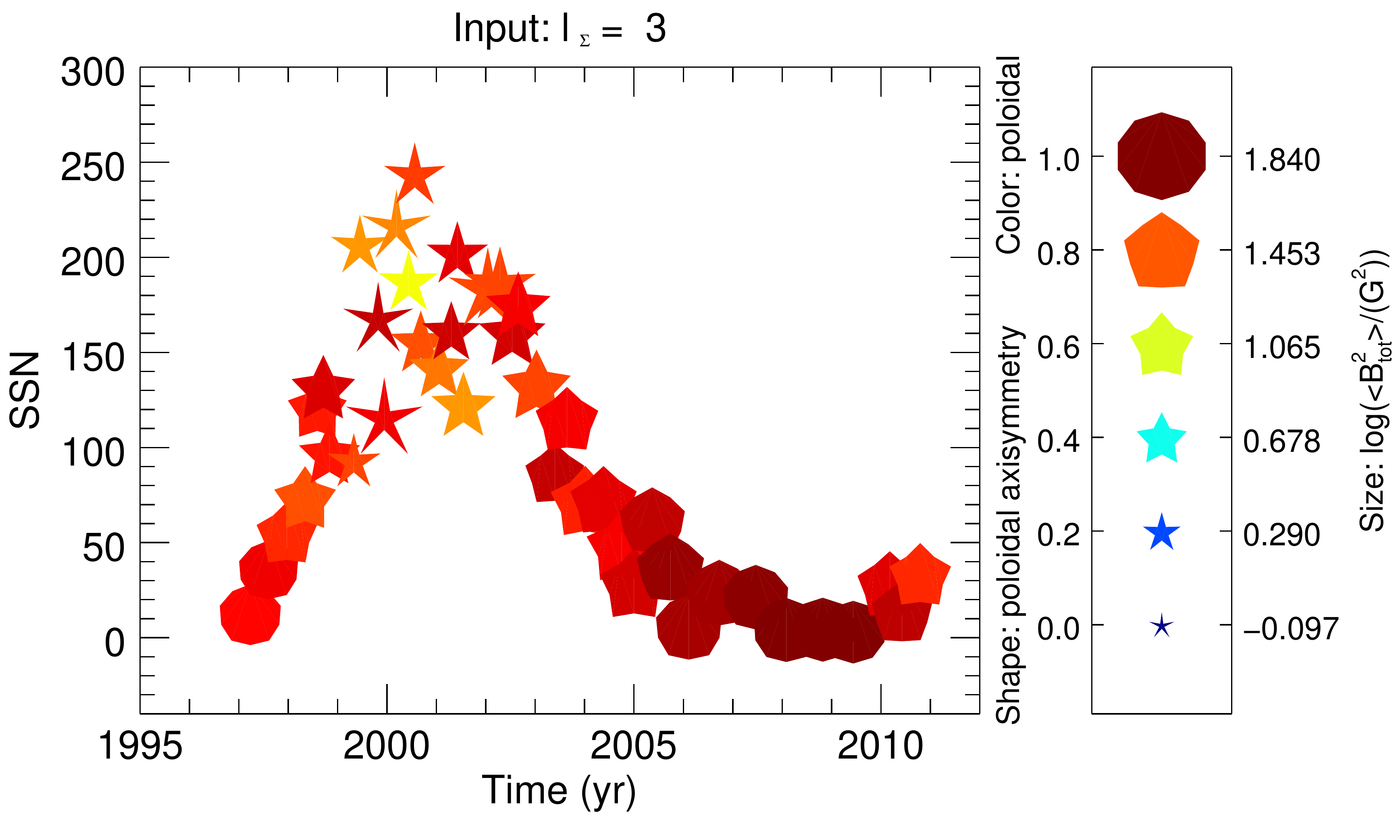}
\includegraphics[height=4.9cm, trim = {0 20 0 0}, clip]{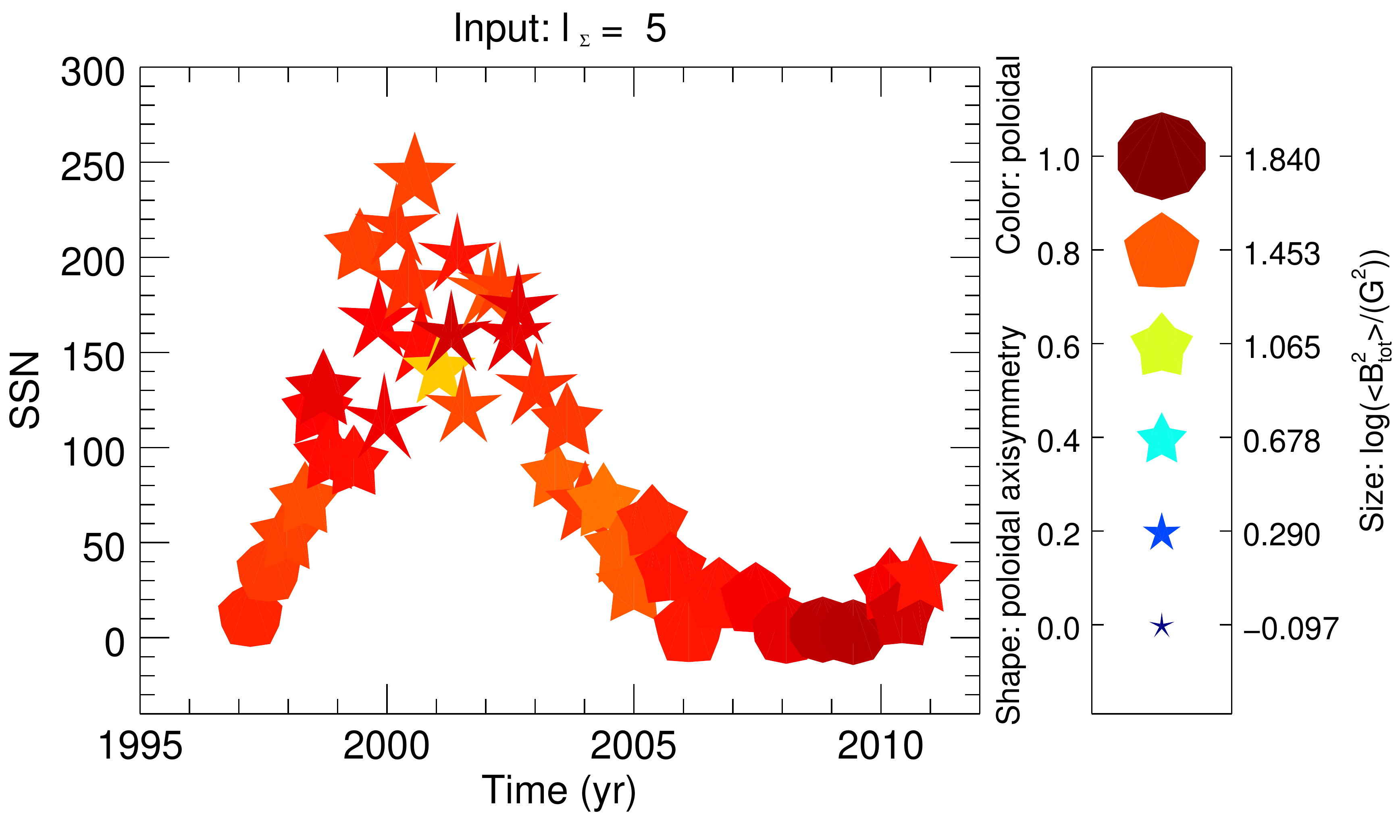}\\
\raggedright
\textbf{a.} \hspace{60ex} \textbf{b.}\\
\centering
\includegraphics[height=4.9cm, trim = {0 20 0 0}, clip]{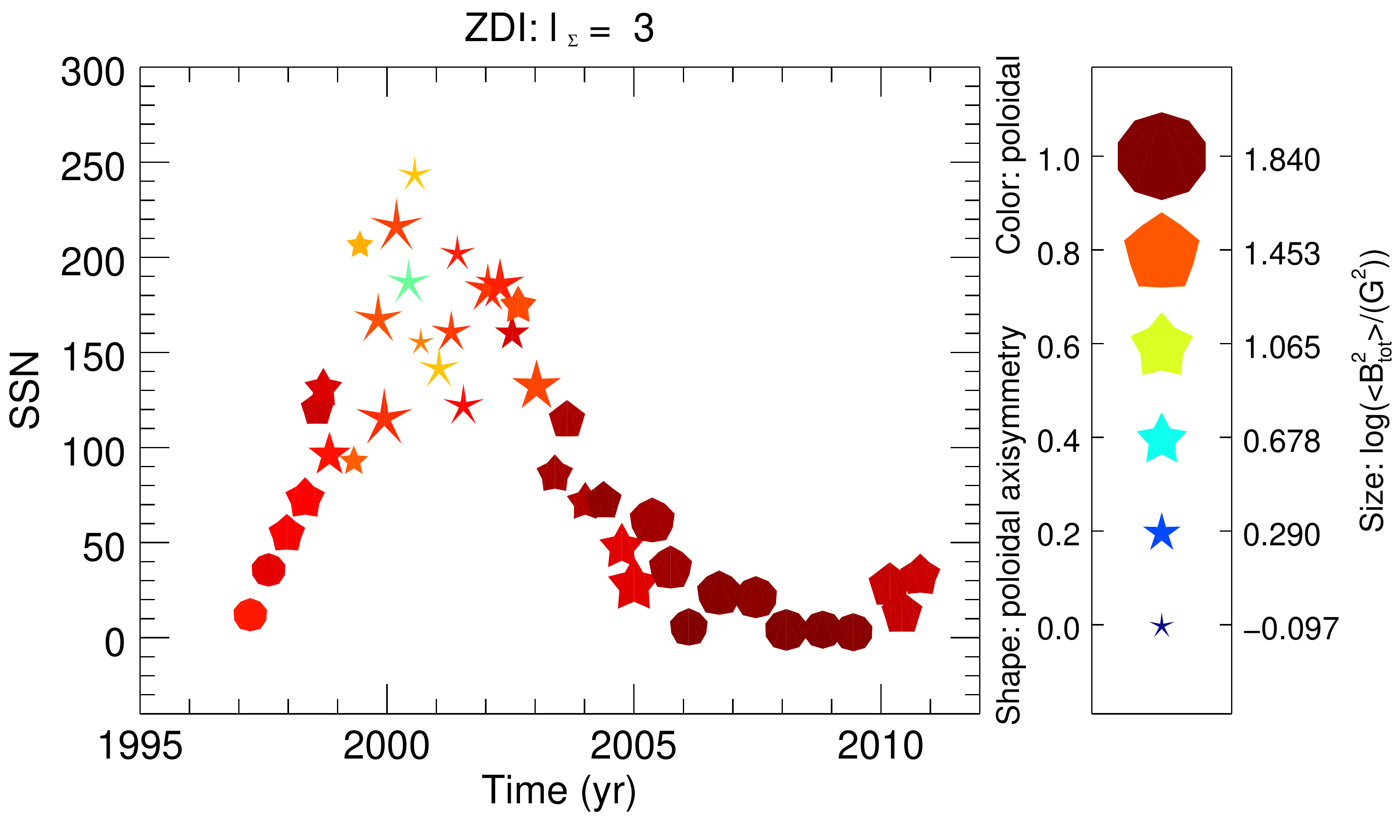}
\includegraphics[height=4.9cm, trim = {0 20 0 0}, clip]{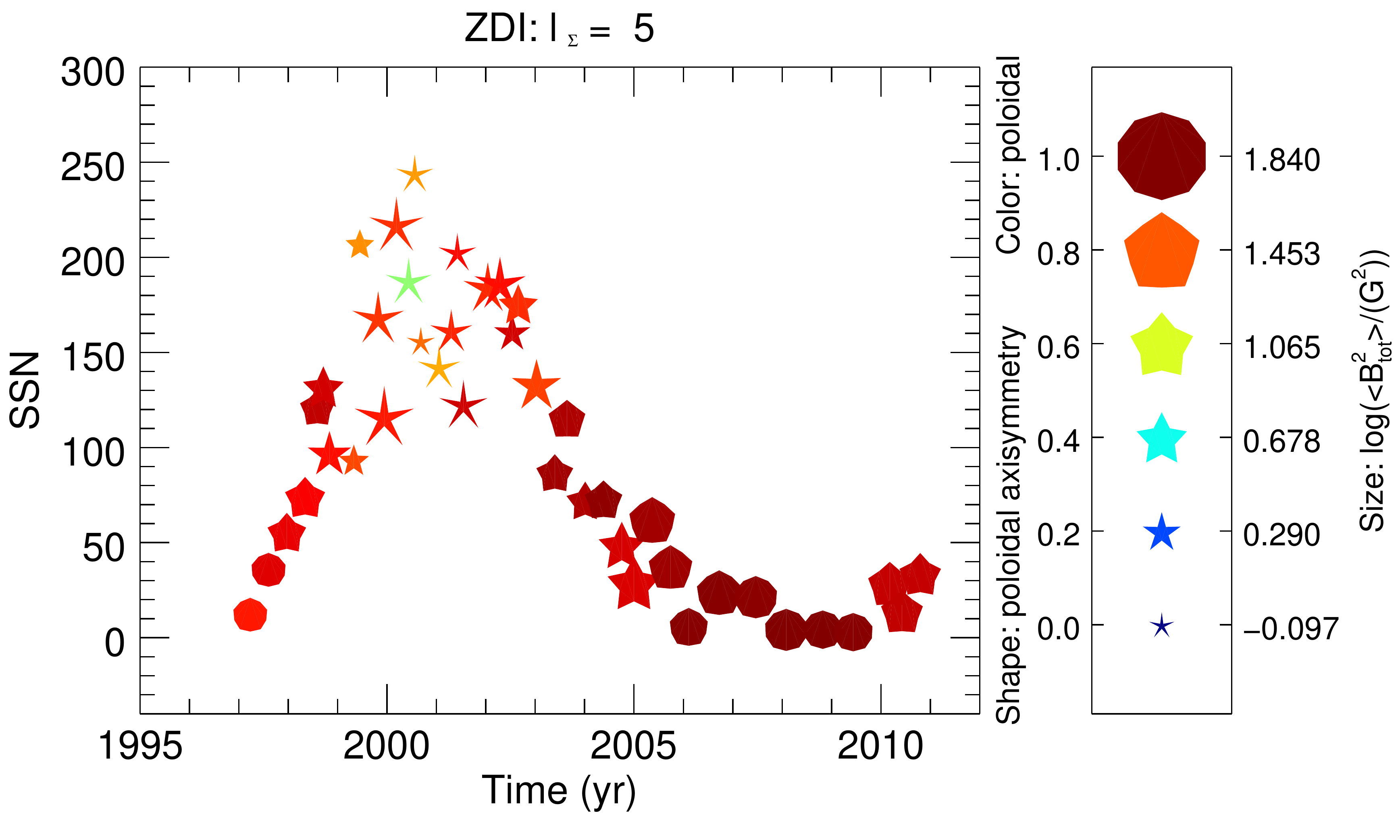}\\
\raggedright
\textbf{c.} \hspace{60ex} \textbf{d.}\\
\caption{The large-scale magnetic field topology for the 41 simulated input (\textbf{a.,b.}) and their ZDI reconstructed maps (\textbf{c.,d.}) for $\ellsum\ = 3$ (\textbf{a.,c.}) corresponding to the cumulative $\ellsum\ $-mode that provides the best match between input simulations and ZDI reconstructions and for $\ellsum\ = 5$ (\textbf{b.,d.}) corresponding to the full resolution of the ZDI maps using the same format as in Fig.~\ref{Fig:ConfusogramInpZDI}.}
\label{Fig:Confusogram_lmode}
\end{figure*}

In this paper and our previous study, we have discovered that the large-scale magnetic field topology recovered by ZDI is most similar to the input simulations for $\ellsum\ = 3$. However, ZDI also recovers magnetic energy in higher modes, $\ell = 4$ and 5, so the full resolution of the reconstructed ZDI maps is higher than $\ell = 3$. By restricting ZDI maps to $\ell = 3$ one should be aware that the magnetic energy values and also the fractions recovered in these maps will be different compared to the fully resolved input map. Nevertheless, for a fair order-of-magnitude comparison one should always compare input and reconstructed maps using the same $\ellsum\ $--modes.
We also strongly suggest to only compare the magnetic field properties for different stars if the ZDI maps include or are restricted to the same number of $\ellsum\ $--modes, which is sadly often not the case in the literature. The literature tends to use the full ZDI maps to compare and to analyse the magnetic field properties of cool stars observations regardless of the different numbers of included $\ellsum\ $--modes. 
For our solar case the differences between $\ellsum\ = 3$ and $\ellsum\ = 5$ are easily noticeable for the input simulations, see Fig.~\ref{Fig:Confusogram_lmode}a,b, but very small for the ZDI reconstructed maps, see Fig.~\ref{Fig:Confusogram_lmode}c,d. For faster rotating or more active stars this difference can be larger. For high SSN, we see small changes in $f_{\mrm{pol}}$ and $f_{\mrm{axi, pol}}$ for the ZDI maps by comparing Fig.~\ref{Fig:Confusogram_lmode}c with Fig.~\ref{Fig:Confusogram_lmode}d. However, the changes in the fractions are still relatively small compared to the spread of the sample in general.

%
%
%
\begin{figure}
\centering
\includegraphics[angle=0,width = 0.5\textwidth ,clip]{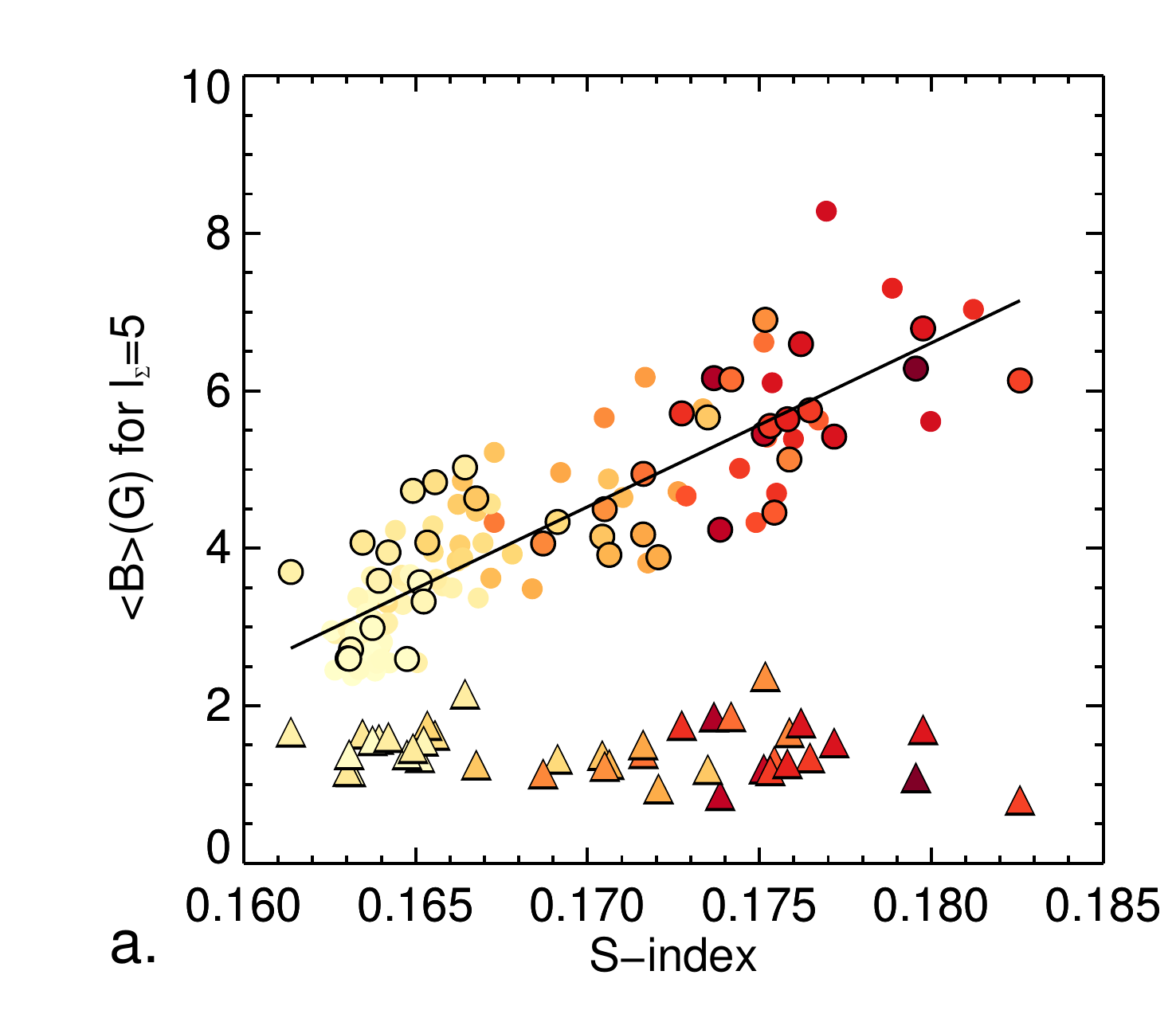}\\
\caption{The surface averaged large-scale magnetic field $\langle B \rangle$ including \lmod\ \ up to $\ellsum\ = 5$ plotted against S-index colour-coded by sunspot number using the same colour bar as in Fig.~\ref{Fig:Bl3vsSIndAndBl5vsSSN}a.}
\label{Fig:Bl5vsSInd}
\end{figure}

Figure~\ref{Fig:Bl5vsSInd} shows the average large-scale field against S-index similar to Fig.~\ref{Fig:Bl3vsSIndAndBl5vsSSN}a but for $\ellsum\ = 5$. The simulations (circles) show now a clear increase with S-index following the slope,
\begin{equation}
\langle B_{\ellsum\ =5} \rangle[\mrm{G}] = (-31\pm2)+ (208\pm12)\cdot \text{S-index} , 
\end{equation}
while the ZDI reconstruction still shows a flat distribution and only slightly higher values for $\langle B \rangle$ compared to $\ellsum\ = 3$.

\section{Sunspot number and S-index}
\label{App:SSNSInd}

%
%
%
\begin{figure}
\centering
\includegraphics[angle=0,width = 0.5\textwidth ,clip]{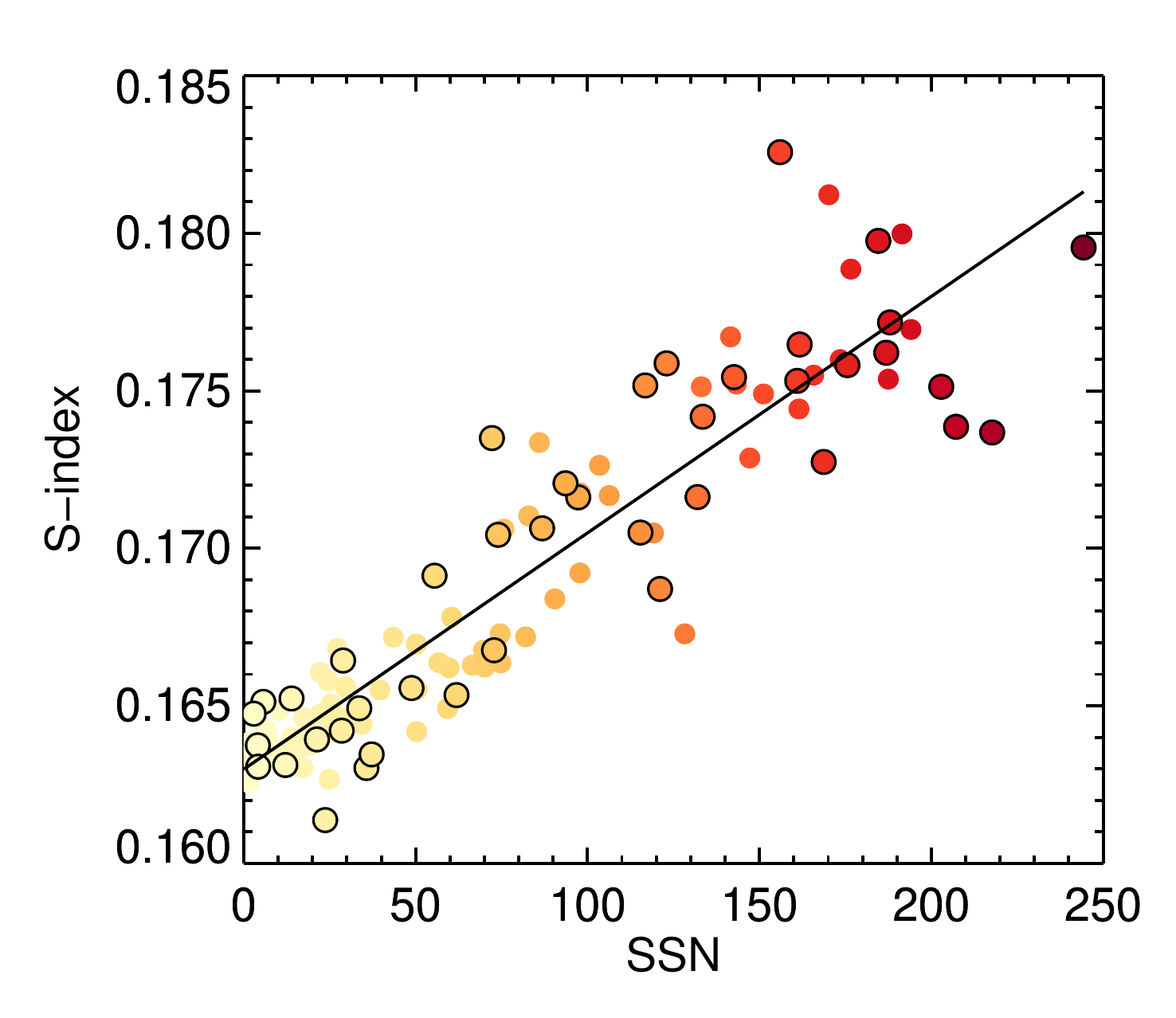}
\caption{The S-index against sunspot number for our data sample of all simulated maps using the same format as for Fig.~\ref{Fig:Bl3vsSIndAndBl5vsSSN}a. The solid line indicates the least-squared fit.}
\label{Fig:SSNvsSInd}
\end{figure}

We find that some of the magnetic field parameters show correlations with sunspot number (SSN). However, to date we can only measure the star spot numbers for the Sun. For other cool stars, the starspot number remains unknown, although there are first approaches to count star spots, e.g., via highly eccentric exoplanetary transits, which are nevertheless incomplete probes in latitude \citep{Llama2012,Morris2017}. 
To allow  comparison with other stars we use the S-index instead of SSN. The S-index describes a flux ratio for the Ca II H\& K line, see Eq.~\ref{Eq:SIndex}, and is a proxy for the chromospheric activity. The S-index is widely observable for  stars other than the Sun \citep{Baliunas1995, Frick2004,Lockwood2007}, and several stars show a periodic behaviour in S-index, e.g., by the Mt Wilson Survey \citep{Wilson1968}.
For the solar S-index, we used the dataset published by \cite{Egeland2017}. We averaged the S-index values over the 27-day rotation period per map and confirmed a linear correlation between S-index and SSN, see Fig.~\ref{Fig:SSNvsSInd},
\begin{equation}
\text{S-index} = (0.163\pm0.001) + (7.5 \pm 0.3) \cdot 10^{-5} \mrm{SSN}.
\end{equation}
The S-index is therefore a good alternative to the SSN as it is widely observable for  stars other than the Sun and due to its linear correlation with SSN. The spread in the S-index increases at higher SSN but the trends of many of the individual magnetic field parameters with SSN are still recoverable for S-index.

However, caution needs to be taken by searching for stellar cycles in young, rapidly rotating stars. The S-index measures the brightness of the chromosphere, while the SSN is related to the photosphere. We know that the photospheric brightness correlates with the chromospheric brightness for older stars like the Sun, but in younger stars these brightness variations are in anti-phase with S-index, \citep{Lockwood2007}. We would not be surprised if the correlation of, e.g., the toroidal energy with S-index, is reversed and shows anti-correlations in stellar activity cycles of younger stars.

%
%
%
\begin{figure}
\centering
\includegraphics[angle=0,width = 0.49\textwidth ,trim = {10 55 20 20},clip]{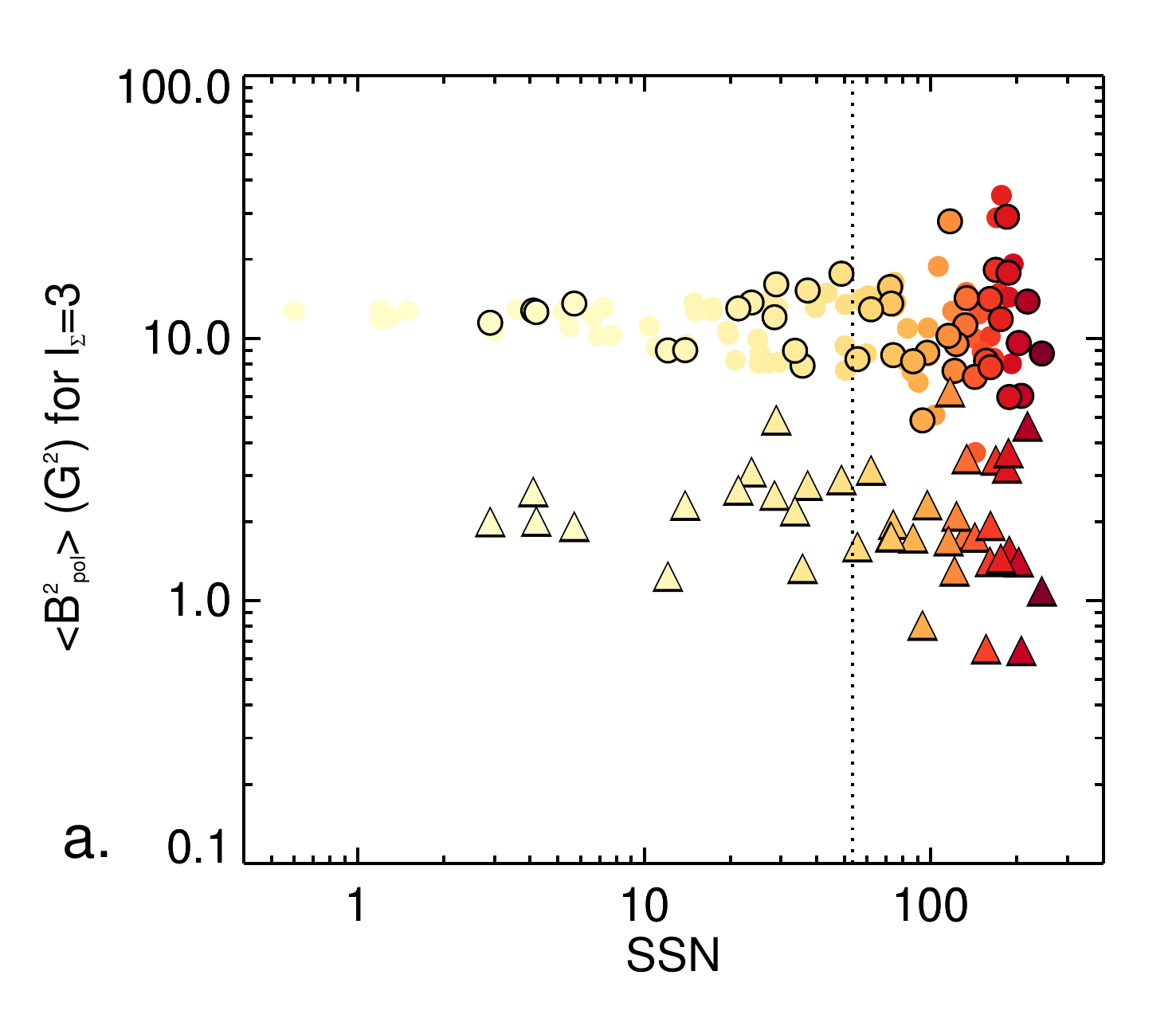}\\
\includegraphics[angle=0,width = 0.49\textwidth ,trim = {10 10 20 27},clip]{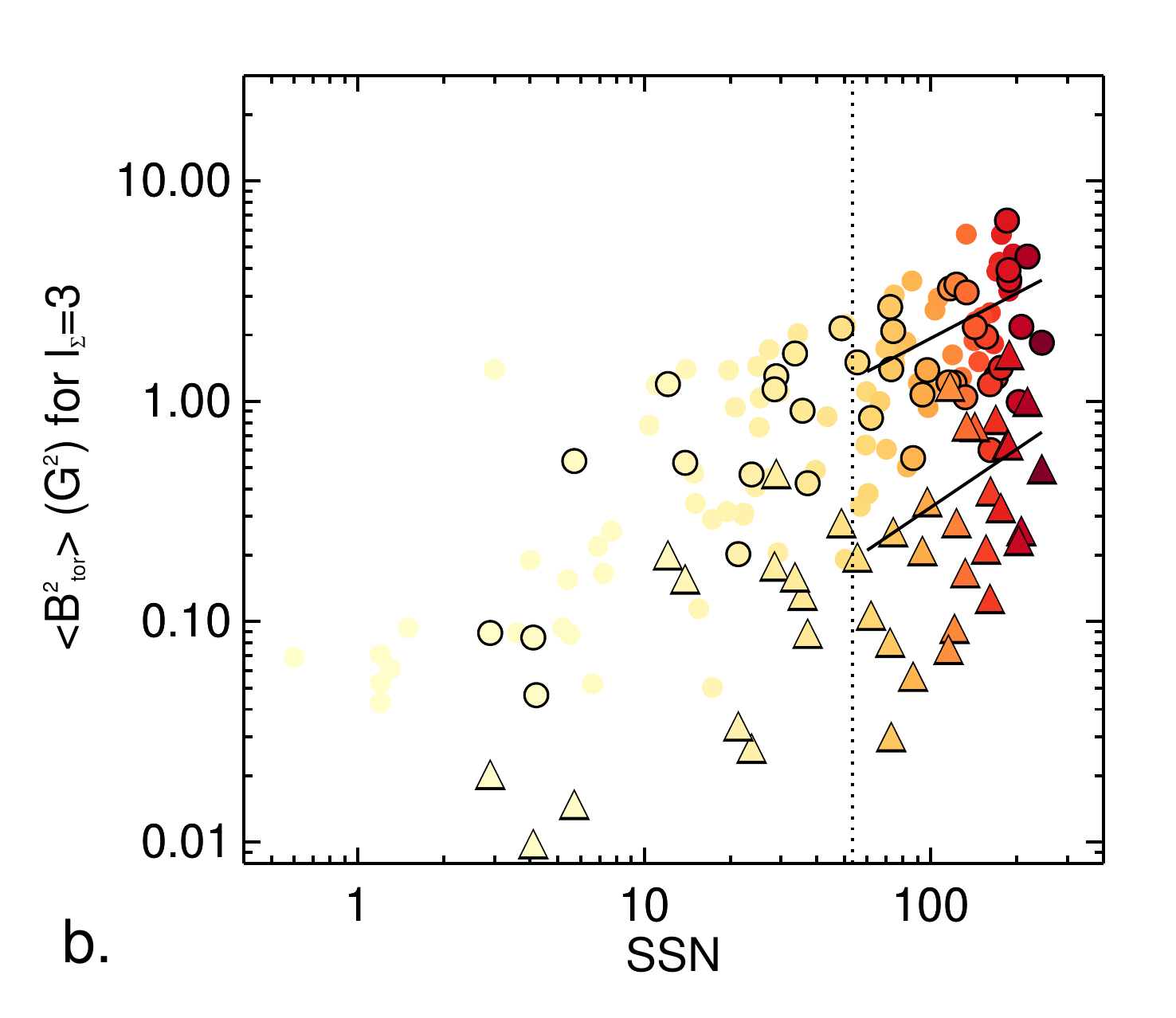}
\caption{The poloidal $\langle B^2_{\mrm{pol}} \rangle$(\textbf{a.}) and toroidal $\langle B^2_{\mrm{tor}} \rangle$ (\textbf{b.}) magnetic energy over sunspot number (SSN) for the cumulative $\ellsum\ = 3$ mode. The dashed line indicates $\mrm{SSN} = 50$ and the solid lines the least square fits for the toroidal energy for the simulated input maps and their ZDI reconstructions, respectively.}
\label{Fig:EpolEtorvsSSN}
\end{figure}

In addition to Fig.~\ref{Fig:EpolEtorvsSInd}, we show the poloidal and toroidal energy against SSN in Fig.~\ref{Fig:EpolEtorvsSSN}. The dashed line indicates a SSN$=50$. The poloidal energy is constant for SSN below 50, while  the spread increases strongly for higher SSN, see Fig.~\ref{Fig:EpolEtorvsSSN}a. The lower limit of the poloidal energy for low SSN is most likely related to the poloidal global dipolar field, which is strong for low SSN. The ZDI results (triangles) recover very well the trend shown in the input simulations  (circles), despite an offset by around one order of magnitude to lower energies.

For the toroidal energy we see an increasing relation between the toroidal energy and SSN, Fig.~\ref{Fig:EpolEtorvsSSN}b. The spread is less than for the corresponding Fig.~\ref{Fig:EpolEtorvsSInd}b with S-index. There is no change in behaviour for low SSN, only a steady increase with SSN, which highlights the driving role of the emerging small-scale field in sunspots for the large-scale toroidal energy. The spread is too large to derive secure conclusions, similar to $\langle B^2_{\mrm{tor}} \rangle$ over S-index, but we could confirm that ZDI recovers a power-law relation with SSN within the error.

\section{Additional Figures}
\label{App:AddFigures}

\begin{figure}
\centering
\includegraphics[angle=0,width = 0.49\textwidth ,trim = {10 10 20 27},clip]{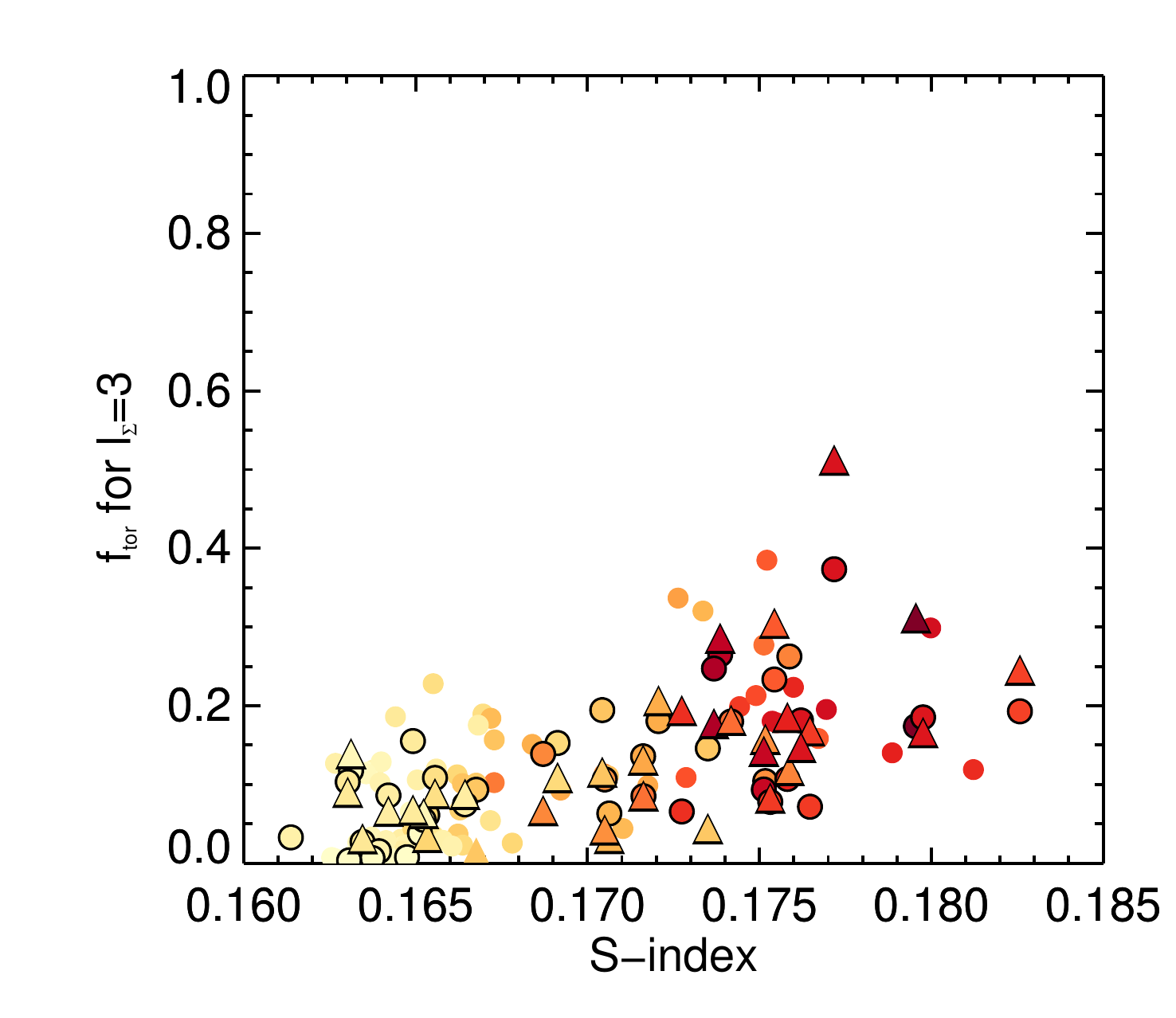}
\caption{The toroidal fraction $f_{\mrm{tor}}$ over S-index using the same format as for Fig.~\ref{Fig:Bl3vsSIndAndBl5vsSSN}a.}
\label{Fig:ftorvsSIndex}
\end{figure}

First, we show how the toroidal fraction $f_{\mrm{tor}}$ behaves with S-index, see Fig.~\ref{Fig:ftorvsSIndex}.

%
%
%

The toroidal fraction shallowly increases with S-index. The ZDI reconstructions (triangles) follow the same trend as the $\ellsum\ = 3$ restricted input simulations (circles). ZDI well recovers the $f_{\mrm{tor}}$ trend with S-index but $f_{\mrm{tor}}$ is still not a good choice to trace solar-like cycles. The $f_{\mrm{tor}}$ values are low for low-active stars like the Sun. The spread is large and the increase with S-index only very shallow. This prevents the identification of certain cycle phases. The toroidal fraction might be a cycle tracer for stars more active than the Sun, but it is not clear what trend $f_{\mrm{tor}}$ will show with S-index for these stars. For low activity stars similar to the Sun an increase in the $f_{\mrm{tor}}$ values and in their spread for high S-index could be seen as a hint of a solar-like cycle behaviour.

We further show here two figures showing the evolution of three further magnetic field parameters with time and its ZDI reconstruction using the same format as for Fig.~\ref{Fig:EtotvsTime}.

%
%
%
\begin{figure}
\centering
\includegraphics[angle=0,width = 0.5\textwidth, trim={0 0 0 0}, clip]{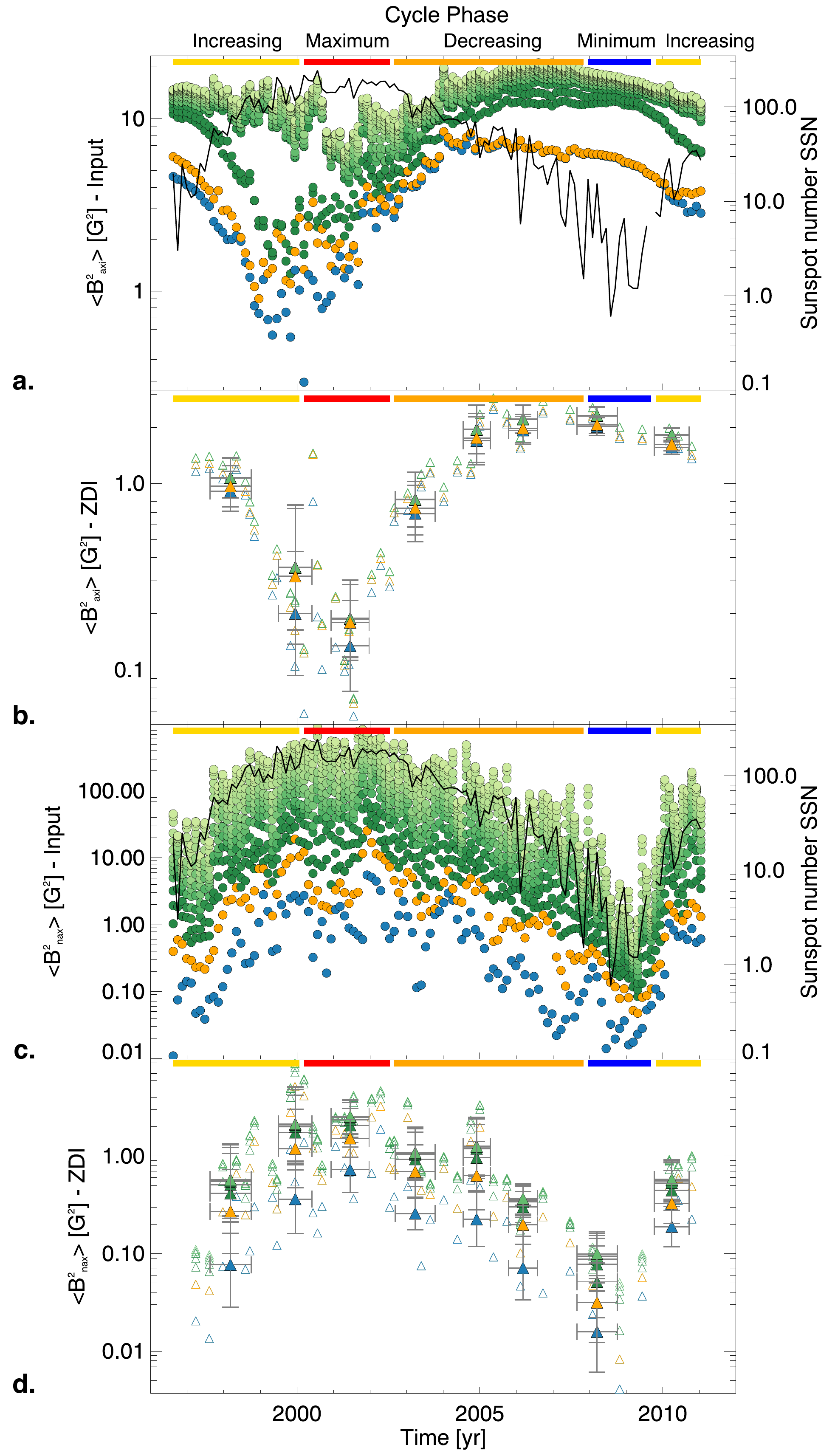}
\caption{The axisymmetric $\langle B^2_{\mrm{axi}} \rangle$ (\textbf{a.}, \textbf{b.}) and the non-axisymmetric $\langle B^2_{\mrm{nax}} \rangle$ magnetic energy (\textbf{c.}, \textbf{d.}) for all 118 simulations (\textbf{a.}, \textbf{c.}) and the 41 ZDI reconstructed maps (\textbf{b.}, \textbf{d.}). The same format as in Fig.~\ref{Fig:EtotvsTime} is used.}
\label{Fig:EaxiEnaxvsTime}
\end{figure}

As well as the evolution of the axisymmetric fraction, see Fig.~\ref{Fig:faxivsTime}, we find that another promising parameter is   the analysis of the axisymmetric  $\langle B^2_{\mrm{axi}} \rangle$ (Fig.~\ref{Fig:EaxiEnaxvsTime}a,b) and non-axisymmetric energy $\langle B^2_{\mrm{nax}} \rangle$ (Fig.~\ref{Fig:EaxiEnaxvsTime}c,d). 

For the axisymmetric energy $\langle B^2_{\mrm{axi}} \rangle$ the low \lmod\ \ of the simulations show a clear and sharp minimum during the activity maximum and a broad maximum at the end of the decreasing phase and during the minimum phase, see Fig.~\ref{Fig:EaxiEnaxvsTime}a. ZDI recovers this trend very well in the large-scale field, see Fig.~\ref{Fig:EaxiEnaxvsTime}b. 
The non-axisymmetric energy $\langle B^2_{\mrm{nax}} \rangle$ shows the opposite trend with a broader maximum around the activity maximum and sharp minimum during the minimum phase, see Fig.~\ref{Fig:EaxiEnaxvsTime}c. All \lmod\ \ of the simulations follow this trend and show also an excellent agreement with the SSN evolution, see Fig.~\ref{Fig:EaxiEnaxvsTime}c. The ZDI-reconstructed large-scale magnetic energy $\langle B^2_{\mrm{nax}} \rangle$ recovers this trend very well, see Fig.~\ref{Fig:EaxiEnaxvsTime}d. By determining and analysing the single magnetic field components of $\langle B^2_{\mrm{axi}} \rangle$ and $\langle B^2_{\mrm{nax}} \rangle$ for the ZDI-reconstructed maps, one receives a more than promising tracer for discovering solar-like cycles. Both parameters show easily distinguishable trends, which are well recovered by ZDI. This is supported by the smaller error bars in Fig.~\ref{Fig:EaxiEnaxvsTime}b,d compared, e.g., to the total energy in Fig.~\ref{Fig:EtotvsTime}b.

%
%
%
\begin{figure}
\centering
\includegraphics[angle=0,width = 0.5\textwidth]{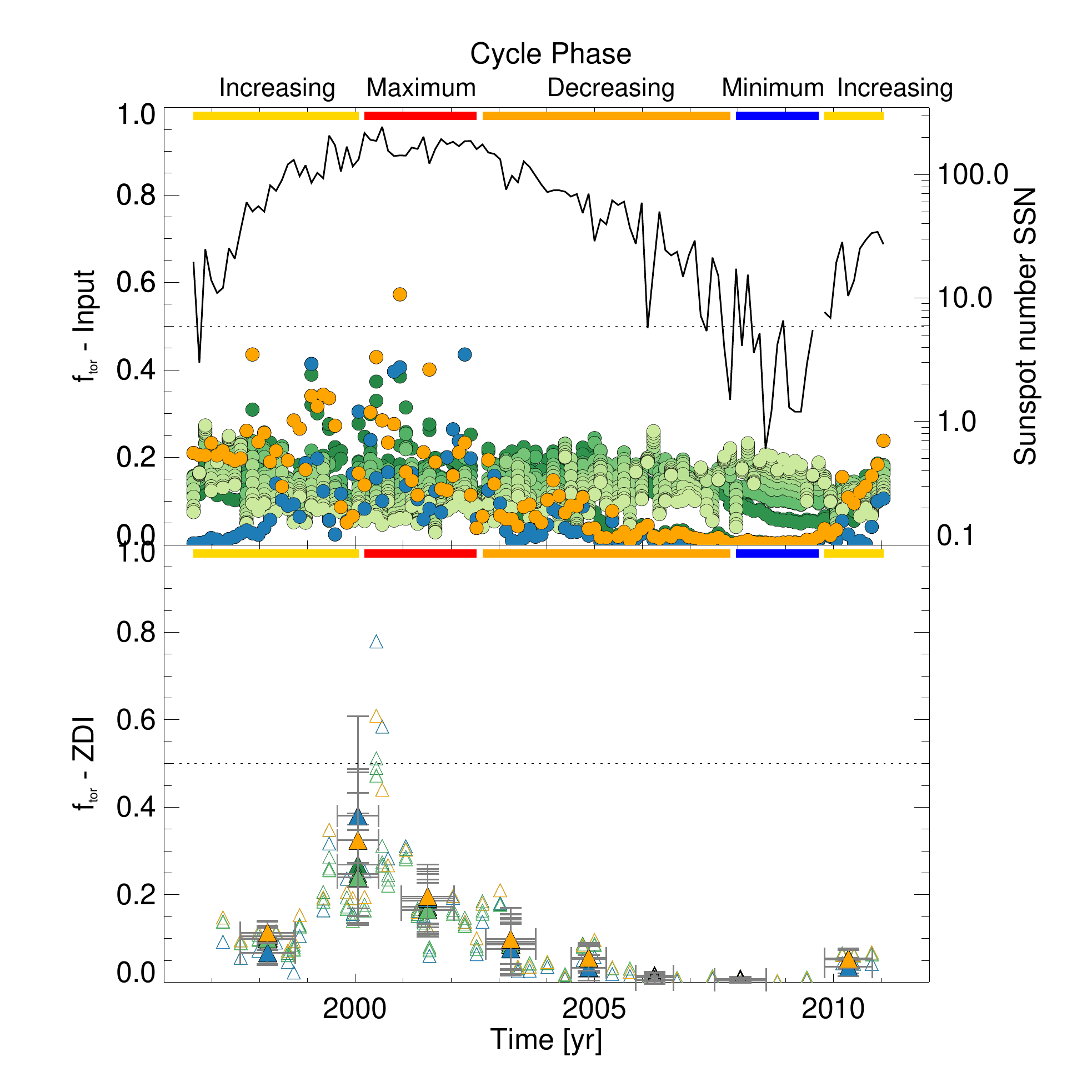}
\caption{The toroidal fraction $f_{\mrm{tor}}$ for the simulations and  ZDI reconstructions. The same format is used as in Fig.~\ref{Fig:EtotvsTime}.}
\label{Fig:ftorvsTime}
\end{figure}

As mentioned before, the energy fraction of the toroidal field $f_{\mrm{tor}}$ is not a suitable tracer for a solar-like cycle but can be used to confirm the detection of an activity maximum for slowly rotating solar-like active stars. Figure~\ref{Fig:ftorvsTime} displays the evolution with time using the same format as in Fig.~\ref{Fig:EtotvsTime}. The higher \lmod\ \ ($\ellsum\ > 3$) show a mostly flat distribution with time, while the dipolar mode (blue circles) and quadrupolar mode (orange circles) show a scattered peak around activity maximum for the simulated maps, see Fig.~\ref{Fig:ftorvsTime}a. The toroidal fraction is widely scattered from 0.05 to 0.6 for $\ellsum\ = 2$ during that time. The ZDI reconstructed maps with all $\ellsum\ \le 7$ modes follow the trend of the quadrupolar mode although showing less spread and partly lower values. Estimating the cycle phase is therefore difficult as the only significant feature is  the low amplitude peak around activity maximum which appears in a small time window. The Sun is too inactive to use $f_{\mrm{tor}}$ as reliable cycle tracer. Nevertheless the maximum of $f_{\mrm{tor}}$ should correlate with the activity maximum for solar-like cycles, which is worth checking by analysing stellar cycles.


\bsp	
\label{lastpage}
\end{document}